\documentclass[twocolumn,apj]{aastex63}
\pdfoutput=1
\usepackage{amsmath,amssymb,gensymb,color,graphicx,array,multirow,soul,rotating,bm}
\usepackage{natbib,multirow,placeins,textcomp,subfigure}
\usepackage[outline]{contour}
\usepackage[utf8]{inputenc}
\usepackage[T1]{fontenc}
\usepackage{booktabs}
\usepackage{microtype}
\usepackage{listings}
\usepackage{hyperref}
\usepackage{wrapfig}

\DeclareFixedFont{\ttb}{T1}{txtt}{bx}{n}{9} 
\DeclareFixedFont{\ttm}{T1}{txtt}{m}{n}{9}  

\newcommand{\dfn}[1]{\textbf{#1}}
\newcommand{\img}[2][]{\raisebox{-0.5\height}{\includegraphics[#1]{#2}}}
\newenvironment{closetabcols}[1][0.5mm]{\setlength{\tabcolsep}{#1}}{}
\newenvironment{closetabrows}[1][0mm]{}{}

\newcommand{\cskip}{\multicolumn{1}{c}{}}
\newcommand{\ukrts}{$\micro$K$\sqrt{\text{s}}$}
\newcommand{\ukam}{$\micro$K arcmin}
\newcommand{\LCDM}{$\Lambda$CDM}
\newcommand{\ttop}{T$\rightarrow$P}
\newcommand\fnsep{\textsuperscript{,}}

\begin{document}

\title{The Atacama Cosmology Telescope: DR6 Maps}

\keywords{}

\begin{abstract}
We present Atacama Cosmology Telescope (ACT) Data Release 6 (DR6)
maps of the Cosmic Microwave Background
temperature and polarization anisotropy at arcminute resolution
over three frequency bands centered on 98, 150 and 220 GHz. The maps
are based on data collected with the AdvancedACT camera
over the period 2017--2022 and cover 19,000 square degrees with a median
combined depth of 10 µK arcmin. We describe the instrument, mapmaking
and map properties and illustrate them with a number of figures and tables.
The ACT DR6 maps and derived products are available on LAMBDA at
\url{https://lambda.gsfc.nasa.gov/product/act/actadv_prod_table.html}.
We also provide an interactive web atlas at \url{https://phy-act1.princeton.edu/public/snaess/actpol/dr6/atlas} and HiPS data sets in Aladin (e.g. \url{https://alasky.cds.unistra.fr/ACT/DR4DR6/color_CMB}).
\clearpage
\end{abstract}

\author[0000-0002-4478-7111]{Sigurd~Naess} \affiliation{Institute of Theoretical Astrophysics, University of Oslo, Norway} \footnote{sigurdkn@astro.uio.no}
\author[0000-0002-1697-3080]{Yilun~Guan} \affiliation{Dunlap Institute for Astronomy and Astrophysics, University of Toronto, 50 St. George St., Toronto, ON M5S 3H4, Canada}
\author[0000-0003-2856-2382]{Adriaan~J.~Duivenvoorden} \affiliation{Max-Planck-Institut fur Astrophysik, Karl-Schwarzschild-Str. 1, 85748 Garching, Germany}
\author[0000-0002-2408-9201]{Matthew~Hasselfield} \affiliation{Flatiron Institute, 162 5th Avenue, New York, NY 10010 USA}
\author[0000-0002-8710-0914]{Yuhan~Wang} \affiliation{Department of Physics, Cornell University, Ithaca, NY, USA 14853}
\author[0000-0003-3230-4589]{Irene~Abril-Cabezas} \affiliation{DAMTP, Centre for Mathematical Sciences, University of Cambridge, Wilberforce Road, Cambridge CB3 OWA, UK} \affiliation{Kavli Institute for Cosmology Cambridge, Madingley Road, Cambridge CB3 0HA, UK}
\author[0000-0002-2147-2248]{Graeme~E.~Addison} \affiliation{Dept. of Physics and Astronomy, The Johns Hopkins University, 3400 N. Charles St., Baltimore, MD, USA 21218-2686}
\author[0000-0002-5127-0401]{Peter~A.~R.~Ade} \affiliation{School of Physics and Astronomy, Cardiff University, The Parade, Cardiff, Wales, UK CF24 3AA}
\author[0000-0002-1035-1854]{Simone~Aiola} \affiliation{Flatiron Institute, 162 5th Avenue, New York, NY 10010 USA} \affiliation{Joseph Henry Laboratories of Physics, Jadwin Hall, Princeton University, Princeton, NJ, USA 08544}
\author{Tommy~Alford} \affiliation{Department of Physics, University of Chicago, Chicago, IL 60637, USA}
\author[0000-0002-4598-9719]{David~Alonso} \affiliation{Department of Physics, University of Oxford, Keble Road, Oxford, UK OX1 3RH}
% ...

\date[]{\emph{Affiliations can be found at the end of the document}}
\keywords{}
\suppressAffiliations

\tableofcontents

\section{Introduction}

The cosmic microwave background (CMB) has been a key cosmological observable for the last three decades, and has been studied in increasing detail
by space-based \citep{cobe-dmr-1992,wmap_final_cosmology,planck_2018_overview}, ground-based \citep[e.g.,][]{bicep_bmode_2018,Koopman/2016,spt3g-2014} and balloon-borne telescopes \citep[e.g.,][]{boomerang-2002,spider-instrument-2018}.
The angular power spectrum of its anisotropies forms the early-universe anchor point for cosmological models and was critical in establishing the \LCDM{} paradigm \citep{spergel/etal/2003}. The gravitational lensing of the CMB probes the later growth of structure, as do the spectral distortions the CMB picks up
as it travels through hot intracluster gas in galaxy clusters on the way to us.

\begin{figure*}[htp]
	\centering
	\includegraphics[width=\textwidth]{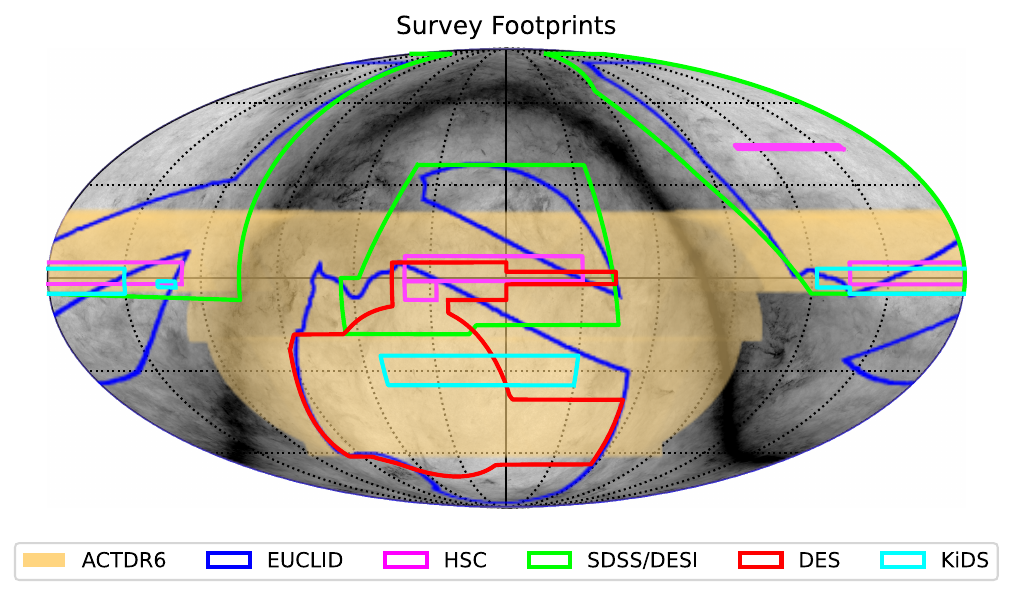}
	\caption{ACT DR6's 19\,000 square degree sky coverage and its overlap with Euclid, HSC,
	DESI, DES and KiDS. The map is in equatorial coordinates, centered on RA=0\degree,
	dec=0\degree. The background map shows galactic dust intensity from Planck. See table~\ref{tab:overlaps} for DR6' overlap with other surveys.}
	\label{fig:footprint}
\end{figure*}

\begin{figure}[htp]
	\centering
	\hspace*{-5mm}\includegraphics[width=1.17\columnwidth]{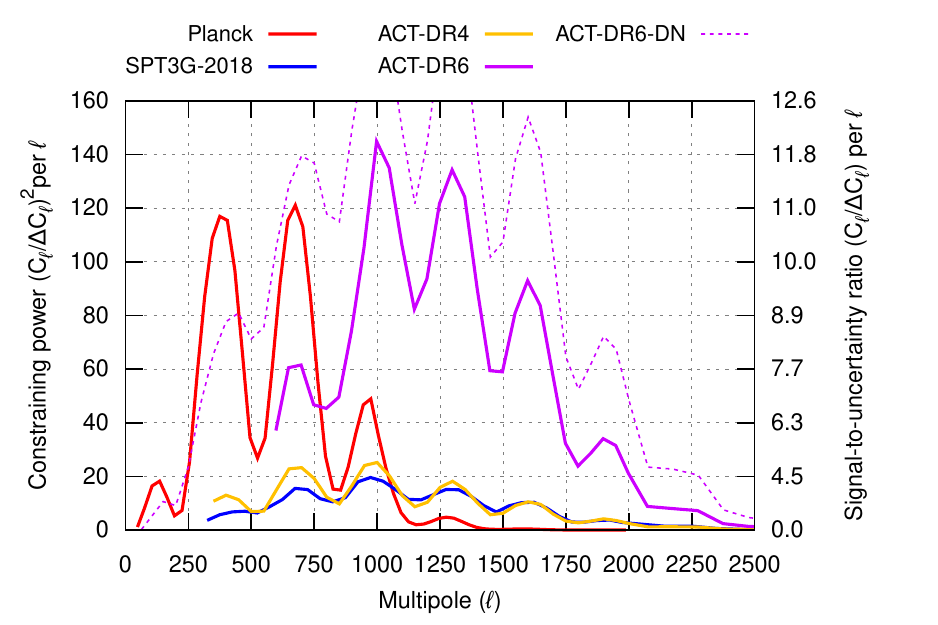}
	\caption{The state of the art for EE power spectrum measurements,
	illustrated using the signal-to-uncertainty ratio per $\Delta \ell=1$ bin
	on the right axis (non-equispaced) and its square, the constraining power,
	on the left axis (equispaced).
	When plotted this way, the areas under the curves are proportional to
	each survey's inverse variance for the overall EE spectrum amplitude.
	Where not cosmic variance limited, the curves grow proportionally with
	integration time, so for example Planck would need to integrate for 3x as long to
	match ACT DR6 at $\ell=1000$. The curves show Planck \citep[red]{planck_2018_overview},
	the SPT-3G 2018 data release \citep[blue]{spt-spectrum-2018}
	and ACT DR4 \citep[yellow]{act-dr4-spectrum}, as well as our new result (violet) and an
	estimate for the potential of a future analysis that adds the DR6 day-time
	data (dotted violet).}
	\label{fig:snr-ratio-ee}
\end{figure}

Recent measurements of the CMB power spectrum have been made by Planck \citep{planck_2018_overview}, BICEP \citep{bicep_bmode_2018}, ACT \citep[DR4]{act-dr4-spectrum}, South Pole Telescope \citep[SPT,][]{spt-spectrum-2018,spt-ge}, CLASS \citep{class-emodes-2025}, Spider \citep{spider-bmodes-2022}, POLARBEAR \citep{polarbear-bmodes-2022} and others. Of these, Planck has been the most constraining for multipoles $\ell<2100$ in total intensity (T) and $\ell < 1100$ in polarization E-modes. Above these multipoles ACT and SPT take over with roughly equal sensitivity. For polarization B-modes, BICEP and SPT have the tightest constraints for $\ell < 320$ and $\ell > 320$ respectively.

This paper is one of a series presenting ACT Data Release 6 (DR6), which represents a substantial increase in sensitivity compared to our previous data releases and define the state of the art for $\ell > 1800$ in T and $\ell > 750$ in E (see figure~\ref{fig:snr-ratio-ee} and \citealp{act-dr6-spectra}). This paper focuses on ACT's multifrequency maps of the CMB and the microwave sky. Other papers present the angular power spectra and fit to \LCDM{} \citep{act-dr6-spectra}, and extensions to \LCDM{} \citep{act-dr6-extensions}. See section~\ref{sec:conclude} and figure~\ref{fig:dr6-papers} for more details and other DR6 papers.

ACT was a 6-meter off-axis Gregorian telescope, located at 5190 meters altitude in the Parque Astronómico Atacama in Chile's Atacama Desert, with access to over half the celestial sphere at arcminute resolution. ACT's most recent data release was DR5 \citep{act-dr5-maps}, but this was an interim data release focused on small-scale science, and was not sufficiently calibrated for, say, power spectrum analysis.  The previous full ACT data release was DR4, which used data collected from 2013 to 2016. Since then ACT upgraded to the AdvancedACT\footnote{Also known as Advanced ACTPol.} camera and conducted a 2017--2022 survey, the results of which we now release as DR6.

\begin{figure}[ht]
	\centering
	\includegraphics[width=\columnwidth]{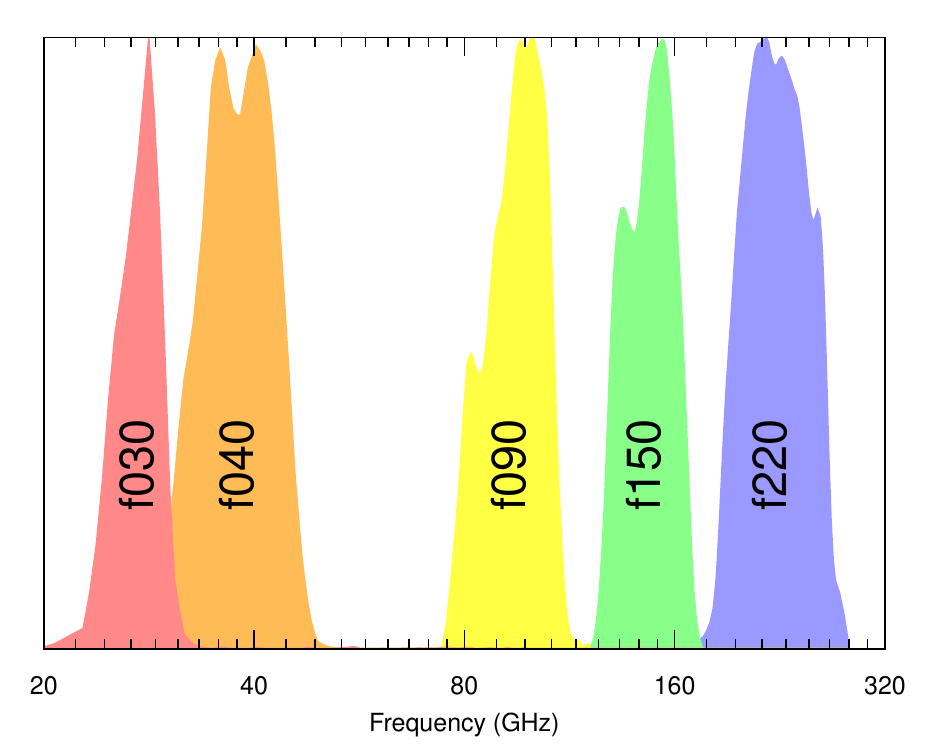}
	\caption{AdvancedACT's five bandpasses normalized to a peak of 1.
	DR6 is based on f090, f150 and f220. We postpone the lowest two
	bands, which were installed in 2020, for future work.}
	\label{fig:bandpass}
\end{figure}

This is the largest step up in data volume of any full ACT data release, with 6--10 times\footnote{It is 6 for night-time observations, 10 when including lower quality day-time observations.} as much data as DR4. DR6 covers about 45\% of the sky with (see figure~\ref{fig:footprint}) relatively even depth, unlike DR4 where most of the time was spent observing small patches.  The AdvancedACT camera also expanded ACT's frequency coverage from two bands, f090 (77 -- 112 GHz) and f150 (124 -- 172 GHz) to five: f030 (21 -- 32 GHz), f040 (29 -- 48 GHz), f090, f150 and f220 (182 -- 277 GHz) (see figure~\ref{fig:bandpass}), though we postpone analysis of f030 and f040 to a future release. Due to differences in systematic effects we do not include DR4 as a subset of DR6, keeping them as independent data sets.

\section{Data selection and characterization}
\label{sec:data}

\subsection{Summary of instrument and observations}
The ACT camera consists of three optics tubes \citep{advact-polsens}, each housing a single dichroic polarized array (PA) with a $0.85\degree$ field-of-view on the sky. The PAs are each equipped with feedhorn-coupled AlMn transition edge sensor polarimeters fabricated on 150-mm diameter silicon wafers, and operate at around 100 mK \citep{advact-detectors,advact-pa456,advact-pa7}. These arrays use a two-stage superconducting quantum interference device (SQUID) system, specifically designed for time-division multiplexing \citep{advact-readout-b}. From 2017 to the end of 2019, ACT operated with two mid-frequency (MF, f090/f150) arrays, PA5 and PA6, and one high-frequency (HF, f150/f220) array, PA4. At the beginning of 2020, the MF array PA6 was replaced by the LF array PA7 (f030/f040).

DR6 is based on 428 TB of data (144 TB compressed; see table~\ref{tab:disk}) collected from 2017-05-05 to 2022-07-02 (excluding PA7). During this 1883 day period we collected 870 days worth of data (46\% observing efficiency), of which 828 days (95\%) are CMB observations (i.e. not of calibration targets like planets).  Of this, 428 (52\%) were observed during the night, and 400 (48\%) during the day. Day-time data are typically of worse quality due to the Sun's heat deforming the mirror surface.

\begin{table}[htp]
	\centering
	\begin{tabular}{rr|rrrrrr}
	band & arr &  ndet &   dur &  srate &  size &  disk & ratio \\
	\cskip & \multicolumn{1}{c|}{} & N &  days &     Hz &    TB &    TB &    \% \\
	\hline
	f090 & PA5 &   852 & 828.0 & 395.25 &  87.7 &  30.5 & 34.8 \\
	f090 & PA6 &   852 & 458.7 & 395.25 &  48.6 &  16.5 & 34.1 \\
	f150 & PA4 &  1006 & 818.5 & 300.48 &  77.8 &  22.5 & 28.9 \\
	f150 & PA5 &   852 & 828.0 & 395.25 &  87.7 &  30.5 & 34.8 \\
	f150 & PA6 &   852 & 458.7 & 395.25 &  48.6 &  16.5 & 34.1 \\
	f220 & PA4 &  1006 & 818.5 & 300.48 &  77.8 &  22.5 & 28.9
	\end{tabular}
	\caption{DR6 data volume. The DR6 raw data are stored in
	\emph{dirfiles} \citep{getdata} and take up 144 TB of disk space after
	being compressed to around 1/3 of their raw 32-bit per
	sample representation using the \emph{libslim}
	lossless compression library \citep{slim}. The columns are:
	\dfn{arr} and \dfn{band}: The detector array and bandpass.
	\dfn{ndet}: The number of detectors for this array/bandpass combination.
	\dfn{dur}: The cumulative time spent taking data with this array; \dfn{srate}: How often a sample is read out per detector; \dfn{size}: The total uncompressed size; \dfn{disk}:
	The compressed size; and \dfn{ratio}: The compression ratio.}
	\label{tab:disk}
\end{table}

\subsection{Data selection}
\label{sec:data-sel}

\begin{table*}[htbp]
	\centering
	\begin{tabular}{rr|rrr|rrr|rrr}
	\cskip & \cskip & \multicolumn{3}{c}{full} & \multicolumn{3}{c}{night} & \multicolumn{3}{c}{day} \\
	band & arr & obs   &  sel  & \%   & obs   &  sel  & \%  & obs   &  sel  & \%   \\
	\hline
	f090 & PA5 & 828.0 & 537.1 & 64.9 & 428.1 & 335.4 & 78.3 & 399.9 & 201.7 & 50.4 \\
	f090 & PA6 & 458.7 & 308.5 & 67.3 & 237.3 & 192.3 & 81.1 & 221.4 & 116.2 & 52.5 \\
	f150 & PA4 & 818.5 & 502.8 & 61.4 & 422.9 & 310.5 & 73.4 & 395.7 & 192.3 & 48.6 \\
	f150 & PA5 & 828.0 & 532.6 & 64.3 & 428.1 & 333.4 & 77.9 & 399.9 & 199.2 & 49.8 \\
	f150 & PA6 & 458.7 & 300.4 & 65.5 & 237.3 & 187.5 & 79.0 & 221.4 & 112.9 & 51.0 \\
	f220 & PA4 & 818.5 & 471.9 & 57.7 & 422.9 & 294.2 & 69.6 & 395.7 & 177.7 & 44.9 \\
	\end{tabular}
	\caption{Days of observing time before and after the
	first data selection step. The columns are: \dfn{band}: the passband;
	\dfn{arr}: the detector array, \dfn{obs}: observing time
	before the first selection step, in days;
	\dfn{sel}: time remaining after TOD-level cuts, in days; and
	\dfn{\%}: the percentage of data surviving.}
	\label{tab:tod-sel}
\end{table*}

Our data selection broadly follows the procedures in \cite{act-dr4-maps},
which are described in more detail in~\cite{dunner/etal:2013}. It proceeds
in three stages. First, data for each array is split into chunks typically 11 minutes
long called TODs,\footnote{TOD is short for Time-Ordered Data. A typical TOD consists
of around 260k samples each for 700 selected detectors, for a total of around 200 million
samples.} and each of these is accepted or rejected as a whole
based on levels of precipitable water vapor (PWV) or the number of well-performing
detectors (appendix~\ref{sec:per-tod-cuts}); or if the day-time beam deformation
gets too large (appendix~\ref{sec:daytime-beam}). The fraction of observing time that passes
this selection for each data set is shown in table~\ref{tab:tod-sel}, but typically
75\% of the TODs are accepted during the night, and 50\% during the day.

\begin{table}[htbp]
	\centering
	\begin{closetabcols}[1.1mm]
	\begin{tabular}{rr|rrr|rrr|rrr}
	\cskip & \cskip & \cskip & \cskip & \cskip & \multicolumn{3}{c}{night} & \multicolumn{3}{c}{day} \\
	band & arr & raw& \multicolumn{2}{c|}{alive} & \multicolumn{2}{c}{sel} & samp & \multicolumn{2}{c}{sel} & samp \\
	\cskip & \multicolumn{1}{c|}{} & {\scriptsize N} & {\scriptsize N} & {\scriptsize \%} & {\scriptsize N} & {\scriptsize \%} & {\scriptsize \%}   &   {\scriptsize N} &  {\scriptsize \%}  &  {\scriptsize \%} \\
	\hline
	f090 & PA5 &  852 & 738 & 87 & 646 & 88 & 99.0 & 618 & 84 & 94.9 \\
	f090 & PA6 &  852 & 637 & 75 & 532 & 84 & 99.1 & 520 & 82 & 97.1 \\
	f150 & PA4 & 1006 & 475 & 47 & 324 & 68 & 98.9 & 291 & 61 & 95.9 \\
	f150 & PA5 &  852 & 745 & 87 & 665 & 89 & 99.2 & 634 & 85 & 96.4 \\
	f150 & PA6 &  852 & 651 & 76 & 576 & 89 & 99.0 & 554 & 85 & 97.3 \\
	f220 & PA4 & 1006 & 502 & 50 & 341 & 68 & 98.5 & 287 & 57 & 97.0 \\
	\end{tabular}
	\end{closetabcols}
	\caption{The detector and sample data selection. The columns are:
	\dfn{band}: the passband;
	\dfn{arr}: the detector array;
	\dfn{raw}: the number of detectors contained in the array;
	\dfn{alive}: the number of detectors that are ever usable, and
	their percentage relative to the raw number;
	\dfn{sel}: the average number of detectors passing data selection,
	and their percentage relative to the alive number;
	\dfn{samp}: the average fraction of samples kept after sample-level
	data selection.}
	\label{tab:det-sel}
\end{table}

Secondly, individual detectors are cut on a TOD-to-TOD
basis, based on metadata availability (e.g., usable bias step\footnote{
	In a bias step, the voltage bias on a detector is stepped by a few percent approximately every 60 mins to track the responsivities and time constants of the detectors.
} calibrations)
or the detector data's statistical properties. In particular, we
cut detectors which correlate poorly with the common mode in the atmosphere-dominated
frequency range 0.01--0.1 Hz; as well as detectors with significant
skewness, kurtosis or abnormal RMS in the white-noise dominated frequency
range 10--20 Hz. This is described in appendix~\ref{sec:per-detector-cuts},
and the result is shown in table~\ref{tab:det-sel}. Around 30\% of the
detectors are ``dead'' (never work). Of the remaining 70\%, on average
around 80\% are usable for each individual TOD, for a total average yield
of around 55\%.

Finally, we also cut sample ranges within each detector timestream
that are affected by short glitches from cosmic rays, scan
speed anomalies, and samples near scan turnarounds, etc. (appendix~\ref{sec:per-sample-cuts}).
We identify glitches as large spikes in flux that are not coincident with
known bright point sources. We also cut
samples that see the Sun or Moon in the far sidelobes (appendix~\ref{sec:sidelobe-cuts}).
In most cases around 1\% of the samples of the selected detectors are
cut during the night and 5\% during the day. See table~\ref{tab:det-sel}
column ``samp'' for detailed numbers.

\subsection{Time Constants}
Our detectors are bolometers with non-zero heat capacity, where the energy deposited by photons decays approximately as $e^{-t/\tau}$, with $\tau$ being the detector time constant that dominates the overall ``time constant''. In DR6 our time constants are quite fast, with
a typical value of $\tau = 1$ ms (see table~\ref{tab:time-constant}).
Uncorrected, this would induce a 0.1\%/1\%/10\% power loss at $\ell = 2000/6000/19000$.
When making sky maps we deconvolve each detector's average time constant
as measured from Uranus observations in the period 2017--2019. This ignores
small day-to-day fluctuations in the time constants as well as a roughly 0.3
ms increase in the 2020-2022
data. This results in a tiny contribution to the effective beam of the map
which is absorbed by our pointing jitter correction (see section~\ref{sec:beam}).

\begin{table}[htb]
	\centering
	\hspace*{-20mm}\begin{tabular}{rr|c|c}
	band & arr & 2017--2019 & 2020--2022 \\
        \multicolumn{2}{c|}{}& {\scriptsize ms} &  {\scriptsize ms}\\
	\hline
	f090 & PA5 & 0.82 & 1.05 \\
	f090 & PA6 & 0.82 & ··· \\
	f150 & PA4 & 1.35 & 1.68 \\
	f150 & PA5 & 0.77 & 1.06 \\
	f150 & PA6 & 0.74 & ··· \\
	f220 & PA4 & 1.16 & 1.45 \\
	\end{tabular}
	\caption{The detector median time constants. The time constants depend on detector optical loading and temperature. In 2020, the time constants increased by approximately 0.3 ms due to a $\sim$30 mK rise in array temperature following the installation of the PA7 array and changes to the cryostat IR blockers.}
	\label{tab:time-constant}
\end{table}

\subsection{Gain calibration}
We calibrate our time-ordered data and sky maps to the usual linearized CMB micro-Kelvin units,\footnote{
	These are defined as $10^6 \cdot I(\nu)/B'(\nu,T_0)$ where $I(\nu)$ is the
	observed spectral radiance at frequency $\nu$, $B'(\nu,T)$ is the derivative of
	Planck's law for Blackbody radiation with resepct to the temperature $T$ in Kelvin,
	and $T_0 = 2.725$ K is the CMB monopole temperature. To first order, these
	express the local deviation from the monopole temperature, in µK. Unless otherwise
stated, all ``µK'' in this paper are in these linearized CMB units.} but data are read
out from the detectors in data aquisition units (DAQ units) and must therefore be translated.
We do this in two steps: ``relative calibration,'' which consists of flat fielding and converting
from DAQ units to pW, and ``absolute calibration'', which uses planet observations to
translate from pW to spectral radiance and then µK.

\subsubsection{Relative gain calibration}
\label{sec:relgain}
During this analysis we discovered that a high quality relative calibration between
detectors, the flat field, is essential for recovering large angular scales in
the maps (see section~\ref{sec:tf} and \citealp{model-error2}).
The atmosphere is a beam filling calibrator and is, in this regard,
similar to the CMB. This makes it a good candidate for flat-fielding,
but some care is required because the atmosphere has a different frequency
spectrum than the CMB \citep{hasselfield/etal/2013}, and because it can
be confused with ground pickup, magnetic contamination and cryostat temperature
drifts.

A method for differentiating atmospheric
modes from others is presented in \cite{morris-atmosphere-2024}.
While the full formalism was
developed after the flat fielding was fixed for DR6, the procedure behind the
formalism -- clearly identifying atmospheric modes for flat fielding -- was
followed. After identifying atmospheric modes, the optical flat field was
measured from each detector's correlation with the atmosphere in the
0.01--0.1 Hz range, under the assumption that all detectors should see
the same atmospheric fluctuations.\footnote{This ``common mode'' assumption
limits the maximum accuracy achievable with this method. Spatially
separated detectors will see slightly different atmospheric fluctuations,
reducing the correlation with the common mode. This reduction will be higher
towards the edge of the array. In ACT this effect is subdominant to errors
from isolating the common mode in the first place.}

In addition to using a cleaner atmospheric signal for flat-fielding,
another change from our previous analysis in \cite{act-dr4-maps}
is the use of a monthly average instead of per-TOD flat fields. This
sacrifice in time resolution was done to reduce measurement error,
and after extensive testing we found a monthly average to be the tradeoff
that minimized the low-$\ell$ power loss.

Despite this improvement, we estimate that the flat field is still only
accurate to a few percent, and is probably the dominant cause of the low-$\ell$
bias that forces us to discard the TT power spectrum for $\ell < 600-1000$
(see section~\ref{sec:tf}). This problem might have been avoided with an
external flat field calibrator like the stimulator built into the Simons
Observatory Large Aperture Telescope's
primary mirror, provided any unmodeled bandpass mismatches between detectors
in the same class are $\ll 1\%$.

Using the bias steps and flat-field,
each sample $d$ from time-step $t$ of detector $i$ is calibrated into
physical pW units as
\begin{align}
	d_{it}^\text{pW} &= R_{it} f_{it} d_{it}^\text{DAQ} . \label{eq:calib}
\end{align}
Here $R_{it}$ is the responsivity obtained from the previous bias step,
and $f_{it}$ is the average optical flat field for that month. Bias steps
are performed at least once per hour, and involve measuring the detector
response as we modulate the detector bias voltage.

\subsubsection{Absolute gain calibration}
The calibration from pW to CMB temperature fluctuation units in µK
is done as in DR4, using observations of Uranus. Uranus is a point source for ACT,
so its observed profile lets us determine the instrument beam, and
its amplitude and beam area determine the calibration when compared to
measurements from Planck. Planck, in turn, has calibrated Uranus'
flux to 1\% accuracy using the CMB dipole \citep{planck-planet-cal:2017},
in agreement with earlier measurements from WMAP \cite[][3\% accuracy at 94 GHz]{wmap-calibration:2011}.

The instrument gain depends on the loading, which is mostly determined
by how much water vapor we look through. We use multiple Uranus
observations at different elevations (el) and weather conditions to fit a
linear model of gain as a function of $\textrm{PWV}/\sin(\textrm{el})$.
This model is then used to predict the calibration for each individual TOD.
The gain is typically around 10 K/pW.

Due to scatter in the gain-loading relation as well as the mismatch
between the beam-filling CMB and the point-like Uranus, this calibration
is only good to around 1--4\% accuracy, depending on the band.
After the maps have been made, we therefore perform a final calibration
against Planck using the CMB perturbations themselves, as described
in \cite{act-dr6-spectra} and section~\ref{sec:poleff}.

\subsection{Polarization angles}
Another critical calibration parameter is the detector polarization angle, which describes the rotation of the polarization signal in the maps relative to the sky. This angle has three components:
\begin{enumerate}
	\item The orientation of the orthogonal pick-up antennas on the wafer. From the fabrication process, this is known to $<0.001\degree$.
	\item The orientation of the wafer in the focal plane. This is relatively straightforward to measure because it also results in a rotation of the individual detectors' pointing on the sky. Using $\sim 150$ Uranus observations, we determine the average effect of this rotation in the DR6 maps to be -0.119/-0.019/-0.028 $\pm$ 0.011/0.004/0.004 degrees for PA4/PA5/PA6.
	\item Polarization rotation from the optical system. The alignment of optical elements with respect to the detectors and the position of the ACT camera relative to the telescope reflectors can introduce a source of rotation of the entire detector array when projected onto the sky. As with DR4 \citep{act-dr4-spectrum}, we model the full optical system, reflectors plus lenses, and quantify the polarization angle rotation across the focal plane using Optics Studio CODEV. We find a smoothly changing 1.2/1.2/0.4 degree change across PA4/PA5/PA6 as one moves away from the optical angle \citep{Koopman/2016}, which is taken into account in our mapmaking. The systematic uncertainty on these numbers is $0.1\degree$ \citep{Murphy/2024}.
\end{enumerate}
The total polarization angle systematic uncertainty is therefore $0.1\degree$, shared between all detectors in each array.

\subsection{Pointing corrections}
The general principle of the pointing correction is the same as for DR4: an initial pointing model constructed from planet observations gives us a
blind pointing accuracy of around an arcminute. This is comparable to our
beam size and therefore unacceptably large, so we construct a per-TOD
correction using observed positions of bright point sources with known
coordinates from external catalogs.

DR6 uses constant elevation scans like DR4, but the typical scanning amplitude has
increased from $26\degree$ to $60\degree$ peak-to-peak,
meaning that each TOD covers a much wider azimuth range. During the course of the
DR6 analysis we discovered that there can be an up to 0.5\arcmin{} pointing error difference
between the left and right ends of our largest $90\degree$ azimuth scans. This led us to
introduce a time-dependent azimuth slope in our pointing model. We also modified
how the fit is done from an expensive direct time-domain fit to a cheaper and
higher time-resolution fit based on short exposure maps (``depth-1 maps'', see appendix~\ref{sec:pointing-details} for details).

Despite these improvements in the model, the average pointing jitter in DR6,
as inferred from the difference between the beam inferred from point sources
in the CMB maps and that measured from individual planet observations,
is $0.15\pm0.04\arcmin$, around 30\% worse than DR4. We interpret this as being
due to the more challenging wide scans in DR6. In any case, our pointing jitter
represents just a 2\% increase in the effective beam FWHM, and is included as
part of our beam model.

\subsection{Beams and polarized beam leakage}
\label{sec:beam}
The instrumental beam for the nighttime observations is determined in a similar way as was done for DR4~\citep{lungu-2022}. Details of the DR6 beam estimation are described in a dedicated paper \citep{act-dr6-beams}. Here, we provide a summary of the methodology and the available beam products.

Our primary beam estimate is based on dedicated observations of Uranus that were taken throughout the observing seasons. To avoid the large-scale power loss described in section~\ref{sec:mapmaking}, we use a custom mapmaking algorithm which eliminates this bias within 12\arcmin{} from the planet location \citep{lungu-2022}. If not corrected, this bias would manifest as negative ``bowling'' around the source, as well as a stronger stripe in the scanning direction. While faint relative to the central peak of the beam, this would be enough to wash out the signal in the wings of the beam.\footnote{See section~\ref{sec:special-source-treatment} for how we avoid this effect around bright point sources in the map.}

A radial profile is computed from each Uranus map. A parameterized model for the radial profile is then jointly fit to all profiles for a given season. The resulting profiles are largely consistent between the seasons, with the exception of the 2017 observing season (s17), which deviates from the later seasons due to a minor refocusing of the telescope at the start of s18. The radial profiles are converted to a harmonic profile using a Legendre polynomial transform and corrected for the non-zero solid angle of Uranus, the pixel window of the maps and other small biases, as described in~\cite{lungu-2022}. Because the sky maps combine all seasons into four split maps made from disjoint observations, the per-season harmonic profiles are linearly transformed into per-split profiles. The weights for this transformation are estimated from the statistical contribution of each season to each of the splits. At this point of the analysis, the beams describe the angular response to a source with the SED of Uranus and have not yet been corrected for any beam-altering effects present in the sky maps.

To color-correct the Uranus beams to beams appropriate for the CMB and other sky components, the frequency-dependence of the beam is inferred using a model of the beam based on physical optics simulations. The parameters of  the model are found by integrating the frequency-dependent beam model weighted by the Uranus SED over the instrumental passband and finding the parameter values that best describe the observed Uranus profile. From the inferred frequency-dependent beam, the beams appropriate for other sky components are then derived. The correction for pointing jitter and other small beam-altering effects that might be present in the sky maps is computed by finding the best-fitting symmetric Gaussian convolution kernel that describes the difference between the beam derived from Uranus to a set of bright point sources in the sky maps.

Unlike the DR4 case~\citep{lungu-2022}, we also make a set of secondary planet maps using the standard CMB mapmaker (section~\ref{sec:mapmaking}), to study the effect of mapmaker nonideality on the beam. In principle this could be used to measure the precise shape of the low-$\ell$ power loss, but in practice even Uranus is not bright enough to get a usable measurement at these multipoles in the presence of atmospheric noise.\footnote{It might be possible to make it work with a brighter source like Saturn, but here the challenge is saturation of the central peak.} We therefore estimate the low-$\ell$ power loss using the CMB itself (section~\ref{sec:tf}). However, the secondary planet maps are still useful for studying temperature-to-polarization (\ttop{}) leakage. This has two advantages compared to the primary planet maps.
\begin{enumerate}
	\item It captures any \ttop{} leakage introduced by the mapmaker itself
	\item It is not limited to $r<12\arcmin$, allowing us to capture $l \lesssim 1500$ where power spectrum null tests indicate significant \ttop{} leakage.
\end{enumerate}
Both types of planet maps are analysed further in \cite{act-dr6-beams}, where we use them to build a beam model including \ttop{} leakage. The effective map beams have a FWHM of 1.42/2.07/1.01 arcminutes at f090/f150/f220 with individual detector arrays in the same band deviating from these averages by $<0.01$ arcmin. When excluding the near sidelobes (see appendix~\ref{sec:buddies}), around 0.01\% and 0.002\% of the total intensity beam power\footnote{Defined as $\sum_\ell B_\ell^2 (2\ell+1)$} leaks into E and B respectively.

\subsection{Data sensitivity}
During DR6, ACT's passbands f090/f150/f220 had
an average instantaneous night-time sensitivity of 8.4/8.9/43 \ukrts{} respectively,
giving a combined sensitivity of 6.1 \ukrts{}.\footnote{The day-time numbers
are about 5\% higher.} The total inverse variance (weight) is 0.95/nK$^2$
(night: 0.61/nK$^2$, day: 0.34/nK$^2$). Hence, DR6 has almost 10x
the weight of DR4's 0.096/nK$^2$ (6.4x for night-only).
See table~\ref{tab:asens} and table~\ref{tab:fsens} for more details.

\begin{table*}[htb]
	\centering
	\hspace*{-20mm}\begin{tabular}{rr|lrrr|lrrr|lrrr}
		\cskip & \cskip & \multicolumn{4}{c}{Full} & \multicolumn{4}{c}{Night} & \multicolumn{4}{c}{Day} \\
		Band & Arr & Weight & \hspace*{-3mm}Dtime & Dsens & Asens & Weight & \hspace*{-3mm}Dtime & Dsens & Asens & Weight & \hspace*{-3mm}Dtime & Dsens & Asens \\
		\cskip & \multicolumn{1}{c|}{} & {\scriptsize $1/\text{nK}^2$} & {\scriptsize years} & {\scriptsize \ukrts} & {\scriptsize \ukrts} & {\scriptsize $1/\text{nK}^2$} & {\scriptsize years} & {\scriptsize \ukrts} & {\scriptsize \ukrts} & {\scriptsize $1/\text{nK}^2$} & {\scriptsize years} & {\scriptsize \ukrts} & {\scriptsize \ukrts} \\
		\hline
		f090 & PA5 & 0.285  & 934 & 321 & 12.8 & 0.184  & 593 & 319 & 12.6 & 0.101  & 342 & 326 & 13.1 \\
		f090 & PA6 & 0.200  & 446 & 266 & 11.6 & 0.128  & 281 & 263 & 11.4 & 0.072  & 166 & 270 & 11.8 \\
		f150 & PA4 & 0.113  & 429 & 346 & 19.6 & 0.073  & 276 & 345 & 19.1 & 0.040  & 153 & 349 & 20.5 \\
		f150 & PA5 & 0.196  & 953 & 392 & 15.3 & 0.127  & 608 & 389 & 15.1 & 0.069  & 346 & 399 & 15.8 \\
		f150 & PA6 & 0.135  & 468 & 331 & 13.9 & 0.088  & 296 & 327 & 13.6 & 0.047  & 172 & 338 & 14.3 \\
		f220 & PA4 & 0.0212 & 415 & 785 & 43.8 & 0.0140 & 275 & 788 & 42.6 & 0.0072 & 140 & 782 & 46.1 \\
	\end{tabular}
	\caption{The total effective exposure and sensitivity of the data sets.
	\dfn{Weight}: The time-integrated inverse white noise variance (after data selection) summed over all
	detectors. \dfn{Dtime}: The total exposure time (after data selection) summed over the detectors.
	\dfn{Dsens}: The mean sensitivity of the selected detectors.
	\dfn{Asens}: The total sensitivity of the detectors in an array (per band).
	The total inverse variance across all the data sets is 0.95/nK$^2$, an order
	of magnitude more than ACT DR4's 0.096/nK$^2$.}
	\label{tab:asens}
\end{table*}

\begin{table}
	\centering
	\hspace*{-20mm}\begin{tabular}{r|lr|lr|lr}
		\cskip & \multicolumn{2}{c}{Full} & \multicolumn{2}{c}{Night} & \multicolumn{2}{c}{Day} \\
		Band & Weight & Sens & Weight & Sens & Weight & Sens \\
		\multicolumn{1}{c|}{} & {\scriptsize $1/\text{nK}^2$} & {\scriptsize \ukrts} & {\scriptsize $1/\text{nK}^2$} & {\scriptsize \ukrts} & {\scriptsize $1/\text{nK}^2$} & {\scriptsize \ukrts} \\
		\hline
		 f090 & 0.485  &  8.6 & 0.311  &  8.4 & 0.173  &  8.8 \\
		 f150 & 0.444  &  9.1 & 0.288  &  8.9 & 0.156  &  9.4 \\
		 f220 & 0.0212 & 43.8 & 0.0140 & 42.6 & 0.0072 & 46.1 \\
		\hline
		total & 0.949  &  6.2 & 0.613  &  6.1 & 0.336  &  6.4 \\
	\end{tabular}
	\caption{The total effective exposure and sensitivity per frequency band.
	\dfn{Weight}: The time-integrated inverse white noise variance (after data selection) summed over all
	detectors in the band.
	\dfn{Sens}: The total sensitivity of the detectors in the band.}
	\label{tab:fsens}
\end{table}

\section{Mapmaking}
\label{sec:mapmaking}
We recover images of the sky from the time-ordered data using
the same general framework as set out in \citet{dunner/etal:2013}
and used in DR2 \citep{dunner/etal:2013}, DR3 \citep{naess/etal:2014},
and DR4 \citep{act-dr4-maps}.
We summarize this below while pointing out differences.

\subsection{Data preparation}
\label{sec:data-preparation}
We prepare each TOD for mapmaking by
\begin{enumerate}
	\item Dividing out the instrumental gain.
	\item Gapfilling glitches to avoid numerical issues from very
		high invalid values, and to avoid having them impact the noise model.
		We gapfill by estimating a detector-detector covariance matrix,
		and for each sample use this to predict the value of a cut detector
		given the uncut ones. If all are cut, we simply gapfill using a
		linear trend. Not all samples that do not pass our data selection
		are gapfilled -- only those flagged as glitches since these are the
		ones at risk of having extreme values. Since only cut regions are
		gapfilled, they do not enter into the maps, but they do slightly
		affect the noise model.
	\item De-sloping each detector by subtracting a linear trend from the
		average of the first 8 samples to the last 8 samples. This is done
		to make the TOD more Fourier-amenable by reducing the implied discontinuity
		between the end and the start of the TOD.
	\item Deconvolving the instrumental antialiasing filter and the detector
		time constants.
	\item For all but the last mapmaking pass (see section~\ref{sec:multipass}),
		Fourier-downsample the TOD to reduce the number of samples and greatly
		speed up the mapmaking.\footnote{Fourier downsampling reduces the
		sample rate by simply truncating in Fourier space. To go from $n$
		samples to $m<n$ samples, one would discard samples above the
		new Nyquist limit: \texttt{irfft(rfft(arr)[:m//2+1])}.
		This method preserves power up to the new Nyquist limit while
		eliminating aliasing, unlike simple averaging of groups of samples
		which suffers from both aliasing and high-frequency power loss.}
	\item Reducing to 32-bit float precision to save memory and improve
		speed. This is enough to handle a per-sample signal-to-noise ratio
		of $10^7$, which is more than enough for our noise-dominated TOD.
		Tests have shown that we end up with the same maps if we use
		64-bit precision.
	\item Subtracting a model of the polarized sidelobes
		(see appendix~\ref{sec:buddies}). This relies on a signal
		estimate from a previous mapmaking pass (see section~\ref{sec:multipass}),
		so it is skipped for the first pass.
\end{enumerate}
Unlike DR4, we no longer perform ground subtraction in time domain.
This was not effective enough to avoid the need for map-space
filtering during power spectrum estimation, and required expensive time-domain
simulations to characterize. In DR6 we leave the ground pickup in the maps
because we find it easier to characterize and subtract there.

\subsection{Noise model}
\label{sec:noise-model}
A noise model is needed to down-weight noisier parts of the data
when projecting it onto the sky. Without noise weighting,\footnote{
	Or similar techniques that fill the same role in other mapmaking
	approaches, e.g. high-pass filtering for filter+bin mapmaking or
	baseline deprojection for destriping.} the presence
of atmospheric noise would make it impossible to map the CMB from
the ground. We use a similar noise model as in
\citet{dunner/etal:2013,naess/etal:2014,act-dr4-maps}. This models
the Fourier-modes of the TOD as being independent, but assumes
correlations between individual detectors:
\begin{align}
	\tilde N_{ff'dd'} &= (D_{b_fd} + V_{di} E_{b_fi} V^T_{id'})\delta_{ff'}
\end{align}
where $b_f$ tells which of around 60 non-equispaced Fourier-bins
contains the frequency $f$, and $d$ and $d'$ are detector indices.
The matrix $D$ represents the part of the noise power that's
uncorrelated between detectors. The matrix $V$ contains noise eigenvectors
(labeled with $i$) built as follows.
We select eigenvectors with at least 16 times the amplitude of the median
eigenvalue from both the atmosphere-dominated frequency range 0.25--4.0 Hz and
the instrumental noise-dominated frequency range 4.0–-200 Hz. This typically
results in around 10 modes being selected. $E$ represents the power of each of
these modes for each frequency bin, $b_f$.

Measuring the noise model from the data itself biases the signal
low, since cases where the noise partially cancels the signal appears
lower-variance than cases where the noise is in phase with the signal.
Overall signal-cancelling noise is therefore given higher weight.
We avoid this effect by using multi-pass mapmaking, where the signal
estimate from a previous mapmaking pass is subtracted before the
noise model is estimated.

\subsection{Data model}
We model the calibrated time-ordered data for sample $t$ of detector $d$
as a linear function of a static sky $m$,
\begin{align}
	d_{dt} &= P_{dtps} m_{ps} + C_{dtj} c_j + S_{dtk} s_k + n_{dt} . \label{eq:data-model}
\end{align}
Here $m_{sp}$ is the value of the Stokes parameter $s \in \{I,Q,U\}$ at pixel $p$
(in units of $\micro\text{K}_\text{CMB}$), and $n_{dt}$ is Gaussian noise with
covariance $N$. $c$ represents the contamination in the cut samples,
which must be included in the equation system to avoid biasing the results;
while $s$ represents the model errors in areas with very high contrast,
such as near bright point sources. The pointing matrix $P$, the cut mapping $C$ and
the model error mapping $S$ are the response of the data to $m$, $c$ and $s$
respectively. See sections~\ref{sec:pmat}, \ref{sec:ml-cuts} and
\ref{sec:special-source-treatment} for these. We can write the data model
in matrix form as
\begin{align}
	d &= \underbrace{[P,C,S]}_{A} \underbrace{\begin{bmatrix}m\\c\\s\end{bmatrix}}_a + n = A a + n \label{eq:data-model-expanded}
\end{align}
which has the maximum-posterior solution
\begin{align}
	\hat a &= (A^TN^{-1}A + \Pi)^{-1}A^TN^{-1} d \label{eq:mapmaking}
\end{align}
where $\Pi$ is a prior that resolves degeneracies between $m$ and $s$
(see section~\ref{sec:special-source-treatment}). For ACT this is a
$\sim 10^9\times 10^9$ equation system that must be solved
using iterative methods like Preconditioned Conjugate Gradients (CG).

\subsubsection{Pixelization}
We represent the sky map $m$ using a Plate Carrée projection in equatorial coordinates,
pixelized with $43200\times10320$ pixels of size $0.5\arcmin\times0.5\arcmin$ covering
$180\degree>\textrm{RA}>-180\degree$ and $-60\degree < \textrm{dec} < 20\degree$.
The north and south pole are not included in the maps to avoid wasting space, but
would have half-integer declination pixel coordinates, making the maps compatible
with Fejer's first integration rule \citep{fejer1+cc}, allowing for efficient ``map2alm'' inverse
spherical harmonics transformations.\footnote{If $Y$ is the matrix of spherical
harmonics basis functions, representing the ``alm2map'' transform, then
the inverse $Y^{-1} \approx Y^TW$, where $W$ is a quadrature weight matrix.
The half-integer declination pixel coordinates for the poles ensures that
$W$ can be evaluated efficiently. These details are irrelevant for the
``alm2map'' transformation.}

This is a change from ACT DR4, where we instead
used whole-integer declination pixel coordinates, corresponding to the
Clenshaw-Curtis integration rule \citep{fejer1+cc}. Clenshaw-Curtis
pixelization has the disadvantage
that it is not robust to simple resolution downgrading. For example, the
simplest way to halve the resolution of a map is to replace each $2\times2$
block of pixels with a single new pixel with the average of their values.
Under this operation, a Fejer-1 map stays a Fejer-1 map, but a Clenshaw-Curtis
map ends up with quarter-integer coordinates for the poles. No integration weights
for such a map are available in the \emph{DUCC} spherical harmonics transform
library we use \citep{ducc}, nor for other commonly used ones like its predecessor
\emph{libsharp} \citep{libsharp}.

The upshot of this is that the DR6 pixelization has a half-pixel declination
shift compared to DR4, and that, unlike DR4, DR6 maps can still be inverse
spherical harmonics transformed after simple downgrading.

\subsubsection{Pointing matrix}
\label{sec:pmat}
The standard practice is to use a nearest-neighbor model for the instrument
response $P$, meaning
that the value in each sample is simply read off from the nearest pixel,
without any interpolation, and this is also what we did in DR4. In practice this means
that if the sample with index $dt$ has nearest pixel $q_{dt}$, then
\begin{align}
	P_{dtps} &= \delta_{pq_{dt}} [1,\cos(2\psi_{dt}),\sin(2\psi_{dt})]_s
\end{align}
where $\delta$ is the Kronecker-delta and
$\psi_{dt}$ is the orientation of the direction of polarization
sensitivity on the sky for detector $d$ at sample $t$, which is used
to handle the spin-2 nature of the Stokes Q and U parameters. This method
is popular because it results in $P$ being extremely sparse and efficient,
with each sample only needing to concern itself with a single pixel, but the
cost of this is that we model our data as a set of sudden jumps in value as
one moves from one pixel to the next.

However, during the DR6 analysis we discovered that despite a nearest-neighbor
$P$ only being an inaccurate description of the data at sub-pixel scales
($<0.5\arcmin$, $\ell>21600$), its use can result in $\mathcal{O}(1)$ bias in
the angular power spectrum at any scale where the noise model $N$
is highly correlated (see section~\ref{sec:tf}). For a ground-based microwave telescope like ACT the
atmosphere causes large amounts of unpolarized correlated noise at low $\ell$,
and the result is that the unphysical sub-pixel treatment in a nearest-neighbor
$P$ propagates into a large power loss in the low-$\ell$ TT power spectrum.
This unintuitive effect is described in detail in \citet{model-error2}.

The most obvious solution for sub-pixel errors is to simply reduce the pixel
size, but this is not practical. Sub-pixel bias is first order in the pixel
size, so to reduce the bias to $1/10$ one would need 100 times as many pixels.
Not only would this be far too big, it would also spread out the samples too
thinly, leaving many pixels unhit. A much better solution is to switch to
a bilinear pointing matrix, which makes the errors second order in the pixel size.
\begin{align}
	P_{dtps} =&
	((1-\Delta x)\delta_{p_x,\lfloor u_x \rfloor} + \Delta x\delta_{p_x-1,\lfloor u_x\rfloor}) \cdot \notag \\
	& ((1-\Delta y)\delta_{p_y,\lfloor u_y \rfloor} + \Delta y\delta_{p_y-1,\lfloor u_y\rfloor}) \cdot \notag \\
	& [1,\cos(2\psi_{dt}),\sin(2\psi_{dt})]_s
\end{align}
where $u_x$ and $u_y$ are the full (not rounded) x and y pixel coordinate
for sample $dt$, $\lfloor \rfloor$ indicates rounding down, and $\Delta x \equiv u_x - \lfloor u_x\rfloor$
and similarly for $\Delta y$. The cost of this approach is that each sample
now touches four pixels rather than one, but we found that the overall
runtime increase was around 50\% rather than the 300\% one might fear.

Bilinear mapmaking produces a different pixel window than nearest neighbor
mapmaking (see figure~16 of \citealp{model-error2}), but to avoid requiring
changes to code that uses our maps, we reconvolve them to the standard
nearest neighbor pixel window.

\subsubsection{Cut sample model}
\label{sec:ml-cuts}
We do not want the values of the samples that do not pass our data
selection criteria to affect our sky map, but we cannot simply skip
them in the likelihood because our noise model is Fourier-based, and
the Fast Fourier Transform requires the samples to be equi-spaced.
On the other hand, using the values as they are would contaminate the map,
and replacing them with a fixed value like zero would bias it. The
standard solution to this problem is to include the values of the cut
samples as degrees of freedom in the likelihood, and this is what the
vector $c$ in equation~\ref{eq:data-model-expanded} represents.

In the simplest form, we allocate one degree of freedom per cut sample,
and use these values instead of the ones read off from the map in
the data model (equation~\ref{eq:data-model}). This is implemented by
zeroing out the corresponding rows in $P$, with these rows becoming
the only non-zero rows in $C$. This is illustrated for a toy example
with a single detector scanning across 4 pixels with constant speed
over the course of 7 samples, with samples 3, 4 and 7 being cut\footnote{
	The $\frac12$ entries in $P$ occur when a sample hits halfway between
	two pixels.}
\begin{align}
	d &=
	\overbrace{\begin{bmatrix}
		1   & 0   & 0   & 0   \\
		\frac12 & \frac12 & 0   & 0   \\
		0   & 0   & 0   & 0   \\
		0   & 0   & 0   & 0   \\
		0   & 0   & 1   & 0   \\
		0   & 0   & \frac12 & \frac12 \\
		0   & 0   & 0   & 0   \\
	\end{bmatrix}}^P m + \overbrace{\begin{bmatrix}
		0 & 0 & 0\\
		0 & 0 & 0\\
		1 & 0 & 0\\
		0 & 1 & 0\\
		0 & 0 & 0\\
		0 & 0 & 0\\
		0 & 0 & 1\\
	\end{bmatrix}}^C c \ .
\end{align}

To save memory, we implement a somewhat more complicated version
of this where our degrees of freedom represent the Legendre polynomial
coefficients of the cut samples instead of the one-to-one mapping shown
above. For cuts of length up to 1/3/6/20 samples we allocate 1/2/3/4
degrees of freedom, and beyond that we use the nearest integer to
$5+T/4$ where $T$ is the duration of the cut in seconds.\footnote{There
are typically around 100 samples per second in the first two mapmaking
passes and 400 samples per second in the final pass.}

The general method was suggested by \citet{patanchon-2008} and used
in DR2--DR4, but was not described in as much detail there.

\subsubsection{Model error mitigation in high-contrast areas}
\label{sec:special-source-treatment}
\begin{figure*}
	\centering
	\begin{closetabcols}
	\hspace*{-8mm}\begin{tabular}{ccc}
		Without & With & Difference \\
		\includegraphics[height=6.4cm,trim=0 0 15mm 0,clip]{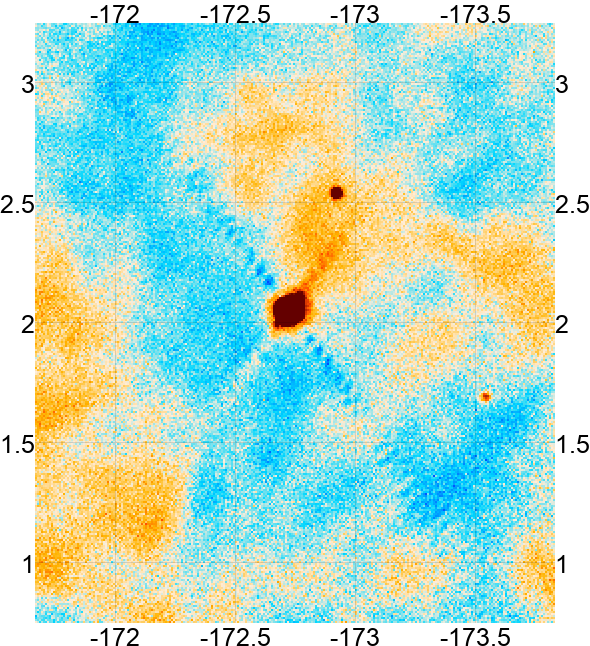} &
		\includegraphics[height=6.4cm,trim=15mm 0 15mm 0,clip]{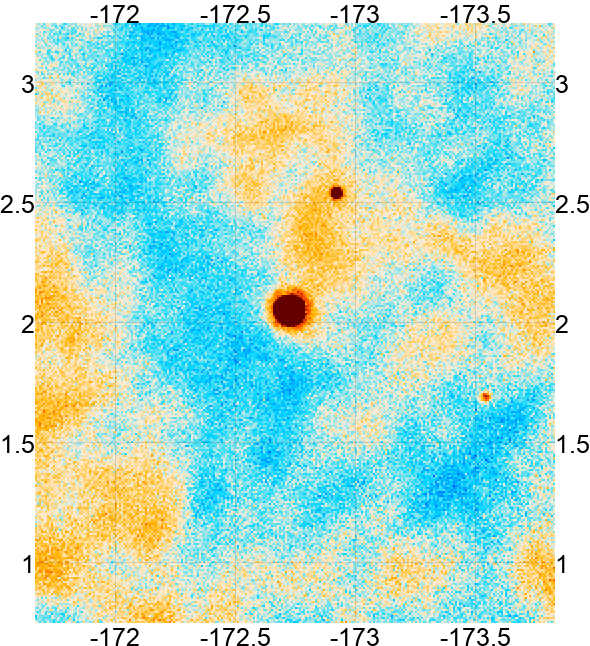} &
		\includegraphics[height=6.4cm,trim=15mm 0 0 0,clip]{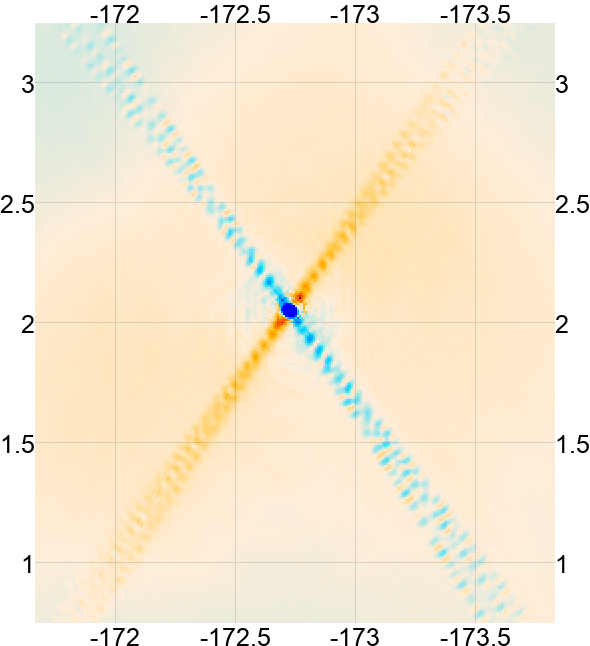}
	\end{tabular}
	\end{closetabcols}
	\caption{PA6 f090 maps without (left) and with (middle) model error mitigation
	in high-contrast areas, along with their difference (right).
	The view is centered on bright quasar 3C 273,
	which has an f090 flux of 12~Jy and a peak amplitude of 110 mK with the
	PA6 beam. The maps are plotted with a $\pm500$ µK color range (dark blue
	to dark red), with RA and dec on the horizontal and vertical axes.
	Without model error mitigation an X-shaped pattern with a typical amplitude
	of 100 µK (0.1\% of the peak) appears around the quasar, extending out
	around $3\degree$ in each direction.}
	\label{fig:ptsrc-x}
\end{figure*}

Even with a bilinear pointing matrix the data model is never
100\% accurate. Tiny sub-pixel errors remain, as well as unmodeled
pointing jitter and gain fluctuations. Furthermore, the sky itself
is variable, especially compact objects like quasars. As described
in \citet{model-error2}, mismatch between the model and data
can manifest as a loss of power at large scale, effectively
introducing a slight high-pass filter in the mapmaking. This is
undesirable as a whole but especially problematic near bright objects,
where it manifests as thin, X-shaped artifacts extending several degrees away
with an amplitude of $\sim0.1\%$ of the peak \citep{model-error}.
An example of this is shown in figure~\ref{fig:ptsrc-x}.

To avoid needing to mask out large areas around each bright source
(or omitting them from the map altogether) we instead completely
eliminate model error for these objects by allocating an extra degree
of freedom for every sample that hits them. This gives the model the
freedom to absorb arbitrary data behavior in this region. We implement
this using the vector $s$ and the corresponding response matrix $S$
which works very similarly to the cuts matrix, $C$, (section~\ref{sec:ml-cuts}), with
two main differences
\begin{enumerate}
	\item We use one degree of freedom per selected sample, not a
		Legendre polynomial.
	\item We do not remove the corresponding rows from the pointing
		matrix $P$.
\end{enumerate}
The last point makes eq.~\ref{eq:mapmaking} underdetermined:
it's ambiguous whether the signal should go into the pixels near
the objects or into $s$. We resolve this using the prior $\Pi$
in the equation, which takes the form:
\begin{align}
	\Pi a &= W R s
\end{align}
where the diagonal matrix $W$ is the uncorrelated
part of the inverse noise variance predicted by $N$ for each sample in $s$,\footnote{$W$ gives the prior the same units as
$A^TN^{-1}A$ and ensures that these terms will have a consistent relative
strength regardless of each TOD's length or noise level.}
and $R$ is an edge weighting factor that interpolates logarithmically
from 10 for samples at the edge of each region to 0.01 for samples 2 arcminutes
away from the edge.\footnote{
	The edge taper avoids noise discontinuities at the edge of the region.}

The effect of this prior is to give a mild preference for putting
the signal in $m$ instead of $s$. Hence, anything that could be
represented by $m$ will end up there, while what cannot (the model
errors) end up in $s$.

One might wonder if this technique could be used to eliminate model errors
everywhere in the map, but this would not work. Aside from needing a
dauntingly large $s$, the very act of solving for so many extra degrees of
freedom degrades the noise properties of the map by spreading the
data too thin. In the absence of the prior $\Pi$ there would be no averaging
at all, and even with it there's a tradeoff between bias and optimality.
The method performs best when restricted to areas small enough that the
correlated noise is still well constrained by the surroundings, and bright
enough that the loss of optimality is not noticeable.

For the DR6 maps, we use this method for pixels that fall within the
mask \texttt{srcsamp\_mask.fits} (see table~\ref{tab:dr6-products}),
which covers areas within
3 arcminutes from a point source detected at $\geq 500$ mJy at f090,
or which is within 3 arcminutes of an area where $|\nabla p|^2$ is
above its 99.99\%-quantile, where $p$ is the Planck 545 GHz map.
This captures the brightest 207 point sources in the map, as well as
the areas of the Galaxy with the brightest dust emission like the
galactic center and the Orion nebula. All in all, 51060 pixels (11.55
square degrees, 0.06\% of our sky coverage) are given this treatment.

\subsection{Multipass mapmaking}
\label{sec:multipass}
There are several reasons why it's useful to split the mapmaking process
into multiple passes.
\begin{enumerate}
	\item As noted in \ref{sec:noise-model} we build the noise model from the data itself, but this contains
		the sky signal we want to measure. Using a noise model contaminated by
		the sky signal to weight the data results in a biased map. We can avoid
		this by subtracting an estimate of the signal before building the noise
		model, but this requires us to already have made a map.
	\item Given an estimate of the sky, we can perform a cheap first-order
		correction for the polarized sidelobes.\footnote{Since these sidelobes are
		an $\mathcal{O}(10^{-3})$ effect, the effect of ignoring higher orders
		is a negligible $\mathcal{O}(10^{-6})$.}
	\item Mapmaking can be greatly sped up by downsampling the time-ordered data,
		but this introduces a small loss of power at high $\ell$. However, high-$\ell$
		converges in just a few Conjugate Gradients steps, so this can be fixed by
		running a few CG steps with no downsampling at the end.
\end{enumerate}
With this in mind, we make maps in three passes, with pass 2 and 3
continuing from the result of the previous pass\footnote{For now
the cut and source sub-sampling degrees of freedom are solved from scratch
in each pass. While not optimal, these degrees of freedom are not bottlenecks
for the CG convergence.} while using it to
debias the noise model and subtract the polarized sidelobes. We use
300 CG steps with 4x downsampling for the first two passes; and
30 CG steps with no downsampling for the last pass.

\subsection{Noise splits}
Like in DR4, we split our data into \emph{noise splits}.
Briefly, each array-band combination for each data set
is split into 4 independent subsets (typically, see table~\ref{tab:dr6-products})
and mapped separately. These give us views of the sky
with the same signal but independent noise realizations,
allowing us to eliminate noise bias in our power spectrum
estimation by using cross-only estimators. See \cite{act-dr6-spectra}
for details on the power spectrum estimation.

\subsection{Null test maps}
Aside from the normal sky map, we also make a special set of
null test maps. The purpose of these is to maximize the effect
of various types of systematics we suspect could be in the data,
and hence maximize our ability to detect these. A disadvantage
of maximum-likelihood mapmaking is that making such null maps
is relatively expensive, so we limited ourselves to the most
well-motivated tests.

The actual null tests performed using these maps are detailed in
\cite{act-dr6-spectra}.

\subsubsection{PWV split}
The amount of precipitable water vapor affects the opacity of the
atmosphere, and therefore both the loading and the amount of correlated
noise. Loading can affect gain and saturation, while more correlated noise
can increase the low-$\ell$ power loss. To compensate for lower sensitivity
at high values of PWV, we split the TODs into two somewhat uneven
subsets: $\textrm{PWV} < 0.70 \textrm{ mm}$ with 33\% of the TODs
and $\textrm{PWV} > 0.70 \textrm{ mm}$ with 67\% of the TODs.

\subsubsection{Elevation split}
The main job of the elevation split is to test for ground pickup, which
is strongly elevation-dependent. To a lesser extent, this test also
tests for loading effects, as lower elevations have longer atmospheric
sightlines. The DR6 scans happened at 3 discrete boresight elevations:
$40\degree$, $45\degree$ and $47\degree$, each with one third of the data.
We mapped these separately for this test.

\subsubsection{Time split}
Some telescope systematics could change over long timescales. For example,
we know that the overall telescope focus changed slightly in May 2018 when
the secondary mirror axes were disabled. A time split makes us maximally sensitive
to long-term changes in telescope behavior like this. For the purpose of this
test we split the TODs into two subsets: one from before February 2019 and one after.
This relatively uneven split was driven by the wish to not dilute the
known focus change in 2018 by too much.

\subsubsection{Inner-outer split}
Unlike the other null tests, this did not split the data by TOD, but by
detector. Each detector array has its own optics tube, and we expect
optical properties to change mainly as a function of the distance from
the tube's central axis. We therefore split the detectors into
an inner and outer subset based on this distance. The cutoff radius
was chosen to give the two subsets equal sensitivity, but in practice
this was close to a 50-50 split.

\subsection{Short-timescale `Depth-1' maps}
\label{sec:depth1}
ACT observes the sky by performing broad-amplitude azimuth scans
with a typical peak-to-peak amplitude of 60$\degree$ at
constant elevation while the sky drifts past. These scans typically
last from 0.5 to 7 hours and cover 100 to 2900 square degrees to
a typical depth of around 250 µK\arcmin{} (roughly 25 mJy) before
repointing to scan a different area of the sky. See table~\ref{tab:depth1-stats}
for details. To explore
the time-variable sky we map each scan by itself in a special mapmaking run.
We call these `depth-1' maps because they are a single scan
deep, though `single scan maps' might have been more descriptive. The maps were
made with the standard maximum-likelihood framework, but since only small scales
are time-variable on human-relevant timescales, only a single pass
with 100 CG steps was used. The number of steps was driven by the need for the
cuts degrees of freedom to converge. Had it not been for this, 10 CG steps would
probably have sufficed.

Due to the large number of these maps (around 30k map-sets, 190k files total; see table~\ref{tab:dr6-products}) and
the expense of updating them, these maps have some caveats that do not
apply to our other maps, including calibration drifts and occasional artifacts and pointing outliers, as well as deviations from our normal beam shape during the subset of observations that happen during the day. See table~\ref{tab:dr6-notes} for details.

\begin{table}[ht]
	\centering
	\begin{closetabcols}[1mm]
	\begin{tabular}{ll|rrr}
		band & arr & maps & depth & depth \\
		& & {\footnotesize N} & {\footnotesize µK\arcmin{}} & {\footnotesize mJy} \\
		\hline
		f090 & PA5 & 5519 & 280 & 25 \\
		f090 & PA6 & 3145 & 230 & 22 \\
		f150 & PA4 & 5339 & 410 & 48 \\
		f150 & PA5 & 5474 & 310 & 32 \\
		f150 & PA6 & 3107 & 260 & 29 \\
		f220 & PA4 & 4928 & 780 & 78
	\end{tabular}
	\end{closetabcols}
	\caption{Depth-1 map statistics. From left to right, band, detector array,
	the number of maps and median depth ($1\sigma$) in µK\arcmin{} and mJy.
	The 10\%/90\% depth quantiles are 0.7/1.4 times the median, which mostly reflects
	differences in scanning amplitude.
	The 10\%/50\%/90\% duration quantiles are 0.28/2.5/7.1 hours.
	The area quantiles are 110/1000/2900 square degrees.
	}
	\label{tab:depth1-stats}
\end{table}

\subsection{Matched filtered maps}
\label{sec:matched-filter}
We also produce a set of matched filtered versions of the depth-1 maps to make
point source analysis easier. The matched filter map is the pixel-by-pixel answer to the
question: ``What is the maximum-likelihood flux density $F$ for a point source at
the center of this pixel, assuming the rest of the map has no signal?'' The answer is
\begin{align}
	F &= \frac{\rho}{\kappa} = \frac{B^T M^{-1} m}{\textrm{diag}(B^TM^{-1}B)} , \label{eq:mf}
\end{align}
where $m$ is the depth-1 map after converting it from µK CMB to
mJy/sr assuming a frequency of 98/150/220 GHz for the f090/f150/f220 bands,\footnote{
	Precision analysis will probably need to rescale these units to the actual
	effective bandcenter based on the bandpass and object's spectral tilt.}
$M$ is the pixel-pixel matrix of $m$, $B$ is the beam covariance matrix,
and the division is done pixel-by-pixel. We label the numerator and denominator
of this estimator $\rho$ and $\kappa$ respectively. These turn out to be
more useful to distribute than $F$ itself, because they are both linear in
the data and therefore easy to coadd over whatever timescale the user is
interested in ($\rho_\textrm{tot} = \sum_i \rho_i$, $\kappa_\textrm{tot} = \sum_i \kappa_i$).
The flux and uncertainty are then recovered as $F = \rho/\kappa$, $\sigma_F = \kappa^{-\frac12}$.

To approximate $M$ (which is not actually available due to being prohibitively
expensive to build for maps of this size) we model it as
\begin{align}
	M \approx C^{\frac12} V C^{\frac12} ,
\end{align}
where $V$ is the per-pixel white noise variance, which is diagonal in pixel
space and available from the \verb|ivar| map, and $C$ is diagonal in 2D
Fourier space, and is measured from $V^{\frac12} m$. This somewhat complicated
model allows us to handle stripy correlated noise and position-dependent depth.

\subsection{ILC maps}
\label{sec:ilc}
\cite{act-dr6-ilc} performed component separation of the DR6.01 maps
using the Needlet Internal Linear Combination NILC method. We repeat
this analysis using the updated DR6.02 maps, with no changes in
methodology.

To summarize, we use the NILC method to isolate the blackbody temperature component, which contains the CMB temperature anisotropies and kinetic Sunyaev Zeldovich effect, the blackbody E-mode polarisation component, and a map of the Compton-y effect. The NILC method minimizes the sum of the foreground and noise variance, and in the presence of noise this can leave residual foregrounds at a level comparable to the noise level, or higher in small areas where the noise or foreground properties change more rapidly than the model can account for.
To mitigate these foreground residuals we provide variations that explicitly remove known contaminants, at the cost of increased noise. To test the effectiveness of the contaminant removal we provide maps with different assumptions about the contaminant signals. For further details see \cite{act-dr6-ilc}.

\subsection{Final gain and polarization efficiency correction}
\label{sec:poleff}
After the maps were made, we performed a final calibration against Planck,
fitting a per-array gain correction and polarization efficiency. This
is described in \cite{act-dr6-spectra}, and the results are
shown in table~\ref{tab:final-cal}. As indicated in table~\ref{tab:dr6-products},
the maps we release are corrected by these factors (all close to unity),
with the exception of the depth-1 maps.

\begin{table}[ht]
	\centering
	\begin{tabular}{l|rrrrrr}
		       &    PA5 &    PA6 &    PA4 &    PA5 &    PA6 &    PA4 \\
		       &   f090 &   f090 &   f150 &   f150 &   f150 &   f220 \\
		       \hline
		gain   & 1.0111 & 1.0086 &      1 & 0.9861 & 0.9702 & 1.0435 \\
		poleff & 0.9534 & 0.9715 &      1 & 0.9545 & 0.9679 & 0.9074
	\end{tabular}
	\caption{Post-mapmaking gain and polarization efficiency
	correction factors from calibration vs. Planck. Our map
	products were corrected by multiplying them by the values
	in the ``gain'' row and dividing the polarization by the values
	in the ``poleff'' (polarization efficiency) row. This was not
	done for the depth-1 maps. PA4 f150 did not have these corrections
	applied since it failed our null tests for the power spectrum analysis.}
	\label{tab:final-cal}
\end{table}

\section{DR6 map products}
\label{sec:maps}
The DR6 map products are summarized in table~\ref{tab:dr6-products}
and its supporting tables~\ref{tab:dr6-types} and \ref{tab:dr6-notes}. The main
products are the DR6 night-time maps \verb|act_dr6.02_std_AA_night| (with supporting null
maps) and the Depth-1 maps
\verb|act_dr6.02_depth1|. We also release a set of DR5-style {\bf coadd maps} aimed
at visualization and cross-correlation science; see \cite{act-dr5-maps} for
details and caveats with these maps.
For the first time we also release our raw {\bf day-time maps}. These were
processed the same way as the night-time maps with additional cuts on
Sun sidelobes and beam deformation. Despite this, the beam in these maps
is less well understood than for the night-time maps. Finally, we
release a few miscellaneous maps targeting small areas of the sky.

\begin{table*}[htp]
	\centering
	\begin{closetabcols}[1mm]
	\begin{tabular}{lrp{12mm}crrrrp{47mm}}
		Name & Types & Notes & Split & Files & Width & Height & Size & Description \\
		\hline
		\hline
		\verb|act_dr6.02_std_AA_night|   &  msvx & NTPWG0 & 4 & 96 & 43200 & 10320 & 399 & Main DR6 data set \\
		\hline
		\verb|act_dr6.02_depth1| & mvti & NTipw dgGH & 1 & 125k & var. & var. & 20000 & Plain Depth-1 maps \\
		\verb|act_dr6.02_depth1| filtered & $\rho\kappa$ & pdgfucsH & 1 & 63k & var. & var. & 24000 & Matched filtered Depth-1 maps \\
		\hline
		Coadd maps & msV & NTp WdGd & 1 & 36 & 43200 & 10320 & 180 & Coadd into single-frequency maps. With/without day-time, with/without DR4, without Planck \\
		Planck coadd maps & msV & pWdBd & 1 & 36 & 43200 & 10320 & 180 & As above, but with Planck \\
		ILC maps & I & RL & 1 & 3 & 43200 & 10320 & 5 & NILC component separated maps \\
		Deprojected ILC maps & d & RL & 1 & 71 & 43200 & 10320 & 118 & NILC maps with explicit deprojection of tSZ/CIB/etc. \\
		\hline
		\verb|act_dr6.02_null:pwv[12]_AA_night| & msvx & NTPWG & 4 & 192 & 43200 & 10320 & 797 & Night PWV split \\
		\verb|act_dr6.02_null:el[123]_AA_night| & msvx & NTPWG & 4 & 288 & 43200 & 10320 & 1196 & Night elevation split \\
		\verb|act_dr6.02_null:t[12]_AA_night| & msvx & NTPWG & 2 & 96 & 43200 & 10320 & 399 & Night time split \\
		\verb|act_dr6.02_null:inout[12]_AA_night| & msvx & NTPWG & 4 & 192 & 43200 & 10320 & 797 & Night in/out split \\
		\hline
		\verb|act_dr6.02_std_AA_day| & msvx & NTPWGd & 4 & 96 & 43200 & 10320 & 200 & Wide day survey \\
		\verb|act_dr6.02_std_DN_day| & msvx & NTPWGd & 4 & 96 & 12911 &  2203 &  13 & North day survey \\
		\verb|act_dr6.02_std_DS_day| & msvx & NTPWGd & 4 & 96 & 13726 &  2849 &  17 & South day survey \\
		\hline
		\verb|act_dr6.02_std_GC_night| & msvx & NTPWGu & 2 & 48 & 1920 & 1560 & 1.3 & Galactic center night \\
		\verb|act_dr6.02_std_BR_night| & msvx & NTPWGu & 2 & 32 & 1080 & 1020 & 0.3 & A399–401 bridge night. No PA6 \\
		\verb|act_dr6.02_std_D5_night| & msvx & NTPWGu & 4 & 96 & 3913 & 1672 & 5.8 & D5 night \\
		\hline
		\verb|srcsamp_mask.fits| & $\bar M$ & - & - & 1 & 10800 & 2580 & 0.0026 & Model error mitigation mask. See section~\ref{sec:special-source-treatment} \\
		\verb|beam_status.txt| & b & - & - & 1 & - & - & 0.025 & Depth-1 beam status \\
		\verb|depth1_index.txt| & j & - & - & 1 & - & - & 0.0032 & Time/pos of each Depth-1 map \\
		\verb|ilc_valid_mask.fits| & M & - & - & 1 & 43200 & 10320 & 0.4 & ILC well tested here \\
		\verb|ilc_inpaint_mask.fits| & M & - & - & 1 & 43200 & 10320 & 0.4 & Strong ILC residuals inpainted here
	\end{tabular}
	\end{closetabcols}
	\caption{Summary of DR6 map products. For each type of product we list its
	name; the map types it contains (see table~\ref{tab:dr6-types});
	applicable notes from table~\ref{tab:dr6-notes};
	the number of noise splits;
	the total number of files; the width and height of each map, in pixels;
	the total size of the files, in GB; and a short description.}
	\label{tab:dr6-products}
\end{table*}

\begin{table*}[htp]
	\centering
	\begin{tabular}{p{5mm}p{16cm}}
		Type & Description \\
		\hline
		\hline
		{\bf m} & \verb|map|: Sky map with shape (3,height,width) corresponding to the Stokes parameters
			I, Q and U, in µK CMB units. The maps have been reconvolved to the standard
			nearest-neighbor pixel window. \\
		\hline
		{\bf s} & \verb|map_srcfree|: Like m, but with all point sources detected at $5\sigma$ in the full ACT
			coadd subtracted. This corresponds to median flux limit of 6.5/8.4/29 mJy
			at f090/f150/f220, but the exact limit is position-dependent. \\
		\hline
		{\bf v} & \verb|ivar|: Inverse variance map with shape (height,width) in units $1/\textrm{µK}^2$.
			Describes the Stokes I noise behavior on small scales, where the noise is
			approximately white. Q and U have half this inverse variance. \\
		\hline
		{\bf V} & \verb|ivar|: Like v, but with all of Stokes I, Q and U present. \\
		\hline
		{\bf x} & \verb|xlink|: Cross-linking information, with shape (3,height,width).
		The three fields in each pixel are given by
		$\text{xlink}_p = \sum_{s\in p} W^{-1}_s [1,\cos(2\zeta_s),\sin(2\zeta_s)]$.
		Here $p$ is a pixel index, $s$ is a TOD sample that hits $p$, $W^{-1}_s$ is the white noise
		inverse variance in sample $s$, and $\zeta_s$ is the angle between the scanning direction
		in sample $s$ and the direction towards the celestial north pole. The inverse variance
		weighted average of the cosine and sine of the scanning direction is then:
		$\langle\cos(2\zeta))\rangle=\text{xlink[1]/xlink[0]}$ and $\langle\sin(2\zeta)\rangle=\text{xlink[2]/xlink[0]}$.
		If a part of the sky were hit equally by scans at all
		angles, both of these would be zero. On the other hand, an area hit only in a single
		direction would have a [cos,sin] vector with length 1.
		In general if the quantity
		$\chi = \sqrt{\langle\cos(2\zeta)\rangle^2+\langle\sin(2\zeta)\rangle^2}$
		is not close to zero, the crosslinking is poor, so this quantity can be
		useful to determine whether to mask pixels. \\
		\hline
		{\bf t} & \verb|time|: Time map with shape (height,width). The value of each
		pixel is the time at which each pixel was hit, in seconds relative to \verb|info.t|.
		Defined as the inverse variance weighted average time of all the samples
		that hit each pixel. This represents the middle of the exposure interval
		of the pixel. The first/last exposure is typically 1.6 minutes before/after. \\
		\hline
		{\bf i} & \verb|info|: HDF5 metadata file with fields:
			\verb|array|: The array and band for this file, e.g. ``pa5\_f090'';
			\verb|box|: The bounding box for the area covered by this map. (2,2)-shaped
				array with form $[[\textrm{dec}_1,\textrm{RA}_1],[\textrm{dec}_2,\textrm{RA}_2]]$,
				where the indices 1 and 2 refer to the bottom-left and top-right corners respectively;
			\verb|ids|: The TOD ids mapped in this map;
			\verb|pid|: Sequential identifier for the scans identified during
				the depth-1 mapmaking;
			\verb|t|: Unix time of the start of the scan. The time map is
				relative to this;
			\verb|period|: Unix time of start and end of the scan.
			\verb|profile|: $(2,N_\textrm{point})$-shape array giving
				dec and RA coordinates tracing out a representative path for
				a single azimuth sweep of the telescope. This can be useful when
				modelling the curvature of the stripy noise. \\
		\hline
		$\mathbf{\rho}$ & \verb|rho|: Matched filter numerator maps with shape (3,height,width)
		corresponding to Stokes I, Q and U, in units of $1/\textrm{mJy}^2$. See section~\ref{sec:matched-filter}. \\
		\hline
		$\mathbf{\kappa}$ & \verb|kappa|: Matched filter denominator maps with the same
		shape as $\rho$, in units of $1/\textrm{mJy}^2$. \\
		\hline
		I & {\bf ILC}: NILC component separated map with shape (height,width). Three components available: Compton-y, CMB blackbody T and CMB blackbody E. These maps were constructed to have a $1.6\arcmin$ Gaussian beam. See \cite{act-dr6-ilc} for details. \\
		\hline
		d & {\bf Deprojected ILC}: As above, but with one or more of the cosmic infrared background (CIB), thermal Sunyaev-Zel'Dovich (tSZ), relativistic Sunyaev-Zel'Dovich (rSZ) and their derivatives. \\
		\hline
		{\bf $\bar M$} & \verb|coarse mask|: Quarter-resolution sky mask with shape $(\textrm{height}/4,\textrm{width}/4)$. One in areas the mask applies to, zero elsewhere. For example, for \verb|srcsamp_mask.fits|, the mask is one in high-contrast areas where special model error mitigation was used. \\
		\hline
		{\bf M} & \verb|fine mask|: Like the coarse mask, but full resolution. \\
		\hline
		{\bf b} & \verb|beam status|: Whether the beam passes our beam deformation cuts or not, per TOD. One line per TOD, with columns TOD ID (e.g. \verb|1494463442.1494478197.ar4:f150|), start and end unix time (UTC seconds since 1970-01-01 00:00:00, e.g. 1494463441) for the TOD, and the status. The status is 0 if the beam passes the cuts and 1 if it fails. See appendix~\ref{sec:daytime-beam}. Combine this with the depth-1 time maps to roughly reject observations with bad beams. \\
		\hline
		{\bf j} & \verb|depth-1 index|: Rough time/position coverage of each Depth-1 map. One line per map, with format \textbf{time start}, \textbf{time end}, \textbf{RA min}, \textbf{RA max}, \textbf{dec min}, \textbf{dec max}, \textbf{name}. Times are Unix time (C time), and coordinates are in degrees. \textbf{name} is e.g. \verb|depth1_1494478923_pa5_f090|, for which a map, ivar, info, rho and kappa file would be available.
	\end{tabular}
	\caption{Description of the file types used in the DR6 map products. The ``type''
	corresponds to the letters in the ``types'' column of table~\ref{tab:dr6-products}.}
	\label{tab:dr6-types}
\end{table*}

\begin{table*}[htp]
	\centering
	\begin{tabular}{p{5mm}p{16cm}}
		Key & Description \\
		\hline
		\textbf{N} & The maps contain correlated noise with a spectrum $N_\ell \sim
		\sigma^2 [1 + (\ell/\ell_\textrm{knee})^\alpha]$ with $\alpha\sim -3$ and
		$\ell_\textrm{knee} \approx 2100/3000/3800$ at f090/f150/f220 total intensity
		and $\sim 500$ in polarization. The noise is stripy, with position-dependent
		amplitude and stripe direction. See section~\ref{sec:corrnoise} and appendix \ref{sec:spatdep}. \\
		\textbf{T} & The maps (except Planck coadd) suffer from an unexpected
		lack of power at $\ell < 500-1000$ in total intensity
		which we belive is a form of \emph{dilution bias} mainly sourced by relative
		gain errors. See section~\ref{sec:tf} and \cite{model-error}. \\
		\textbf{i} & The depth-1 maps' CG iteration was stopped early, and only
		one mapmaking pass was performed, resulting in an additional lack of power for
		$\ell \lesssim 1000$. \\
		\textbf{p} & The maps have an effective polarization
		efficiency of $\approx 95\%$. See section~\ref{sec:poleff}. \\
		\textbf{P} & The maps have been corrected for an effective polarization
		efficiency of $\approx 95\%$. See section~\ref{sec:poleff}. \\
		\textbf{0} & The PA4 f150 maps failed the null tests for the angular power spectrum part of our analysis. They suffer from higher levels of leakage and a larger transfer function than our other maps. Care should be taken when using these for precision analysis, e.g. by checking for consistency with the other, more reliable arrays. \\
		\textbf{w} & The maps were built using a nearest neighbor pointing matrix,
		resulting in a standard sinc pixel window \\
		\textbf{W} & The maps were built using a bilinear pointing matrix, but have
		been reconvolved to a standard sinc pixel window \\
		\textbf{d} & The day-time maps have a poorly characterized beam \\
		\textbf{B} & The map has a scale-dependent bandpass (around 2-5\%) due to combining
		data from different telescopes. See~\cite{act-dr5-maps}. \\
		\textbf{g} & The maps may be subject to $\mathcal{O}(10\%)$ drifts in
		gain on month-to-year timescales. \\
		\textbf{G} & The maps are contaminated by pickup (ground and other sources).
		This is relatively more important in polarization, and mostly manifests as low-$\ell$
		horizontal stripes. See figure~\ref{fig:map-raw-and-filtered}. \\
		\textbf{f} & The flat sky approximation was used locally when building
		the matched filter, resulting in a flux bias up to 1.5\% furthest from the
		equator (dec = 60$\degree$) \\
		\textbf{u} & Occasional artifacts and temporary changes in noise properties
		are not captured by the matched filter noise model, resulting in false positives
		that must be worked around in a transient search \\
		\textbf{c} & Curvature in the stripy noise was ignored in the matched filter,
		making it slightly suboptimal \\
		\textbf{s} & Areas near bright sources are contaminated by ringing from the
		matched filter \\
		\textbf{H} & These maps were not postprocessed for release to add
		detailed FITS keywords and convert from the cosmology/HEALPix
		polarization convention to the IAU one, due to the large data volume
		involved. In practice this means the Stokes U sign is flipped.
		(You will not need to worry about this if you only read the maps with
		\verb|pixell.enmap.read_map|. It automatically converts back to the
		cosmology convention when reading, if necessary, based on the POLCCONV
		FITS entry.) \\
		\textbf{L} & The plain ILC maps and those with one component explicitly deprojected are band-limited at $\ell>17\,000$. For two/three deprojected components, this number is reduced to $11\,000/4\,000$.
	\end{tabular}
	\caption{Caveats and limitations of the maps. The key column corresponds to
	letters in the ``notes'' column of table~\ref{tab:dr6-products}.}
	\label{tab:dr6-notes}
\end{table*}

\begin{figure*}
	\centering
	\hspace*{-11mm}\begin{tabular}{cc}
		\rotatebox[origin=c]{90}{\bf full} & \hspace*{-2mm}\raisebox{-0.5\height}{\includegraphics[width=\textwidth,trim=0 8.8mm 0 0,clip]{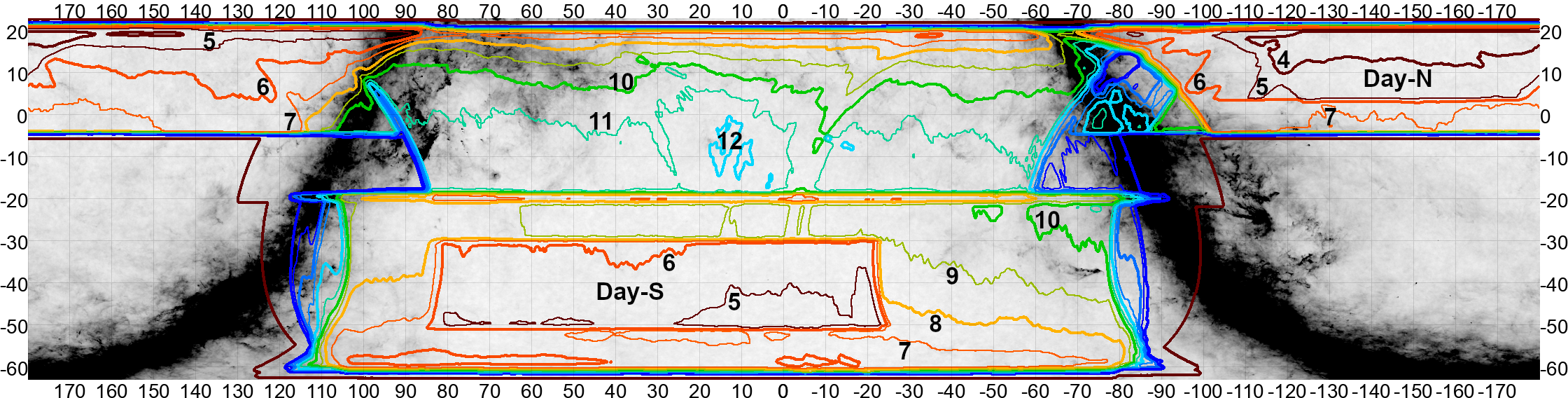}}\vspace*{1mm} \\
		\rotatebox[origin=c]{90}{\bf night}& \hspace*{-2mm}\raisebox{-0.5\height}{\includegraphics[width=\textwidth,trim=0 0 0 9.3mm,clip]{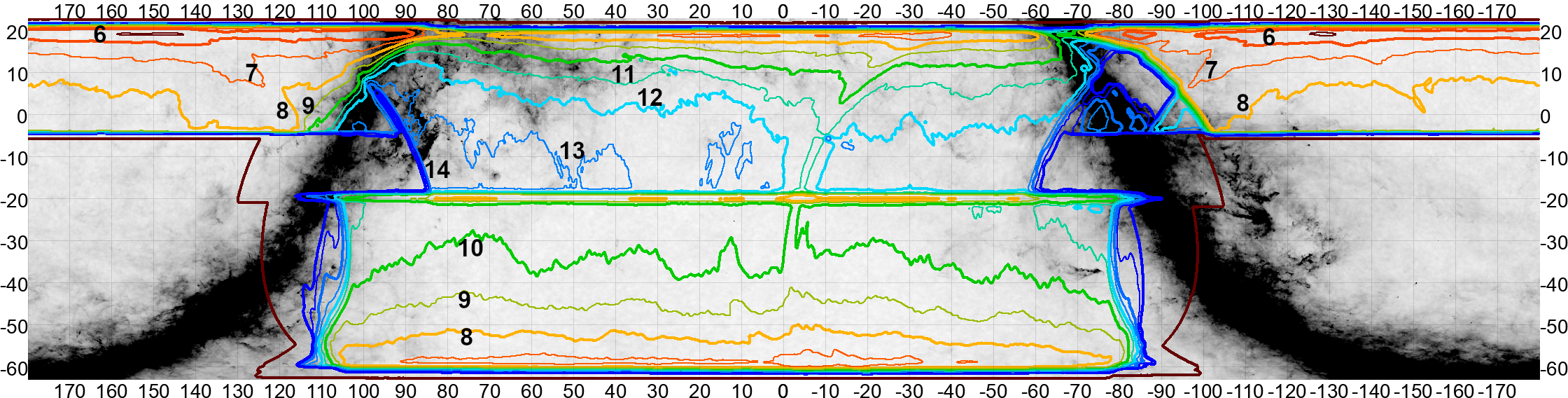}} \\
	\end{tabular}
	\caption{The ACT DR6 survey area in equatorial coordinates, with RA on the horizontal axis and dec on the vertical. The colored contours show the
	total day+night (top) and night-only (bottom) pan-frequency depth in steps of 1 \ukam{},
	going from 4 \ukam{} (dark red) to 17 \ukam{} (dark blue).
	Some contour lines are labeled for convenience.
	About half the day-time data are concentrated in the regions
	labeled ``Day-N'' and ``Day-S''. The rest are spread out over
	the whole area like the night-time data are. See figure~\ref{fig:cumrms-T-perfreq}
	for how the individual frequencies compare, but typically
	f090/f150/f220 values are 1.41/1.45/6.6 times
	higher. The outermost dark red line shows the edge of the exposed area.
	The dust-dominated Planck 353 GHz map is shown in grayscale in the
	background.}
	\label{fig:depthmap}
\end{figure*}

\begin{table*}[htp]
	\centering
	\begin{tabular}{lrrrrrll}
		Survey & Area & $f_\text{sky}$ & $\cap$ & In DR6 & Of DR6 & Type & Ref \\
		& deg$^2$ & \% & deg$^2$ & \% & \% & & \\
	\toprule
		ACT DR6 & 19400 &  47.0 & 19400 & 100.0 & 100.0 & CMB & This work \\
		BICEP3  &  1890 &   4.6 &  1510 &  80.0 &   7.8 & CMB & \cite{BK-XV} \\
		SPT-3G  &  4400 &  10.7 &  3000 &  69.0 &  15.7 & CMB & \cite{spt3g-coverage-guidi} \\
		Planck  & 41000 & 100.0 & 19400 &  47.0 & 100.0 & CMB & \cite{planck_2018_overview} \\
	\midrule
	4MOST  & 22000 & 54.0 & 13600 & 61.0 & 70.0 & Spect. & \cite{4most-coverage-web}      \\
	BOSS   & 17600 & 43.0 &  6700 & 38.0 & 35.0 & Spect. & \cite{sdss-dr16-eboss}         \\
	DESI   & 14300 & 35.0 &  7400 & 52.0 & 38.0 & Spect. & \cite{desi-bgs-2023}           \\
	Euclid & 17200 & 42.0 &  8800 & 51.0 & 46.0 & Spect. & \cite{euclid-wide-survey-2022} \\
	\midrule
	DES  &  5300 &  12.7 &  5100 & 96.0 &  26.0 & Photo. & \cite{des-dr2}           \\
	HSC  &  1500 &   3.6 &  1390 & 92.0 &   7.2 & Photo. & \cite{hsc-coverage-web}  \\
	LSST & 28000 &  68.0 & 17900 & 64.0 &  92.0 & Photo. & \cite{lsst-coverage-web} \\
	WISE & 41000 & 100.0 & 19400 & 47.0 & 100.0 & Photo. & \cite{wise-mission}      \\
	\midrule
	EMU   & 31000 & 75.0 & 19400 & 62.0 & 100.0 & Radio & \cite{emu-survey}   \\
	RACS  & 34000 & 83.0 & 19400 & 57.0 & 100.0 & Radio & \cite{racs-survey}  \\
	VLASS & 34000 & 82.0 & 16400 & 48.0 &  85.0 & Radio & \cite{vlass-survey} \\
	\end{tabular}
	\caption{ACT DR6's overlap with some current and upcoming surveys,
	as calculated from declination bounds or digitized coverage plots The columns are
	\dfn{Area}: the sky coverage in square degrees; \dfn{$f_\text{sky}$}:
	the fraction of the full sky covered; \dfn{$\pmb{\cap}$}: the overlap with
	ACT DR6, in square degrees; \dfn{In DR6}: the fraction of the survey
	inside the ACT DR6 area; \dfn{Of DR6}: the fraction of ACT DR6 covered by the survey;
	\dfn{Type}: the rough category the survey falls into;
	\dfn{Ref}: source for the coverage information.}
	\label{tab:overlaps}
\end{table*}

The DR6 sky coverage and depth is shown in figure~\ref{fig:depthmap}. We cover
$-60\degree \lesssim \text{dec} \lesssim 20\degree$ for $-80\degree \lesssim \text{RA}
\lesssim 110$ and $-5\degree \lesssim \text{dec} \lesssim 20\degree$ for RA outside this range.
When combining f090, f150 and f220 the typical night-time white noise level is 6-12 $\micro$K arcmin in total intensity, with f090/f150/f220 being on average 1.41/1.45/6.6 times higher.
Adding day-time data improves this by about 20\%. The Q and U polarization white noise is $\sqrt 2$ times as high.

\begin{figure}[htp]
	\centering
	{\bf Intensity}\vspace*{-3.5mm} \\
	\hspace*{-5mm}\includegraphics[width=1.1\columnwidth]{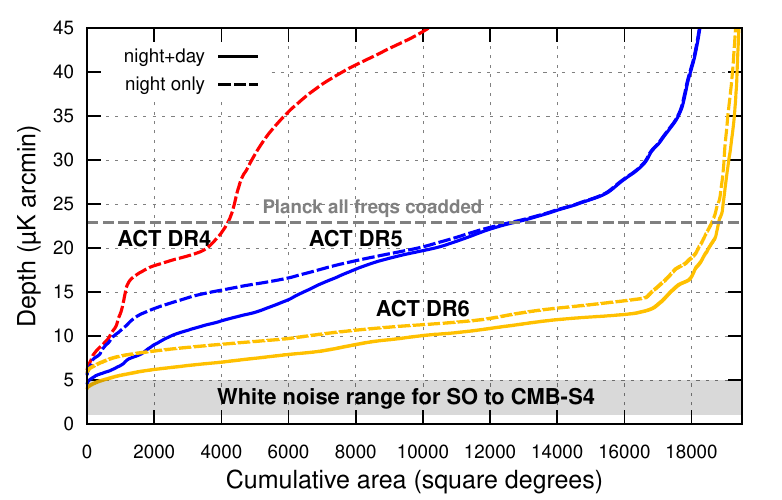}
	{\bf Polarization}\vspace*{-2.8mm} \\
	\hspace*{-5mm}\includegraphics[width=1.1\columnwidth]{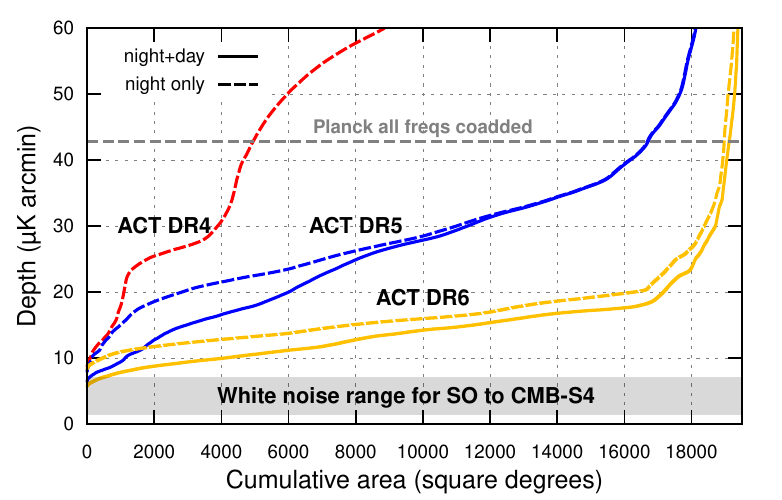}
	\caption{\dfn{Top}: Frequency-combined total intensity map depth
	distributions for ACT DR4 (red), DR5 (blue) and DR6 (yellow, this work)
	along with typical depth achieved by Planck HFI (dashed gray).
	The depths anticipated for Simons Observatory and CMB-S4 fall inside
	the shaded gray band.
	DR6 is almost twice as deep (in RMS) as the non-cosmology-calibrated DR5, and more than four
	times as deep as DR4 (our previous cosmology data release) over most of
	the sky. The typical DR6 combined depth is $9.6\micro$K arcmin, and 19\,000
	square degrees are deeper than $20\micro$K arcmin. For Planck the
	sky average noise level is shown; the actual depth varies by a factor
	of $1.4^{\pm 1}$ at 95\%. \dfn{Bottom}: As top, but for polarization.
	Here the improvement over Planck is even larger.}
	\label{fig:cumrms}
\end{figure}

\begin{figure}[htp]
	\centering
	\hspace*{-5mm}\includegraphics[width=1.05\columnwidth]{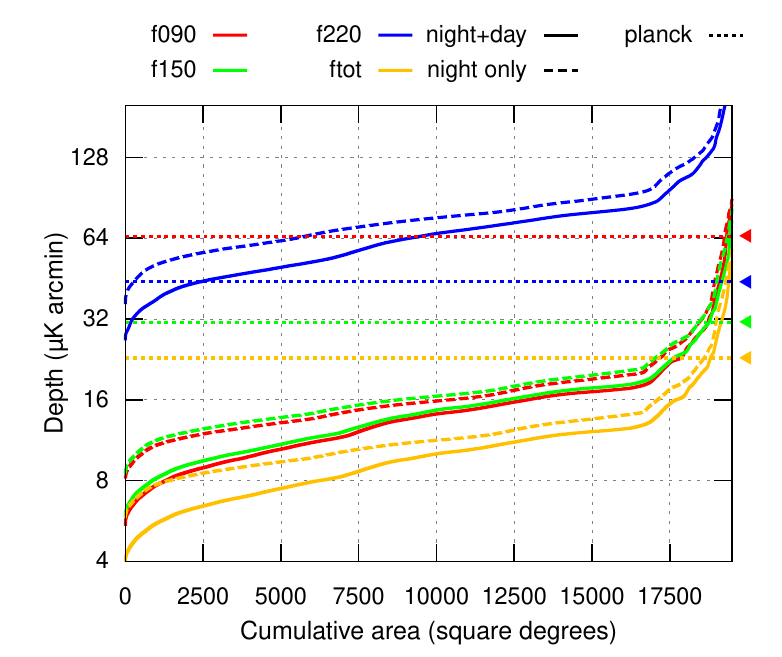}
	\caption{ACT DR6 total intensity map depth distribution for the individual frequency
	bands. The solid curves represent the combined day+night data sets, while
	the dashed curves use only night-time data. The median depth inside the
	exposed area of about 19\,000 square degrees is
	14/14/64/9.6 $\micro$K arcmin for day+night at f090/f150/f220/ftot and
	15/16/74/11 $\micro$K arcmin for night alone.
	The dotted lines show the mean Planck
	depth at these frequencies (65/31/44/23) $\micro$K arcmin).}
	\label{fig:cumrms-T-perfreq}
\end{figure}

\begin{table}[htp]
	\centering
	\begin{tabular}{lrrr}

	\end{tabular}
\end{table}

Figure~\ref{fig:cumrms} shows the depth distribution and compares it with DR4, DR5 and Planck.
DR6 is almost twice as deep (in RMS) as the non-cosmology-calibrated DR5, and more than four
times as deep as DR4 (our previous cosmology data release) over most of
the sky. The typical DR6 combined depth is $9.6\micro$K arcmin, and 19\,400
square degrees (47\% of the sky) are deeper than $20\micro$K arcmin.
Figure~\ref{fig:cumrms-T-perfreq}
splits this into our individual bands. The median depth inside the
exposed area is 14/14/64/9.6 $\micro$K arcmin for day+night at f090/f150/f220/ftot and
15/16/74/11 $\micro$K arcmin for night alone. Here ``ftot'' refers to the coadd
of f090, f150 and f220.

This large sky coverage gives us good overlap with several surveys relevant for
cross-correlation studies, as shown in table~\ref{tab:overlaps} and
figure~\ref{fig:footprint}. We cover
most of BICEP3, SPT-3G, 4MOST, DES, DESI and LSST, to name a few other surveys.

\begin{figure*}[p]
	\centering
	\hspace*{-1cm}\includegraphics[width=20cm]{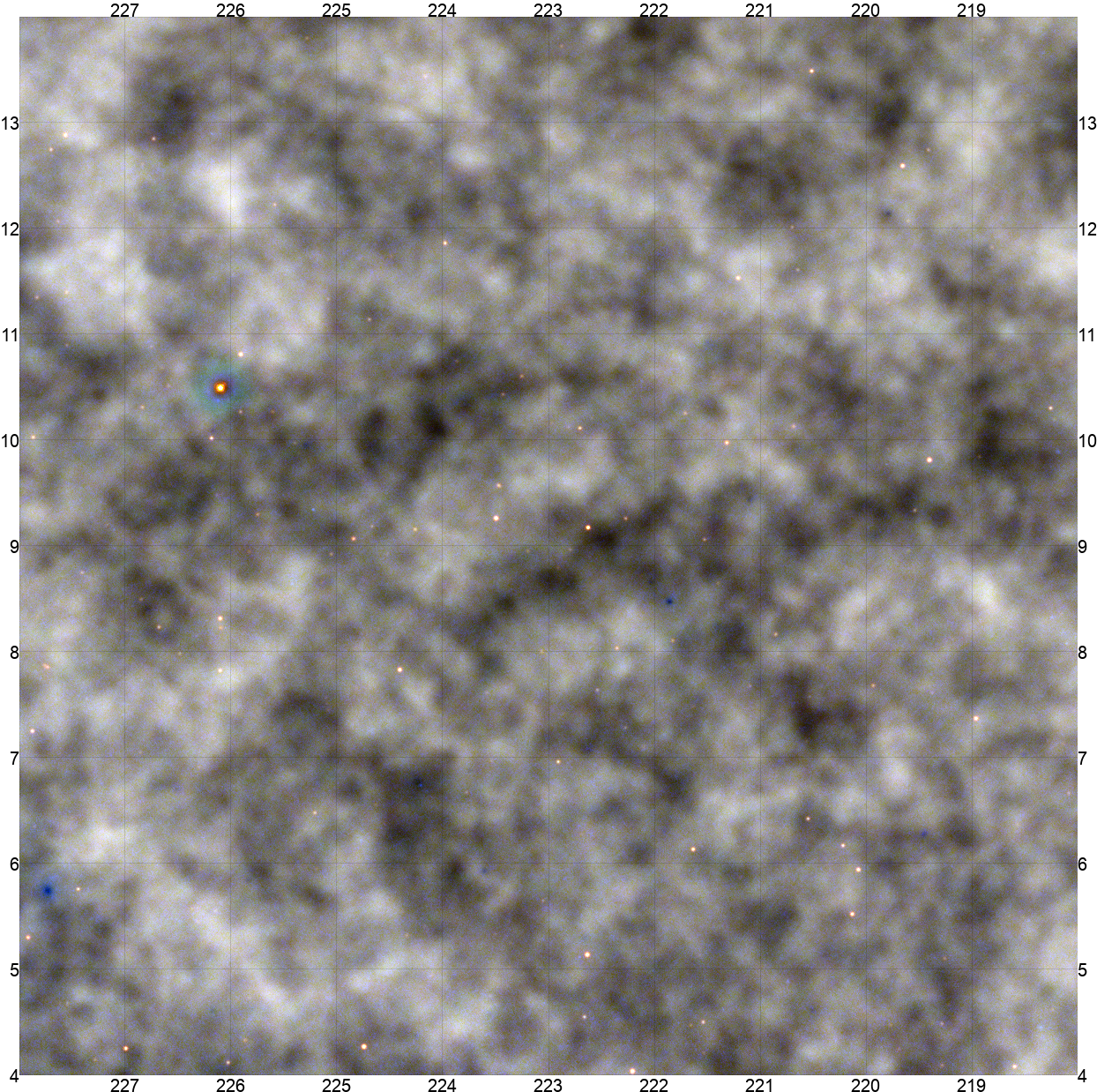}
	\caption{Multifrequency image from ACT DR6 day+night coadded with Planck, showing a 100 square degree (0.5\%)
	subset of the full DR6 area. Here total intensity in the f090/f150/f220 bands is mapped to
	red/green/blue with a 400 $\micro$K color range. In the coadd, Planck dominates on scales
	larger than around $0.3\degree$, and ACT on smaller scales. Active galactic nuclei appear
	bright orange; galaxy clusters are dark blue due to the thermal Sunyaev Zel'dovich effect;
	and nearby dusty galaxies are light blue (but only a few are faintly visible here).
	The CMB itself is gray since its power is frequency-independent in these units.
	Galactic dust emission is faintly visible in blue on large scales.
	The image is signal-dominated on most scales, but some noise is still faintly visible, especially
	in the blue f220 band where ACT is less sensitive. The full map can be seen at \url{https://phy-act1.princeton.edu/public/snaess/actpol/dr6/atlas}, or in the Aladin sky viewer.}
	\label{fig:map-rgb}
\end{figure*}

\begin{figure*}[htp]
	\centering
	\begin{closetabcols}
	\begin{tabular}{cccc}
		{\bf \large Q} &
		\img[width=7.95cm,trim=0 8mm 9.5mm 0]{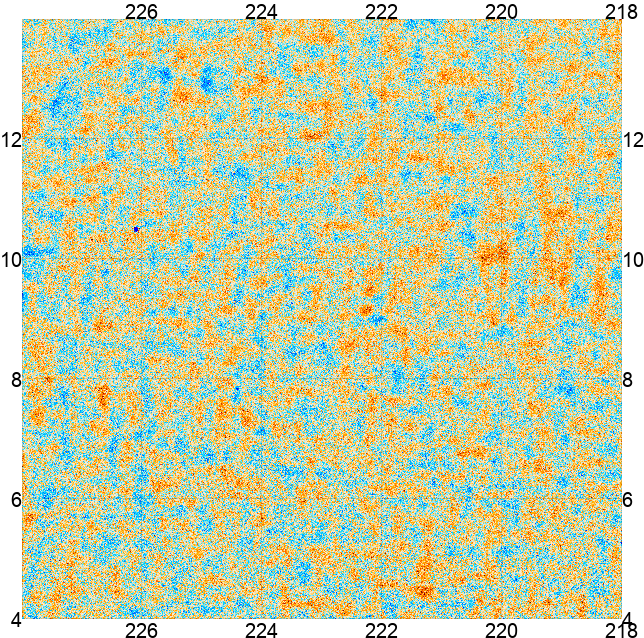} &
		\img[width=8cm,trim=8mm 8mm 0 0]{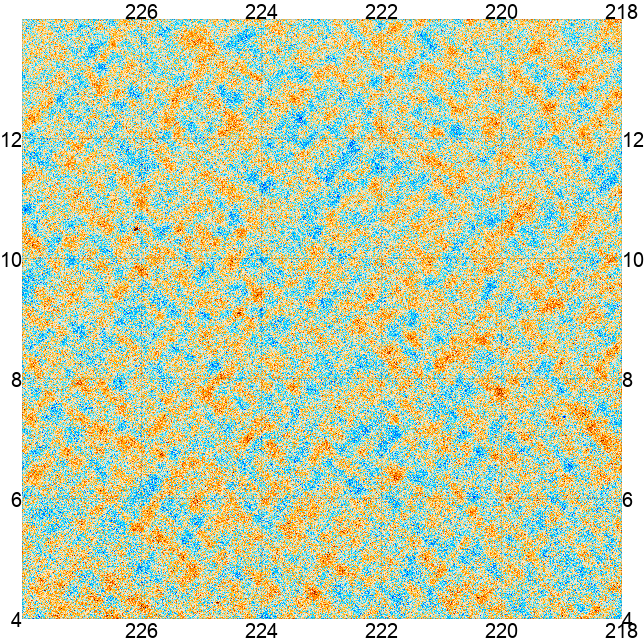} &
		{\bf \large U} \\
		{\bf \large E} &
		\img[width=7.95cm,trim=0 0 9.5mm 8mm]{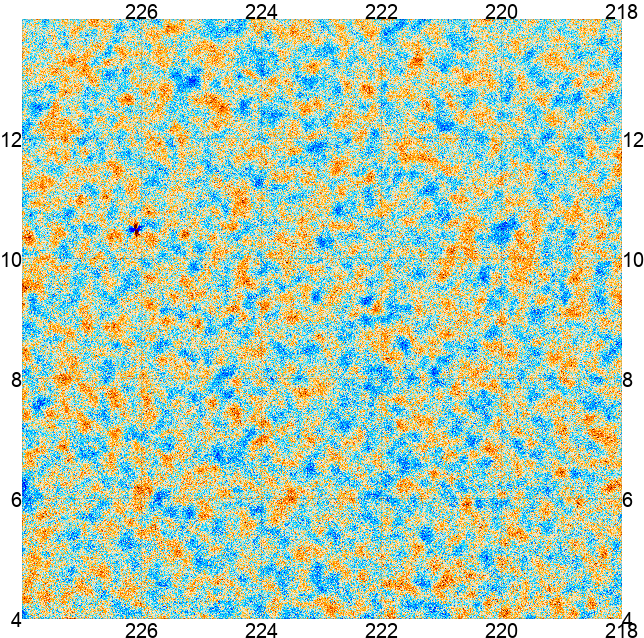} &
		\img[width=8cm,trim=8mm 0 0 8mm]{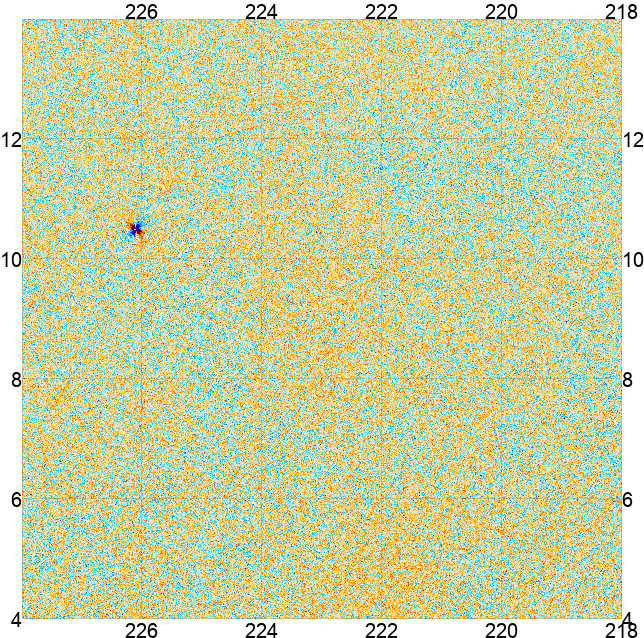} &
		{\bf \large B}
	\end{tabular}
	\end{closetabcols}
	\caption{The same area as in figure~\ref{fig:map-rgb}, but this time showing f090+f150 polarization
	with a $\pm 30 \micro$K color scale. As shown in figure~\ref{fig:depthmap} the white noise
	level is around 4 $\micro$K arcmin in this area. \dfn{Top}: The Stokes Q and U parameters.
	We recognize the + pattern in Q and $\times$ pattern in U characteristic of E-modes.
	\dfn{Bottom}: E and B polarization maps. The E-modes are signal-dominated while the
	B-modes are visually indistinguishable from our noise, except for a polarized point source.
	(The point source appears as a small quadrupole in E and B due to the non-local relationship
	between Q,U and E,B.)
	}
	\label{fig:map-EB}
\end{figure*}

\begin{figure*}[htp]
	\centering
	\begin{closetabcols}
	\begin{tabular}{cccccc}
		& $\mathbf{T}$ & $\mathbf{Q_r}$ & $\mathbf{U_r}$ & $\mathbf{E}$ & $\mathbf{B}$ \\
		\rotatebox[origin=c]{90}{\bf Planck} &
		\img[width=0.19\textwidth]{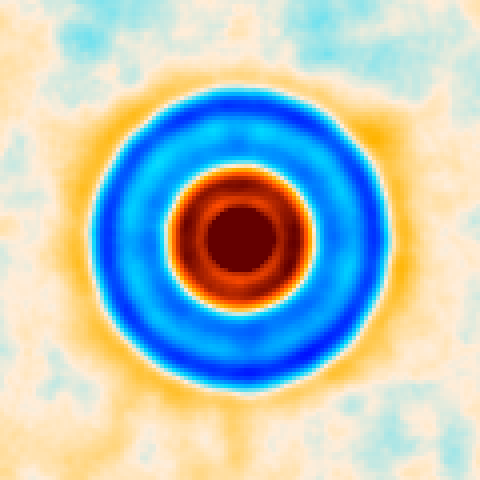} &
		\img[width=0.19\textwidth]{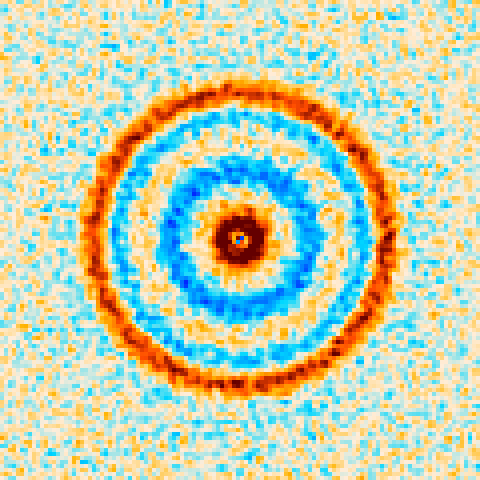} &
		\img[width=0.19\textwidth]{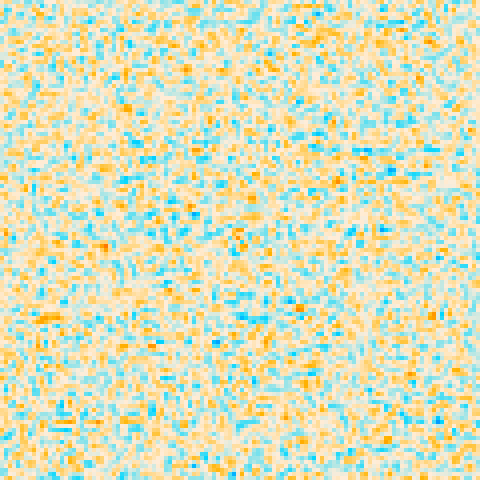} &
		\img[width=0.19\textwidth]{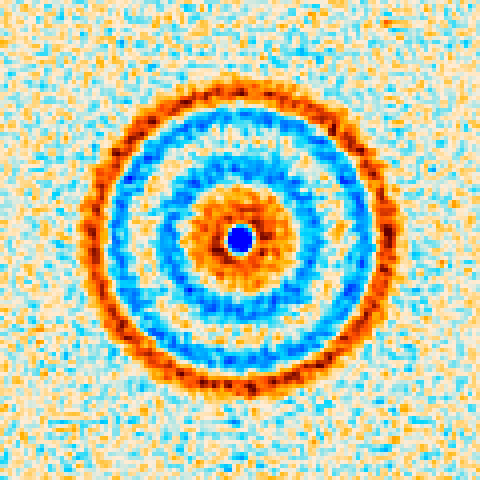} &
		\img[width=0.19\textwidth]{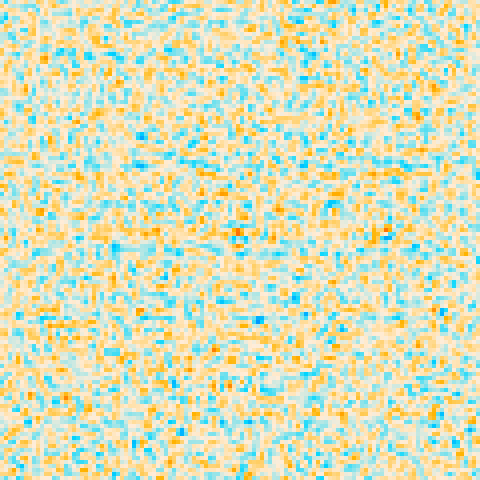} \\
		\rotatebox[origin=c]{90}{\bf ACT} &
		\img[width=0.19\textwidth]{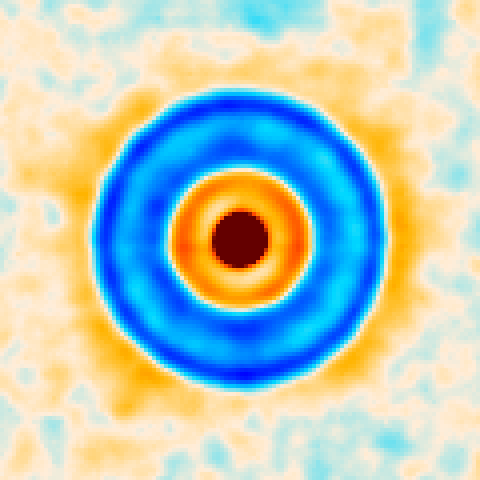} &
		\img[width=0.19\textwidth]{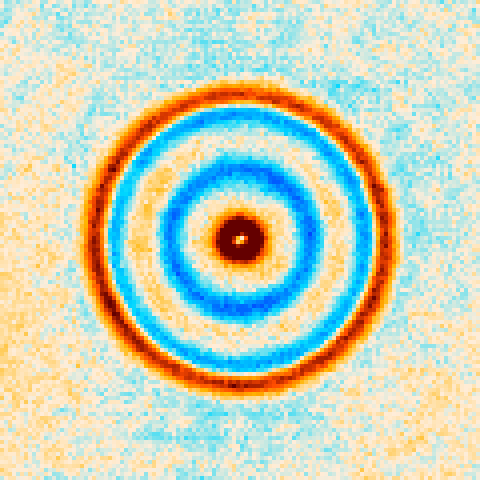} &
		\img[width=0.19\textwidth]{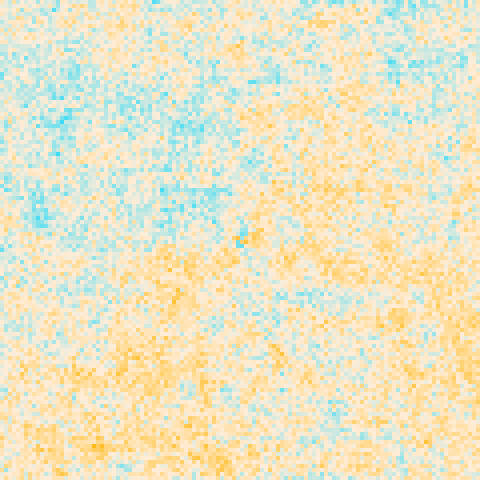} &
		\img[width=0.19\textwidth]{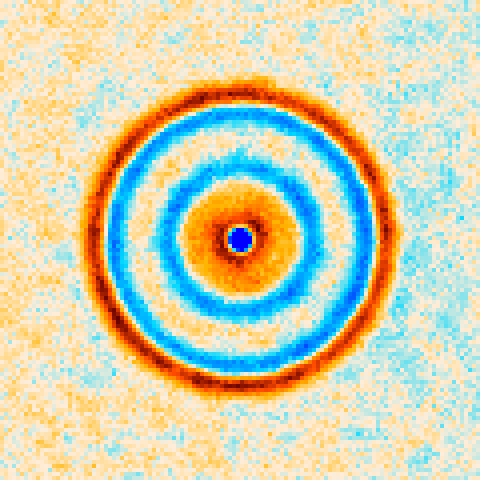} &
		\img[width=0.19\textwidth]{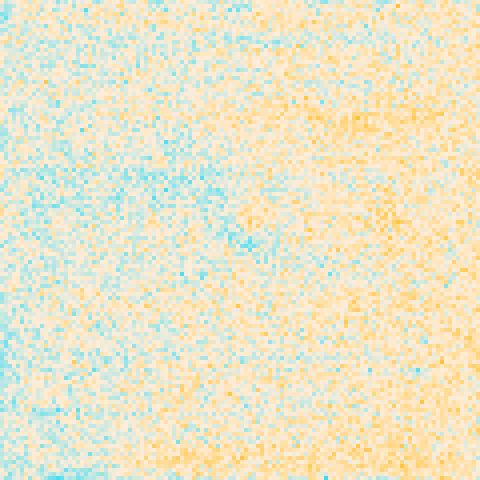} \\
		\rotatebox[origin=c]{90}{\bf ACT+Planck} &
		\img[width=0.19\textwidth]{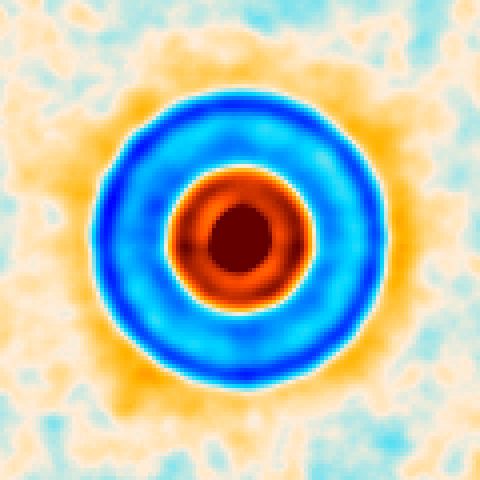} &
		\img[width=0.19\textwidth]{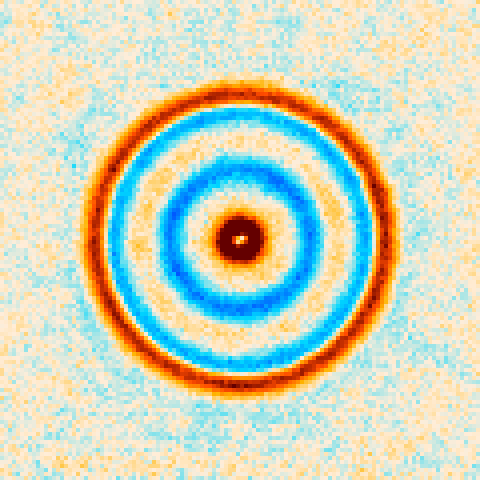} &
		\img[width=0.19\textwidth]{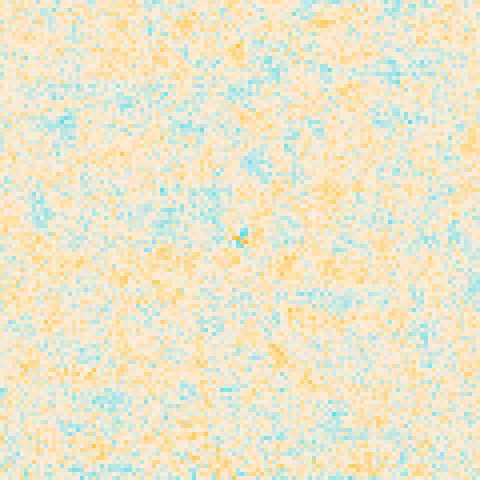} &
		\img[width=0.19\textwidth]{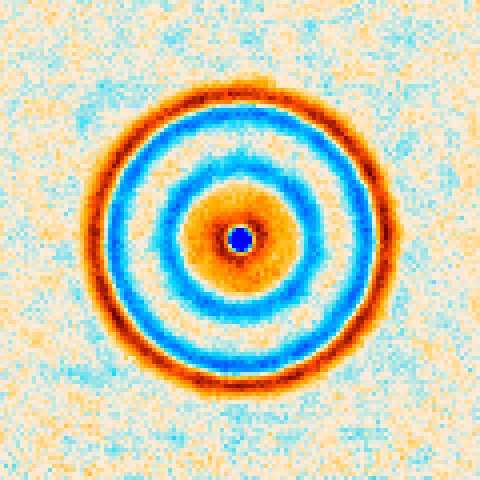} &
		\img[width=0.19\textwidth]{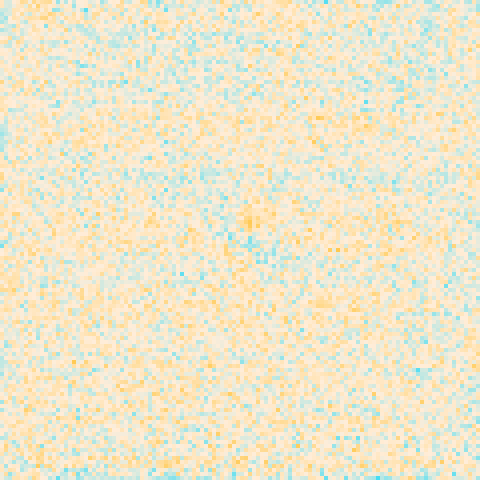} \\
		\rotatebox[origin=c]{90}{\bf Simulation} &
		\img[width=0.19\textwidth]{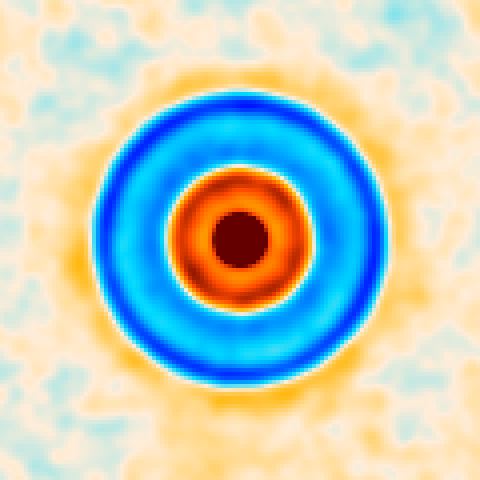} &
		\img[width=0.19\textwidth]{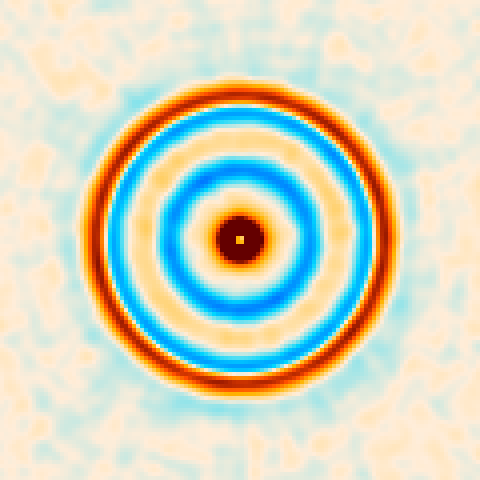} &
		\img[width=0.19\textwidth]{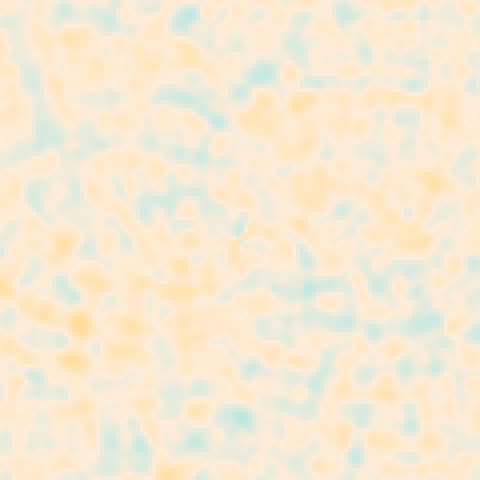} &
		\img[width=0.19\textwidth]{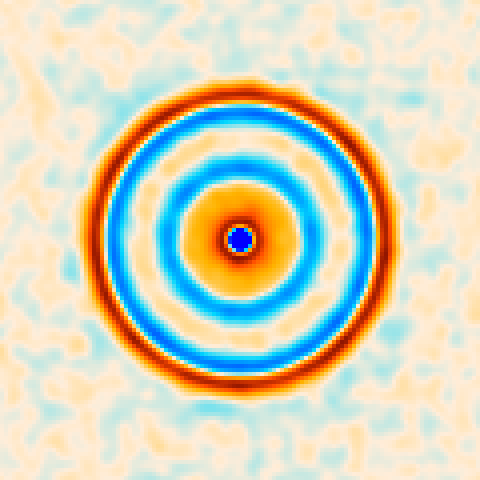} &
		\img[width=0.19\textwidth]{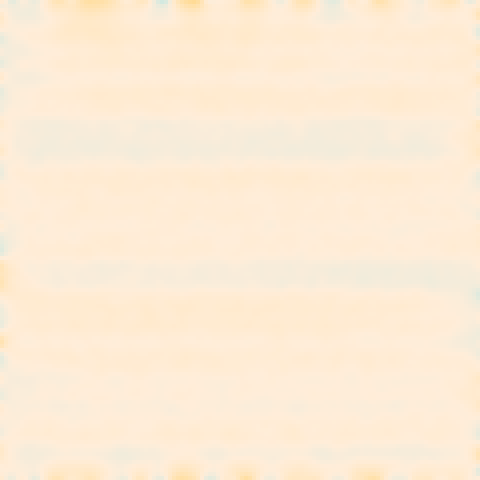}
	\end{tabular}
	\end{closetabcols}
	\caption{$4\deg\times4\deg$ stacks of $T$, $Q_r$, $U_r$, $E$ and $B$ on peaks in $T$
	for Planck SMICA (row 1), an ACT night-only coadd map (row 2),
	an ACT+Planck day+night coadd map (row 3), and a beam-less, noise-less simulation (row 4).
	These stacks are closely related to the CMB autocorrelation function, and give
	a simple way to illustrate the causal structure of the surface of last scattering.
	The outermost circle at $1.2\deg$ represents the autocorrelation of the $0.6\deg$
	sound horizon at the surface of last scattering.
	The images are normalized to give the outer ring an amplitude of 1 ($U_r$ and $B$
	where there is no ring use the same normalization as ($Q_r$ and $E$
	respectively). Peak detection was done separately for each map, and
	beam deconvolution was performed. See appendix~\ref{sec:stacking} for details on the
	peak detection and stacking procedure.
	}
	\label{fig:stack-T}
\end{figure*}

\begin{figure*}[htp]
	\centering
	\begin{closetabcols}
	\begin{tabular}{cccccc}
		& $\mathbf{T}$ & $\mathbf{Q_r}$ & $\mathbf{U_r}$ & $\mathbf{E}$ & $\mathbf{B}$ \\
		\rotatebox[origin=c]{90}{\bf Planck} &
		\img[width=0.19\textwidth]{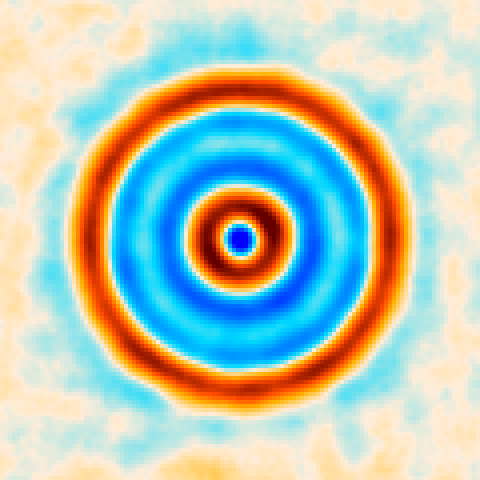} &
		\img[width=0.19\textwidth]{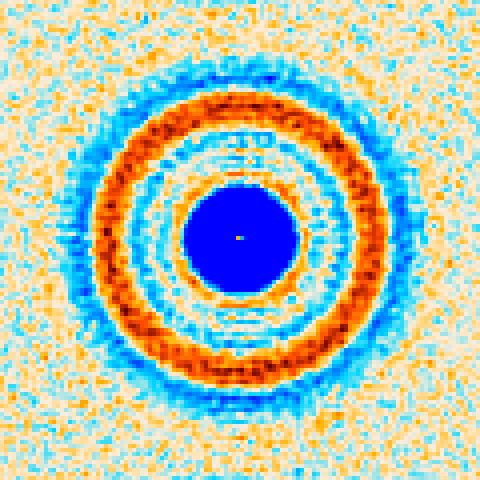} &
		\img[width=0.19\textwidth]{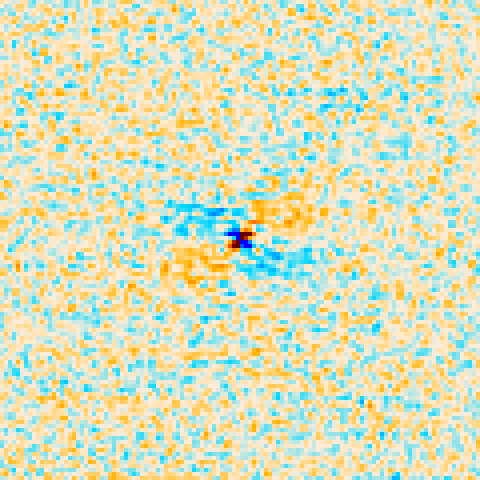} &
		\img[width=0.19\textwidth]{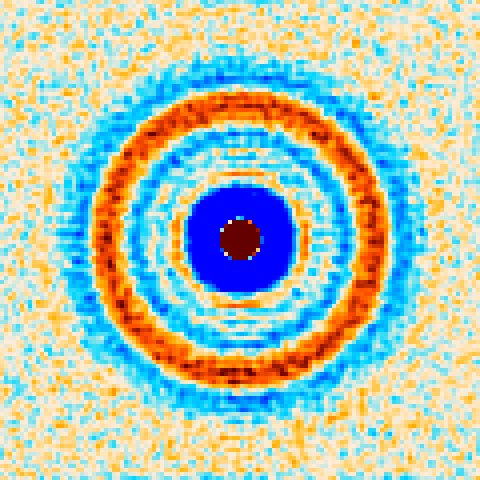} &
		\img[width=0.19\textwidth]{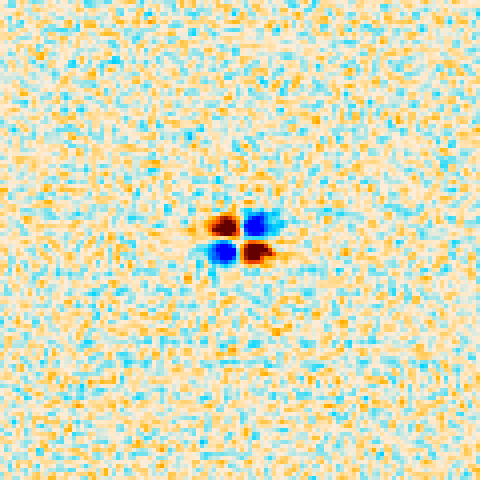} \\
		\rotatebox[origin=c]{90}{\bf ACT} &
		\img[width=0.19\textwidth]{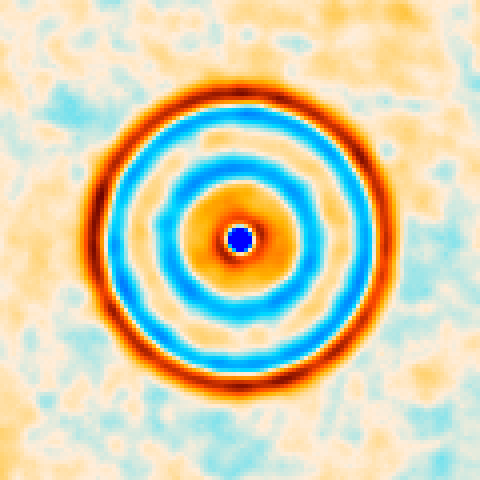} &
		\img[width=0.19\textwidth]{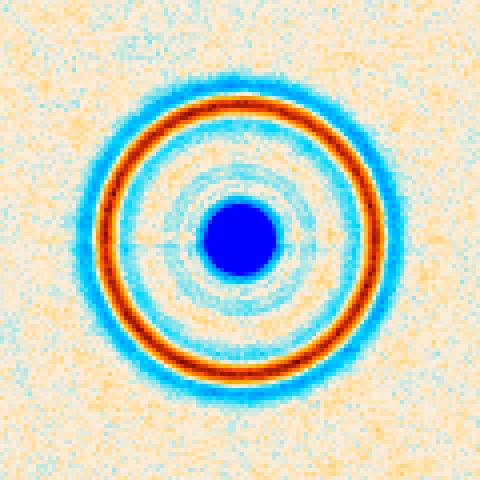} &
		\img[width=0.19\textwidth]{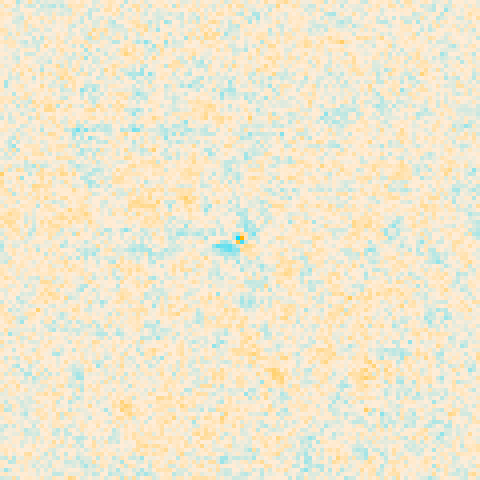} &
		\img[width=0.19\textwidth]{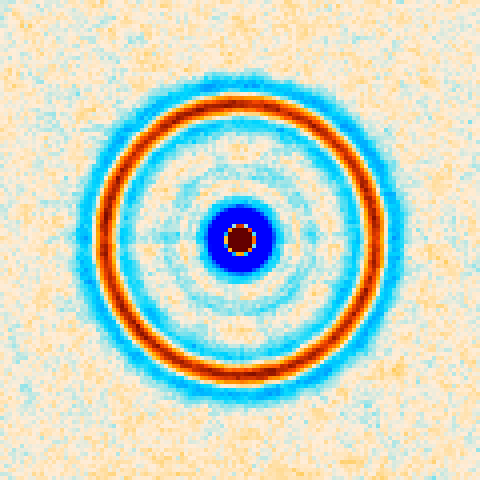} &
		\img[width=0.19\textwidth]{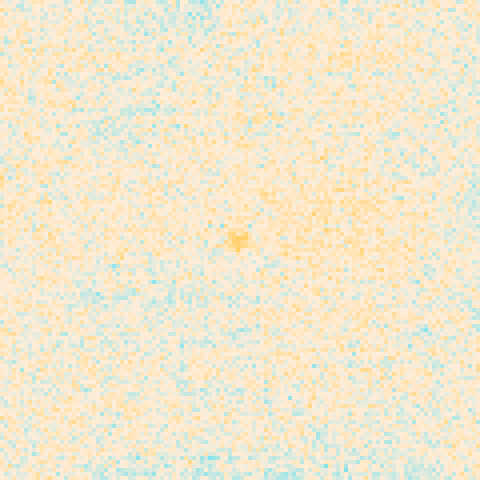} \\
		\rotatebox[origin=c]{90}{\bf ACT+Planck} &
		\img[width=0.19\textwidth]{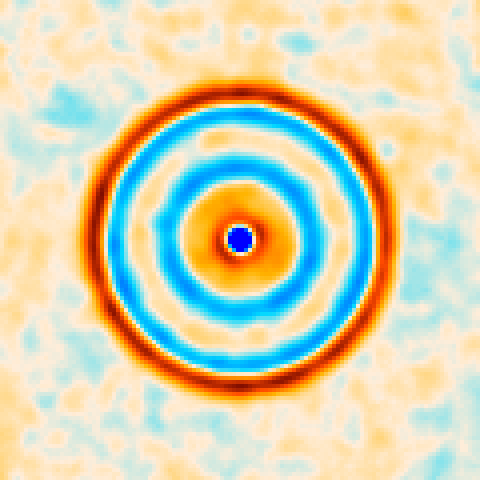} &
		\img[width=0.19\textwidth]{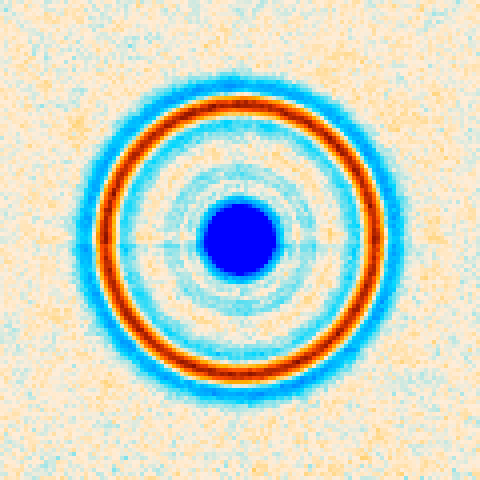} &
		\img[width=0.19\textwidth]{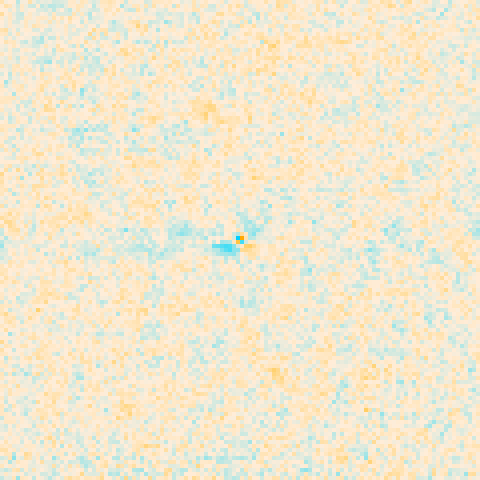} &
		\img[width=0.19\textwidth]{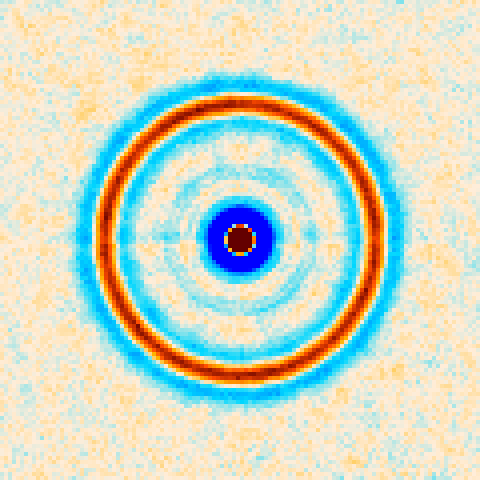} &
		\img[width=0.19\textwidth]{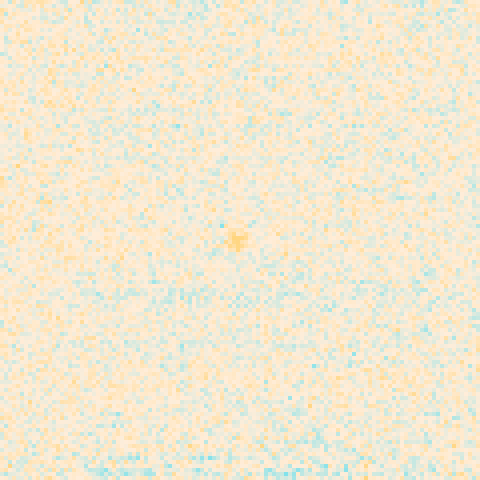} \\
		\rotatebox[origin=c]{90}{\bf Simulation} &
		\img[width=0.19\textwidth]{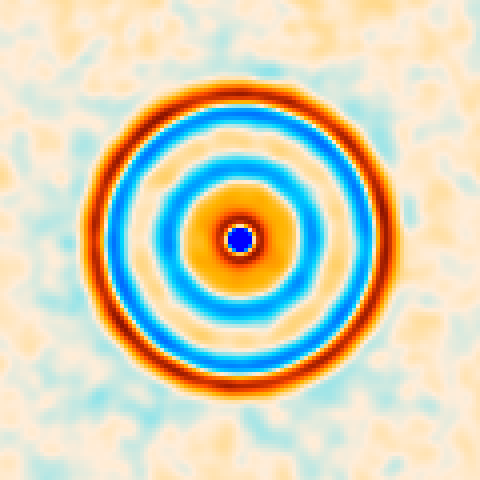} &
		\img[width=0.19\textwidth]{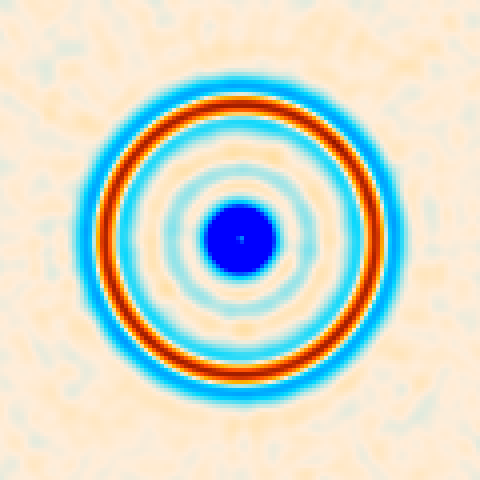} &
		\img[width=0.19\textwidth]{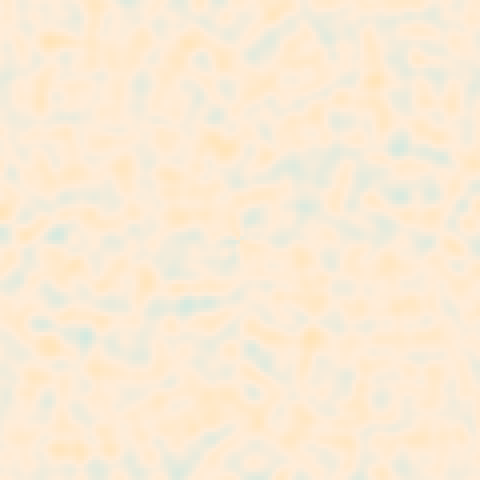} &
		\img[width=0.19\textwidth]{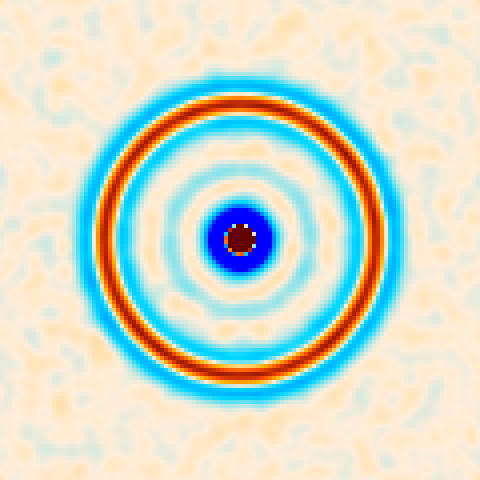} &
		\img[width=0.19\textwidth]{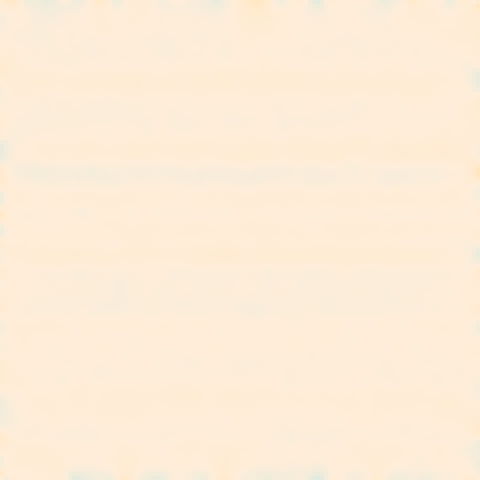}
	\end{tabular}
	\end{closetabcols}
	\caption{Like figure~\ref{fig:stack-T}, but stacks on peaks in $E$
	instead of $T$. The broader features in the Planck stacks are
	caused by it's low S/N for E-modes resulting in lower quality
	peak detection. The faint signal in the ACT B-on-E stack is consistent
	with our $\sim 0.2\deg$ EB power spectrum nulling angle (see \cite{act-dr6-spectra}).
	The B-on-E feature in Planck was previously seen in \cite{planck-parity}.
	}
	\label{fig:stack-E}
\end{figure*}

\begin{figure*}[p]
	\centering
	\hspace*{-1cm}\includegraphics[width=20cm]{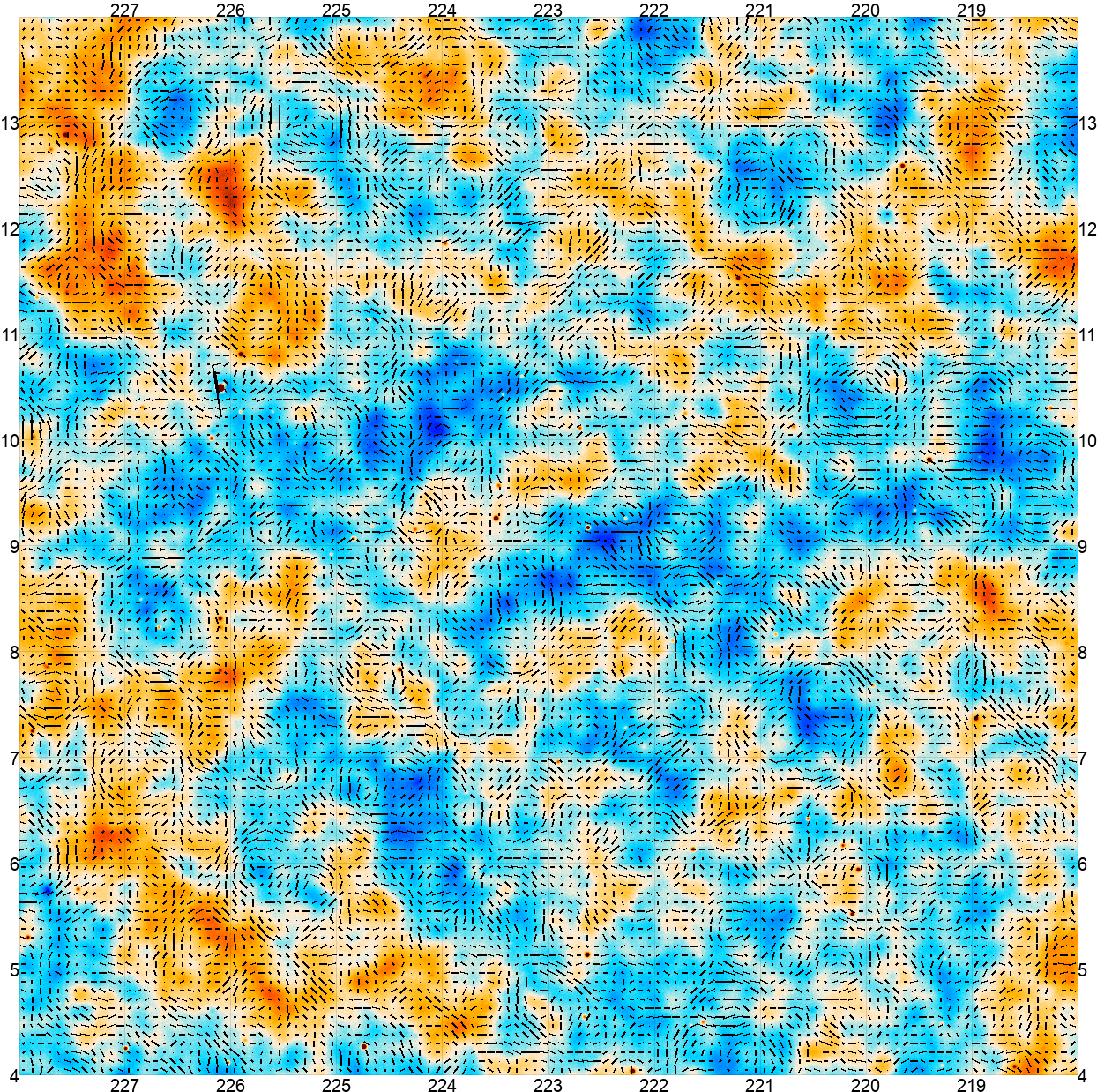}
	\caption{The same area as in figure~\ref{fig:map-rgb}, but this time showing polarization vectors
	overlaid on a total intensity map. Both the map and vectors are from an ACT DR6+Planck f090+f150
	coadd to maximize S/N, but in polarization ACT completely dominates. The theoretical TE correlation is quite low and has a scale-dependent sign, so no clear visual correspondence between the intensity and polarization fields is expected.}
	\label{fig:map-polvecs}
\end{figure*}

\begin{figure*}[htp]
	\centering
	\includegraphics[width=16cm]{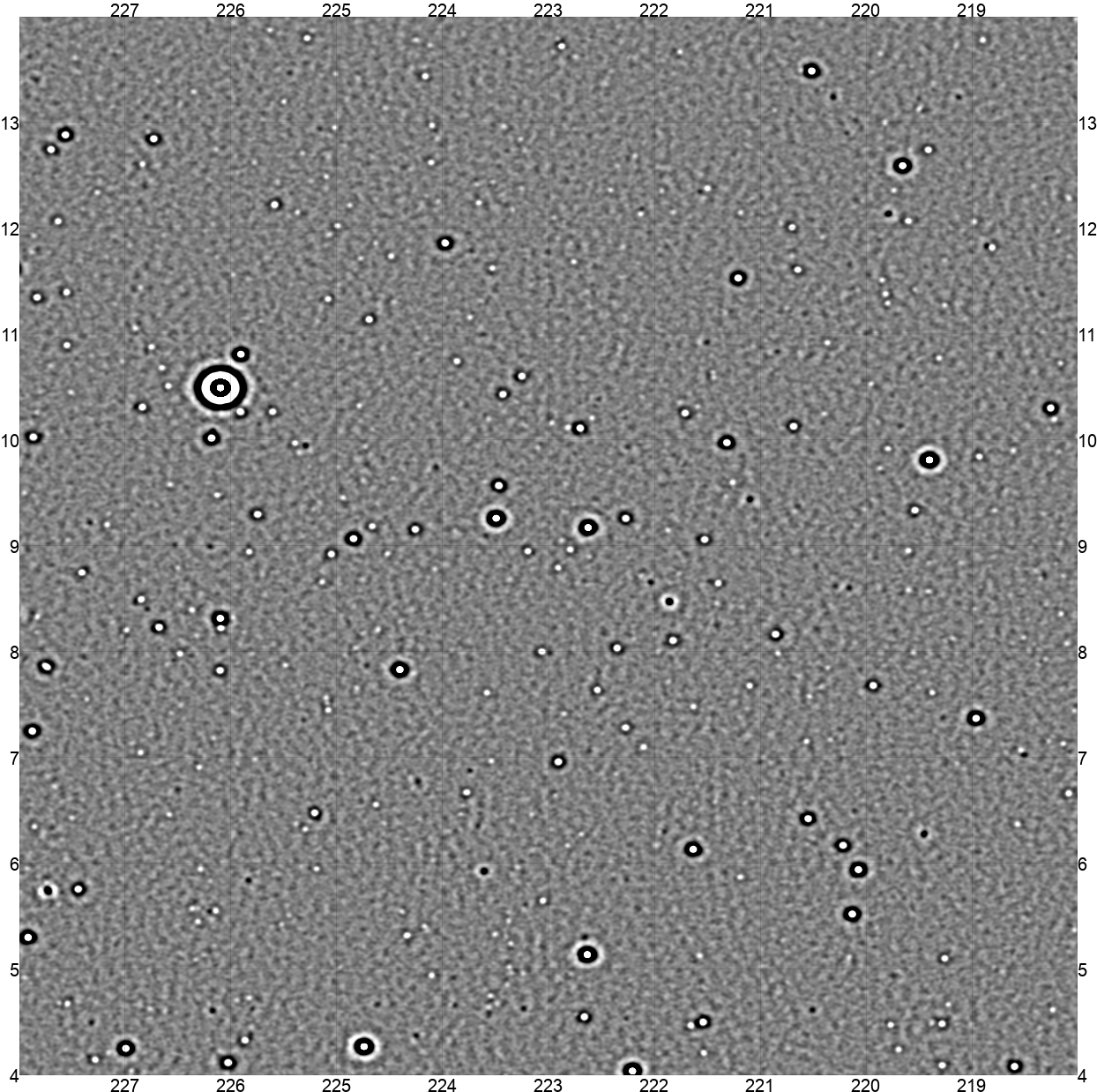}
	\caption{The same area as in figure~\ref{fig:map-rgb}, but this time showing f090 total intensity
	after applying a matched filter optimized for point sources. We detect $>300$ point sources (white dots)
	and $>50$ clusters (black dots, about $\frac12$ per square degree) at $>5\sigma$ in this 100 square degree area. Over the full ACT area, we detect 30\,000 point sources and 6\,000 clusters at $>5\sigma$ (1.9 per square degree and 0.40 per square degree over 16\,000 square degrees after masking).}
	\label{fig:map-srcs}
\end{figure*}

Figure~\ref{fig:map-rgb} shows a multifrequency view of a 100 square degree subset of
the ACT DR6 data, with the f090/f150/f220
bands mapped to the red/green/blue color channels of the image. This paints the CMB as a gray
fog (due to having the same amplitude at all frequencies in these units); synchrotron-dominated
active galactic nuclei have a falling spectrum and therefore show up as bright orange;
the thermal Sunyaev Zel'dovich effect in galaxy clusters causes a power deficit in f090 and f150
but not in f220, and therefore shows up as dark blue spots; while a few nearby dusty galaxies
are faintly visible in light blue. To avoid large-scale atmospheric noise visually dominating the
image, we have coadded it with Planck for this plot, with Planck dominating on scales larger
than about 1/3 of a degree.

Polarization for the same 100 square degree area can be found in
figure~\ref{fig:map-EB}, where we see signal-dominated E-modes and B-modes consistent with noise.
While this is one of the deepest areas of the DR6 day+night map, with a frequency-combined
white noise level of 4 $\micro$K arcmin, there are signal-dominated E-modes over the entire
19\,000 square degree DR6 area.

Figures~\ref{fig:stack-T} and \ref{fig:stack-E} stack our maps on peaks in T and E respectively, as first done for WMAP in \cite{komatsu/etal/2011}, and compare them with Planck and simulations. The baryon-acoustic feature stands out with high signal-to-noise, giving a striking illustration of the causal structure at the surface of last scattering.

Finally, figure~\ref{fig:map-polvecs} overplots polarization
vectors on total intensity to illustrate the correlation between the two, while figure~\ref{fig:map-srcs}
filters total intensity to highlight point sources, revealing $>300$ point sources and $>50$ clusters
at $>5 \sigma$ in this 100 square degree area of the sky. Over the full ACT area, we detect 30\,000 point sources and 6\,000 clusters at $>5\sigma$ (1.9 per square degree and 0.40 per square degree over 16\,000 square degrees after masking).

\begin{figure*}[htp]
	\centering
	\textbf{T map}\\
	\img[width=15cm]{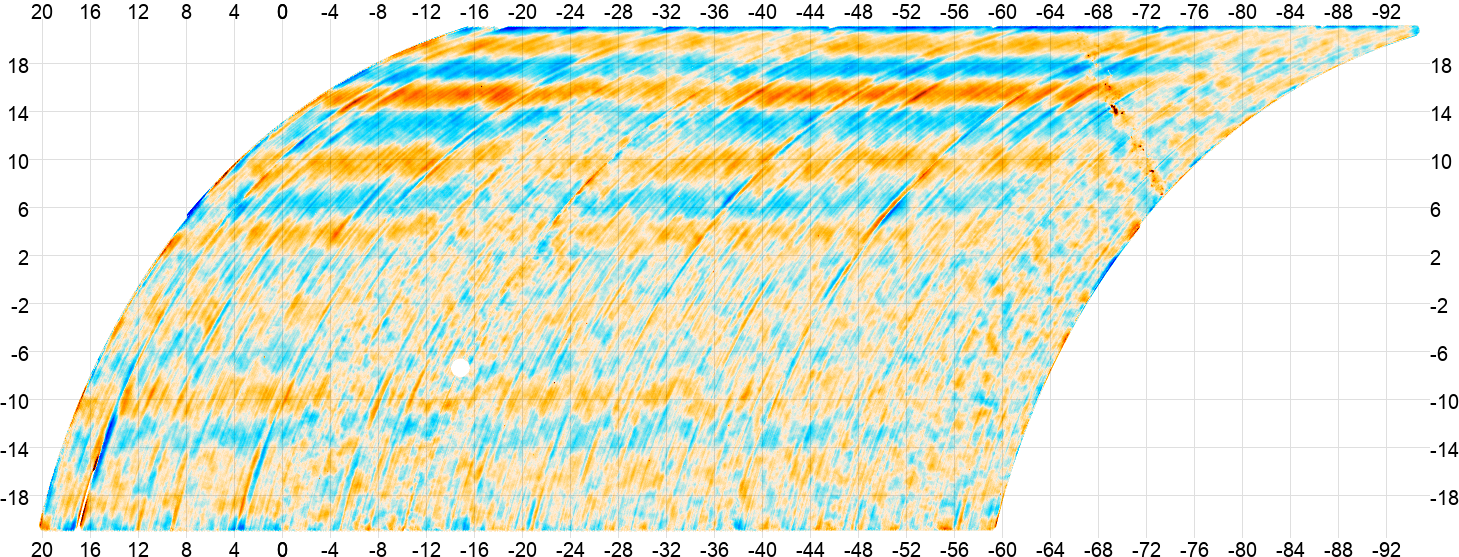} \\
	\medskip
	\textbf{Time map} \\
	\img[width=15cm]{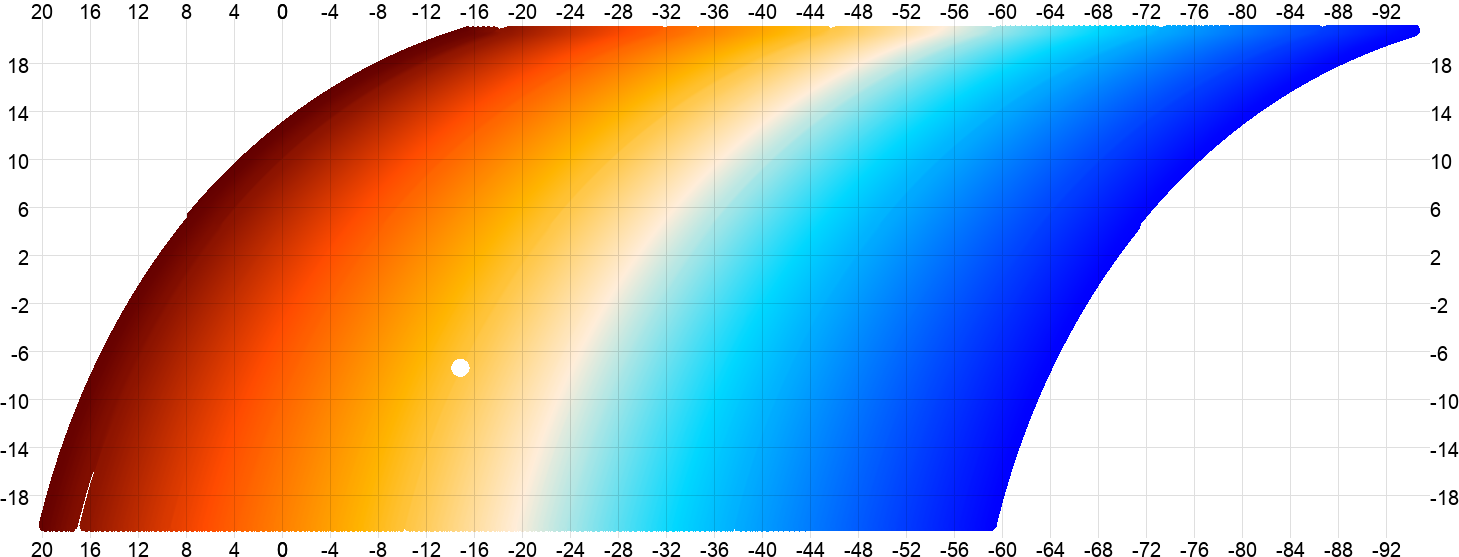}
	\caption{Depth-1 map for PA5 f090 for 5.3 hours of continuous scanning
	starting at unix time 1501030967, corresponding to 2017-07-26 01:02:47 UTC.
	RA and dec are on the horizontal and vertical axes respectively.
	\dfn{Top}: The total intensity map (T map) on a $\pm 5$ mK color scale. The map is
	noise-dominated, except for a few point sources and parts of the Milky Way.
	The circular hole is a planet cut. The horizontal bands are ground pickup,
	while the curvy stripes parallel to the left and right map edge are correlated noise. The levels of these
	vary considerably from map to map, see figure~\ref{fig:pointing-map} for
	a less stripy example. \dfn{Bottom}: Associated map showing when
	each pixel was hit. This is an average over the time at which each detector
	hits the pixel, which varies by a few minutes. Dark blue is 0 hours after
	the start, dark red is 5.3 hours after the start.}
	\label{fig:depth1-map}
\end{figure*}

Figure~\ref{fig:depth1-map} is a low-resolution plot of a typical depth-1 map.
No CMB is visible in these shallow maps, but bright point
sources and parts of the Milky Way are still visible. The time at which each
pixel was hit is also shown.

\clearpage

\section{Technical issues}
\subsection{Correlated noise}
\label{sec:corrnoise}
\begin{figure}[htp]
	\centering
	\begin{closetabcols}
	\begin{tabular}{c}
		TT \\
		\includegraphics[width=1.0\columnwidth,trim=12mm 0 12mm 0]{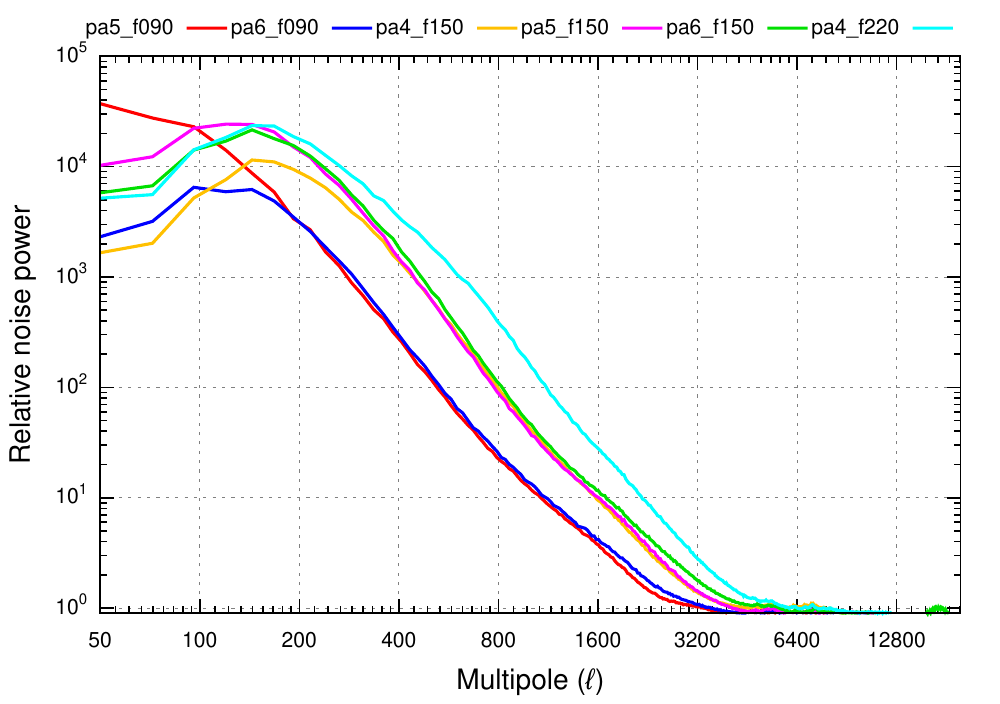} \\
		EE \\
		\includegraphics[width=1.0\columnwidth,trim=12mm 0 12mm 0]{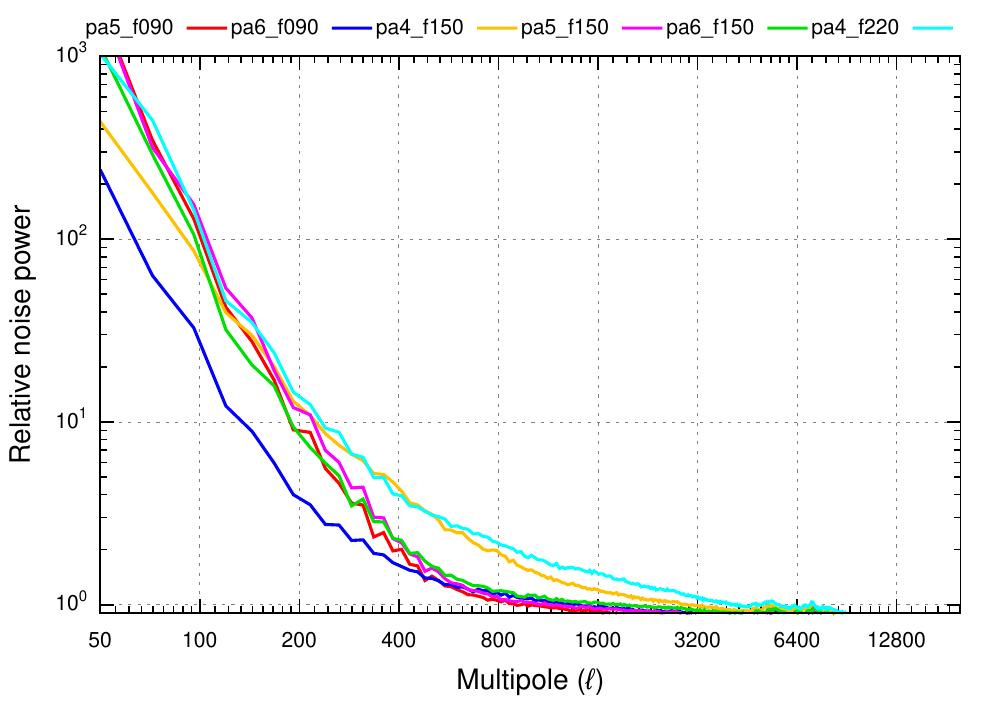}
	\end{tabular}
	\end{closetabcols}
	\caption{TT and EE noise spectra (not beam-deconvolved) measured
	from the \texttt{act\_dr6\_night} data set as the median over the area
	with $-40\degree<\textrm{RA}<60\degree$ for the different detector
	arrays. The power is measured relative to each array's white noise
	floor. The turn-around in the TT spectrum for $\ell \approx 150$
	is due to the transfer function, which was not deconvolved here.
	See section~\ref{sec:tf} for details.}
	\label{fig:nspec-1d-avg}
\end{figure}

Unlike space-based CMB telescopes like Planck and WMAP, ground-based
CMB telescopes have to look through the atmosphere. At these
frequencies the atmosphere is only partially transparent due to
spectral lines from water vapor and oxygen. Water vapor is present
and clumpily distributed even when no visible clouds are present.
Emission from this inhomogeneous water vapor, and to a lesser extent temperature
variations in the atmosphere, are the dominant noise sources on large scales in ACT,
and are responsible for the several orders of magnitude increase in noise power
at low $\ell$ in figure~\ref{fig:nspec-1d-avg}.

The atmospheric noise can be approximated as a power law both in time domain
and map space, with a slope of $\alpha \sim -3$ in both cases.\footnote{
The intrinsic power spectrum of atmospheric turbulence has a slope
of -8/3. This is steepened at small scales by near-field effects,
and further modified when projected onto the sky with the telescope's
scanning pattern. The result is a spectrum with a more complicated shape
that can still be approximately described as a power law with an
exponent of around -3.
}
The
total map noise power spectrum is then roughly a sum of this power law
and a white noise floor:
\begin{align}
	N(\ell) \approx \sigma^2(1 + (\ell/\ell_\textrm{knee})^\alpha) ,
\end{align}
$\ell_\textrm{knee}$ is the multipole where the white noise
and correlated noise have equal power. As we move down from
this (to larger scales) the atmospheric noise grows rapidly,
and by $\ell = 0.1\ell_\textrm{knee}$ the atmospheric noise
power is $\approx 1000$ times higher than the white noise.
In our total intensity maps, $\ell_\textrm{knee}$ is mainly
a function of bandpass, being around 2100/3000/3800 at f090/f150/f220.
In polarization the situation is more complicated, depending both
on the individual array and the declination in the map. See appendix~\ref{sec:spatdep}
for details. Typical numbers here are 300/450/600/500/450/640 for
PA5~f090/\allowbreak{}PA6~f090/\allowbreak{}PA4~f150/\allowbreak{}PA5~f150/\allowbreak{}PA6~f150/\allowbreak{}PA4~f220.

These numbers are similar to those found in our earlier data releases and by other
ground-based CMB telescopes\footnote{SPT has $\ell_\textrm{knee} = 1200/1900/2100$
at f090/f150/f220 for T and 200 for E, but a slope closer to
-4 \citep{spt-lknee-2022}. The SPT site typically has 1/3 the PWV
of the ACT site.}, but are much higher
than for space telescopes. For example, Planck has $\ell_\textrm{knee}\sim50$
and a slope closer to $-1$. This means that while it's a decent approximation
to treat the Planck noise as white, there are hardly any cases where this
is a good approximation for our maps!

\subsection{Transfer function}
\label{sec:tf}
The maximum-likelihood mapmaking estimator $\hat m$ for an instrument
that observes some sky $m$ with response $P$ is
$\hat m = (P^TN^{-1}P)^{-1}P^TN^{-1}d$.
This estimator is unbiased as long as one hasn't done any filtering of the data
$d$ beyond what is captured in the weighting matrix $N^{-1}$.
If $N$ describes the covariance
of the noise, then the solution is additionally optimal. When inversion of
$P^TN^{-1}P$ must be done using iterative methods like CG, as is the case
for us, then stopping the iteration early will introduce some bias even
when the data is unfiltered.

\begin{figure}[htp]
	\centering
	\hspace*{-6mm}\includegraphics[width=1.15\columnwidth]{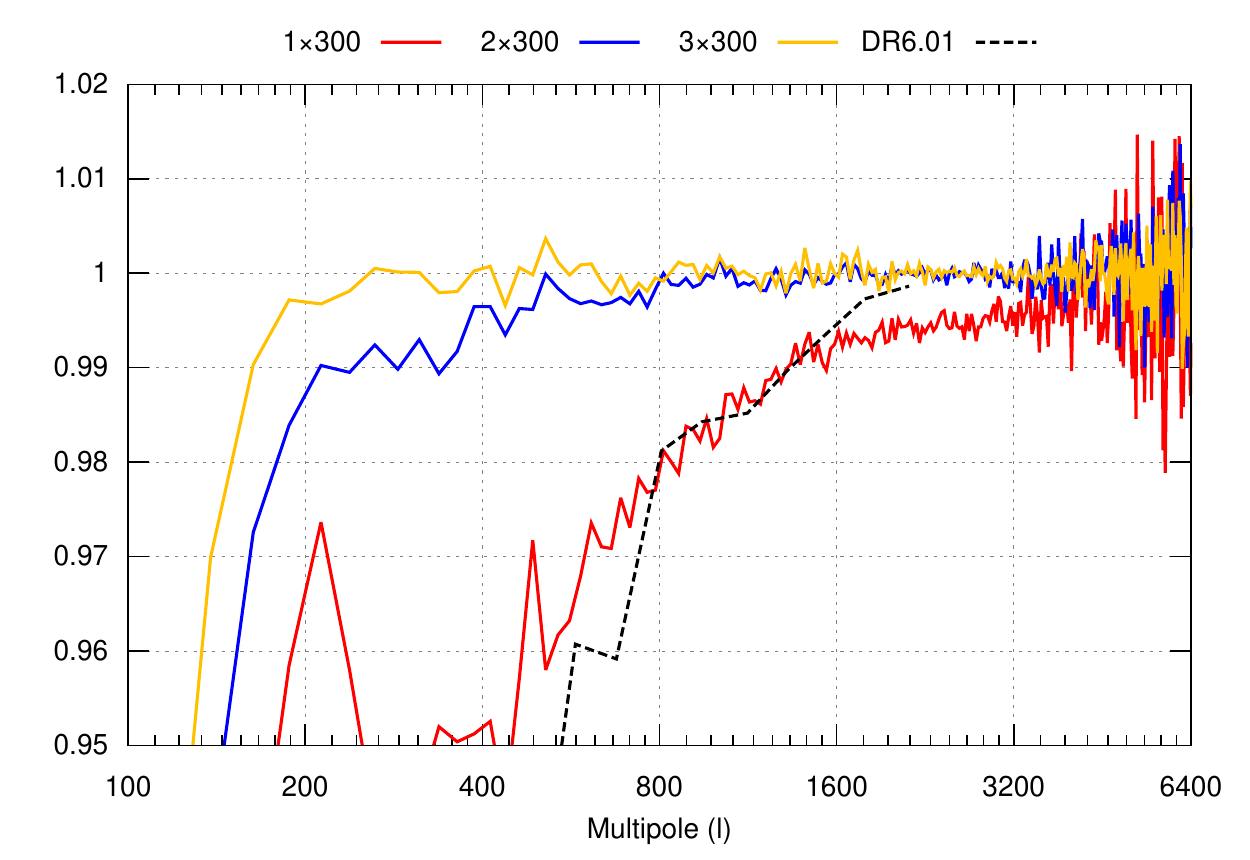}
	\caption{Illustration of the mismatch between the expected total
	intensity transfer function from end-to-end simulations and the one
	measured vs. Planck in the early parts of the DR6 analysis. Each
	curve measures the fraction of the true signal that is recovered
	in the map as a function of multipole $\ell$. The colored curves
	show the result of simulations. According to these
	(red = 1st pass, blue = 2nd pass, yellow = 3rd pass, each with
	300 CG steps) the mapmaker converges towards a unbiased results.
	But to our surprise, the same mapmaker converged towards a biased
	result (relative to Planck) when run on the actual data (black).
	We eventually tracked this down to model error bias, and improved it
	from what is shown here (see figures~\ref{fig:tf-sim} and \ref{fig:tf}).
	The simulations followed the pointing and gain models exactly, while the data do not.}
	\label{fig:tf-naive}
\end{figure}

For DR6 our goal was to recover as low $\ell$ as possible, so unlike DR4 we
chose to forego all filtering of $d$ to minimize mapmaker bias.\footnote{
	Experience from DR4 had already taught us that ground pickup
	could be effectively subtracted in map-space.} We expected
that this would leave conjugate gradient convergence as the only relevant
source of bias in the mapmaking, and we confirmed this using end-to-end
simulations where we estimated the mapmaker transfer function as
\begin{align}
	\textrm{TF}(\ell) &= \frac{\textrm{crossspec}(\textrm{map}(d+P\textrm{sim})-\textrm{map}(d),\textrm{sim})}{ \textrm{powspec}(\textrm{sim}) } .
\end{align}
Here \emph{sim} is a simulated sky map, $P$ is the same response matrix as used in the
mapmaking, $\textrm{map}()$ represents the full multi-pass mapmaking process,
$\textrm{crossspec}(a,b)$ is the cross-spectrum of two maps, and $\textrm{powspec}(a)
= \textrm{crossspec}(a,a)$ is the power spectrum of a single map. The data $d$
needs to enter into this expression since the mapmaker needs it to build the
noise model, but to first order the data cancels when the data-only map is
subtracted.

The result of these end-to-end simulations is shown in figure~\ref{fig:tf-naive}.
As expected this showed that our mapmaker output converged towards an unbiased
result, with the number of conjugate gradient iterations being the only relevant
source of bias. We also confirmed that this convergence continues beyond the 900 CG
steps shown in the figure.

We were therefore surprised when we saw that our actual data
converged towards a result that deviated strongly from Planck at low $\ell$
in total intensity, as seen in the black curve in figure~\ref{fig:tf-naive}.
This was especially so since the bias took the form of a lack of power, which cannot
be explained with additive systematics like ground pickup. It seemed baffling that
the mapmaker should be able to treat our injected signal differently from the
real data when all we were giving it was a sum of the two!

\begin{figure}[htp]
	\centering
	\hspace*{-6mm}\includegraphics[width=1.13\columnwidth]{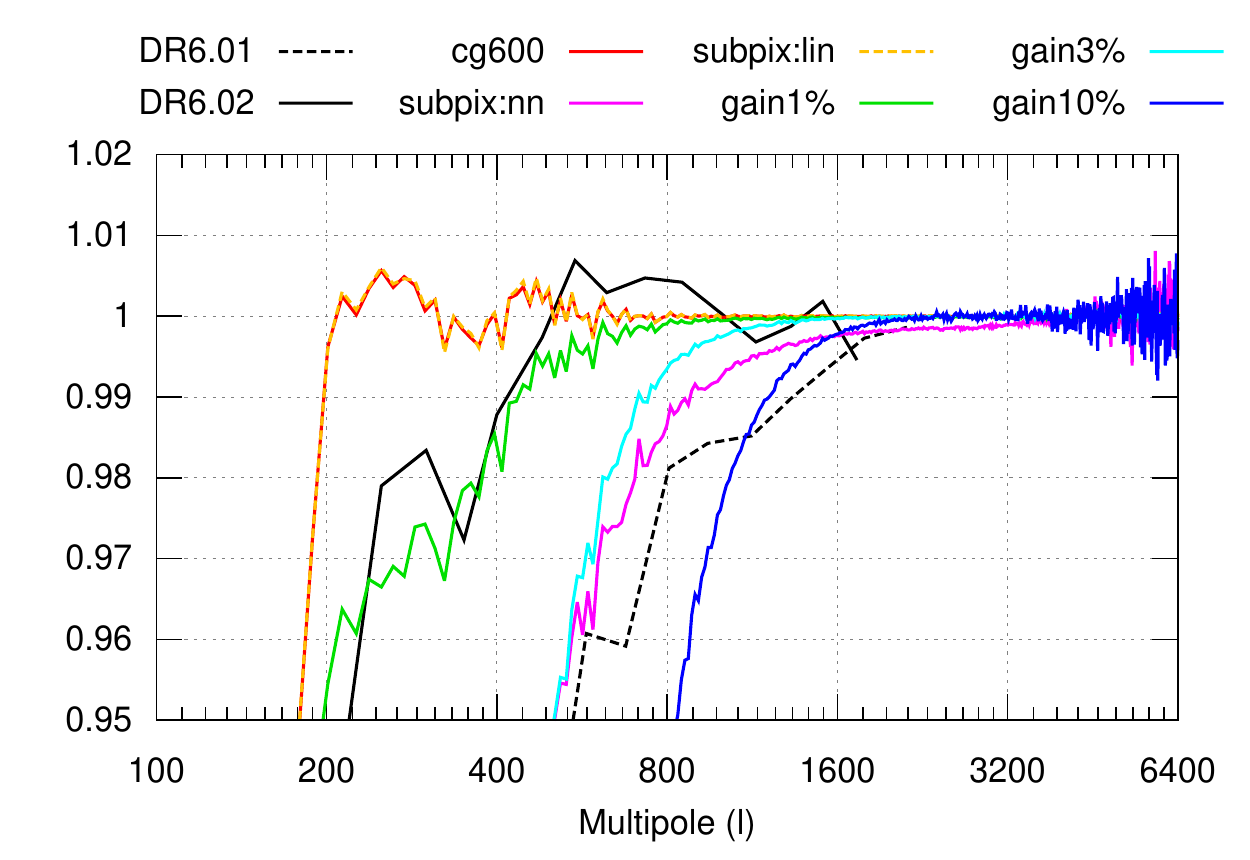}
	\caption{The measured transfer function of PA6 f150 relative to Planck
	in black (see figure~\ref{fig:tf}) compared to the expected transfer function
	from convergence after 600 CG steps (red), subpixel errors with nearest neighbor
	and bilinear mapmaking (magenta and dashed yellow), and three levels of per-detector gain
	errors: 1\%, 3\% and 10\% in green, light blue and dark blue respectively.
	The dashed/solid black curve shows the ACT transfer function before/after switching
	from nearest-neighbor mapmaking and atmospheric flatfielding.
	The dashed yellow curve almost perfectly overlaps the red curve, showing that
	bilinear mapmaking is enough to eliminate subpixel bias for our purposes.
	We do not filter in the mapmaker and use bilinear mapmaking, so gain errors
	are the main suspect for the loss of power at low $\ell$. The PA6 f150 transfer
	function is consistent with 1\% per-detector gain errors.
	}
	\label{fig:tf-sim}
\end{figure}

Of course, we knew that the simulations differed from the data in one small way: the beam-convolved sky we observe is smooth, but the simulation was piecewise
constant inside each pixel due to the nearest-neighbor approximation used in $P$.
We had already seen that this could produce artifacts in high-contrast
areas like those near bright point sources \citep{model-error}, but surely
these sub-pixel details would only matter on the smallest scales in the map?
But no, what we saw when we repeated the simulations with smoother,
high-resolution inputs\footnote{We simulated a map at two times the target resolution
and used a bilinear pointing matrix to read it off. We also tested even higher
resolution input maps. These differed slightly in the effective pixel window at high
$\ell$, but were robust at low $\ell$. The high-$\ell$ behavior is well described by
the ratio of the input to output pixel windows. E.g. a 0.25\arcmin{} resolution
bilinear simulation mapped with a 0.5\arcmin{} nearest neighbor pointing matrix would result
in a total pixel window of
$\textrm{win}^{\textrm{nn}}_{0.5\arcmin}(\ell)/\textrm{win}^{\textrm{lin}}_{0.25\arcmin}(\ell)$.
When taking this into account, the measured transfer function was robust to the
simulation resolution over all multipoles. We therefore stuck with two times
the target resolution for these simulations.}
was that tiny sub-pixel effects could indeed cause a large loss of power at
low $\ell$ in the mapmaker. This is shown in the magenta curve in figure~\ref{fig:tf-sim}.
With the noise weighting ($N^{-1}$) used for PA6 f150, nearest neighbor mapmaking
with a pixel size of 0.5 arcminutes results in a 0.2\%/1\%/5\%/25\% loss of
T amplitude (twice that in power) by $\ell = 1750/870/490/150$. The effect
is qualitatively similar for the other arrays, but it scales with $\ell_\textrm{knee}$.

\cite{model-error2} analyse and describe this effect in detail, but in summary
this is a manifestation of the \emph{regression dilution bias} that occurs in a linear
least-squares estimator when both the independent and dependent variables are
noisy, not just the dependent ones. In our case, the response matrix used in the
analysis is perturbed away from the true response, $P\rightarrow P+\delta$. This
leads to a net-positive $\delta^2$ term in the denominator.
\begin{align}
	\hat m &\approx (P^TN^{-1}P + \delta^TN^{-1}\delta)^{-1}(P^TN^{-1}d) \Rightarrow \notag \\
	\langle \hat m\rangle &\approx \frac{P^TN^{-1}P}{P^TN^{-1}P + \epsilon} m \label{eq:tf} .
\end{align}
Here we have omitted $\mathcal{O}(\delta)$ terms that average down as more
data are added, unlike the squared terms. When expressed in harmonic space,
$P^TN^{-1}P$ and $\epsilon = \langle \delta^TN^{-1}\delta\rangle$ are
nearly diagonal, and so can be approximated as functions of $\ell$.
The $(P^TN^{-1}P)(\ell)$ inherits the $\sim-3$ slope of $N^{-1}(f)$, but non-obviously
the tiny $\epsilon(\ell)$ is much shallower.\footnote{It is practically constant
for gain errors, while its slope is around half that of $N^{-1}$ for
nearest neighbor subpixel errors.} As one goes to lower $\ell$, there
eventually comes a point where $(P^TN^{-1}P)(\ell)$ becomes small enough that
$\epsilon(\ell)$ is no longer negligible in comparison.
At this point the denominator of equation~\ref{eq:tf} becomes larger than
its numerator, and the amplitude of $\langle \hat m\rangle$ starts attenuating.
This is the cause of the loss of power at low $\ell$. See appendix~\ref{sec:epsilon}
for more details.

Satisfied that we had found the cause, we modified the mapmaker to use
a bilinear response matrix as described in section~\ref{sec:pmat}, but
to our disappointment part of the power loss still remained. Further
investigation pointed to a relative gain miscalibration between
detectors in the array as the likely culprit.\footnote{We also investigated
time constant and polarization angle errors, but these did not have an
appreciable effect for reasonable error sizes.} The green, light blue and
dark blue curves in figure~\ref{fig:tf-sim} show the simulated transfer
function for 1\%, 3\% and 10\% standard deviation per-detector gain errors.\footnote{
When simulating this it is essential
that the same gain errors are present \emph{both} when building the noise
matrix $N$ and when solving for the map, otherwise the effect will be missed!}
For PA6 f150 these lead to a 1\% loss of signal at $\ell\approx 400/750/1200$.
This motivated the gain calibration changes in section~\ref{sec:relgain}.\footnote{
	This consisted of reverting to the atmospheric flatfielding we used in DR4,
	plus some work improving these.}
After these changes, the measured transfer function improved from the
dashed black curve in figure~\ref{fig:tf-sim} to the solid one, with
the point of 1\% signal loss moving from $\ell\sim1300$ to 400 for PA6 f150.

\begin{figure}[htp]
	\centering
	\hspace*{-5mm}\includegraphics[width=1.1\columnwidth]{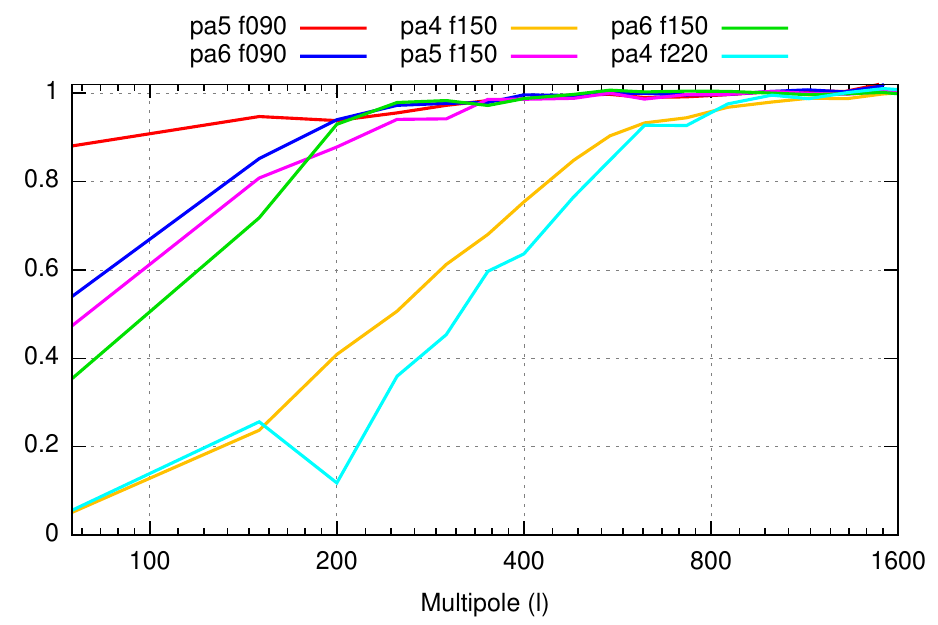}
	\caption{ACT amplitude in total intensity relative to Planck as a function of multipole.
	This is the square root of the power spectrum ratio, and is measured as
	$\textrm{cov}(\textrm{ACT},\textrm{Planck})/\textrm{cov}(\textrm{Planck1},\textrm{Planck2})$,
	where Planck1 and Planck2 are two half-data splits of Planck NPIPE. If
	Planck is unbiased, then this is a measurement of the ACT transfer function.
	All arrays are consistent with 1 at high $\ell$, but eventually fall in amplitude
	as we move to larger scales.}
	\label{fig:tf}
\end{figure}

The final transfer functions after these improvements are shown in
figure~\ref{fig:tf} for all the arrays in this data release.
A more detailed estimate with uncertainties is reported in \cite{act-dr6-spectra}.
There
is still considerable room for improvement: the transfer functions now
reach a 1\% signal loss at $\ell\approx 400$, except for PA4 where it happens
around $\ell\approx 1000$. We believe this is still mainly driven by
gain miscalibration, but the lack of a good calibrator makes it difficult
to improve further.\footnote{
We expect that the Simons Observatory Large Aperture Telescope, which in
many respects is ACT's successor, will be much less impacted by this effect
due to the gain calibrator built into its primary mirror.}
In principle, one could solve jointly for both the sky
and the per-detector, per-TOD\footnote{Or some other suitable timescale
over which the gain are hopefully stable.} gains. We have demonstrated that
this works in small toy examples, but convergence is hopelessly slow for
realistic data sets. Solving this efficiently is an open problem in CMB
mapmaking, but perhaps of low importance since low-$\ell$ total intensity
has already been exquisitely measured by Planck and WMAP.

\begin{figure}[htp]
	\centering
	\hspace*{-3mm}\includegraphics[width=1.05\columnwidth,trim=10mm 0 8mm 0]{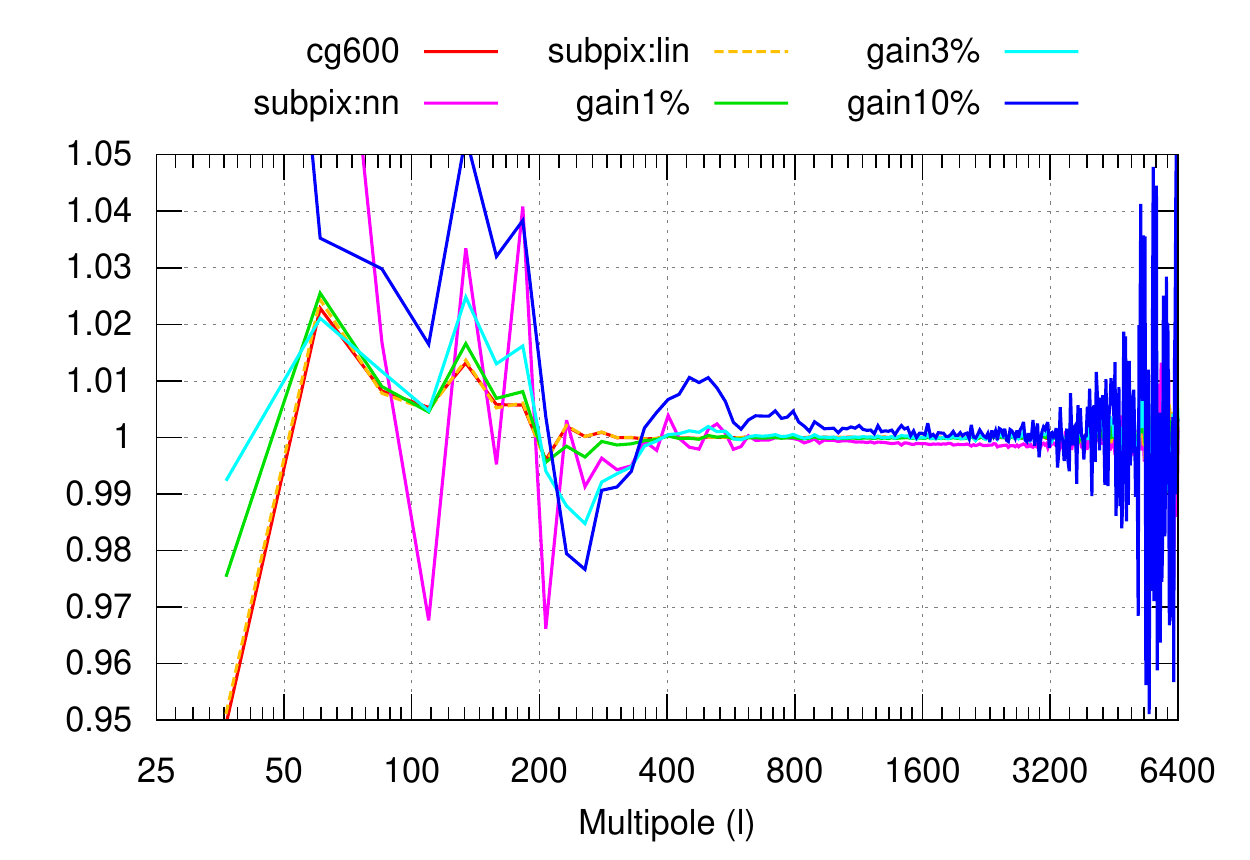}
	\caption{PA6 f150 polarization transfer functions from the same simulations
	as figure~\ref{fig:tf-sim}. The case of 1\% gain errors (green) is consistent
	with our measured transfer function in total intensity, so we expect no appreciable
	transfer function in polarization.
	}
	\label{fig:tf-E}
\end{figure}

As we saw in section~\ref{sec:corrnoise}, the polarized $\ell_\textrm{knee}$
is $\sim 1/5$ as high as in total intensity. The $P^TN^{-1}P$ from
equation~\ref{eq:tf} will therefore reach levels where $\epsilon$ is relevant
at $\sim 1/5$ times as high multipole, so a priori we would expect the
polarized transfer function to deviate less than 1\% from unity for
$\ell \gtrsim 80$. Figure~\ref{fig:tf-E} shows the polarized transfer
function from simulations. Unlike total intensity there
are no clear trends, but the larger scatter makes it hard to quantify the
behavior for $\ell<200$. For the 1\% gain error case that best matches our
T transfer function, the polarization transfer function deviation from 1 is
less than 0.005\% for $\ell>400$ and less than 0.2\% for $\ell>200$.
The scatter could be reduced with more simulations, but since we do not consider
$\ell<600$ in our cosmological likelihood, this is sufficient for DR6.

\subsection{Pickup contamination}
\begin{figure*}[htp]
	\centering
	\dfn{Raw map}
	\begin{closetabrows}
		\hspace*{-10mm}\begin{tabular}{m{3mm}m{19cm}}
			\bf T & \raisebox{-0.5\height}{\includegraphics[width=17cm]{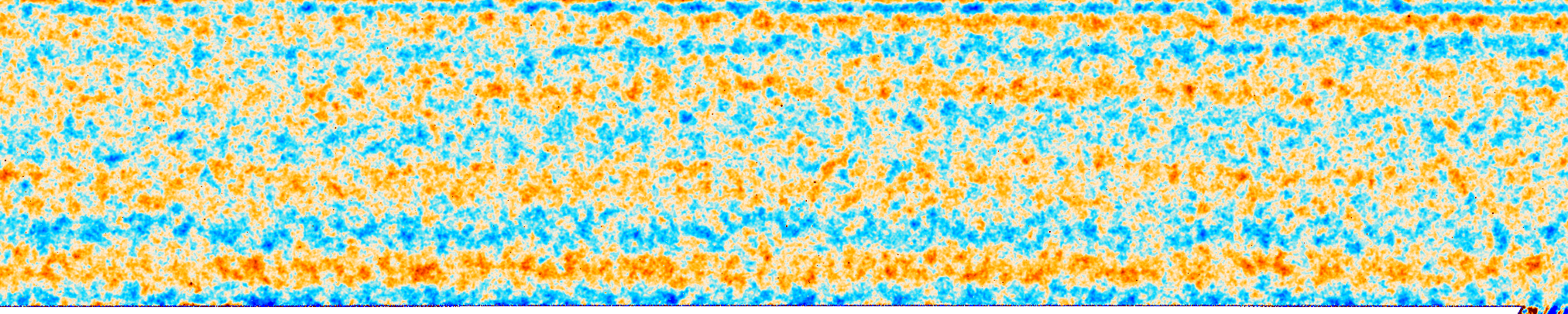}} \\
			\bf Q & \raisebox{-0.5\height}{\includegraphics[width=17cm]{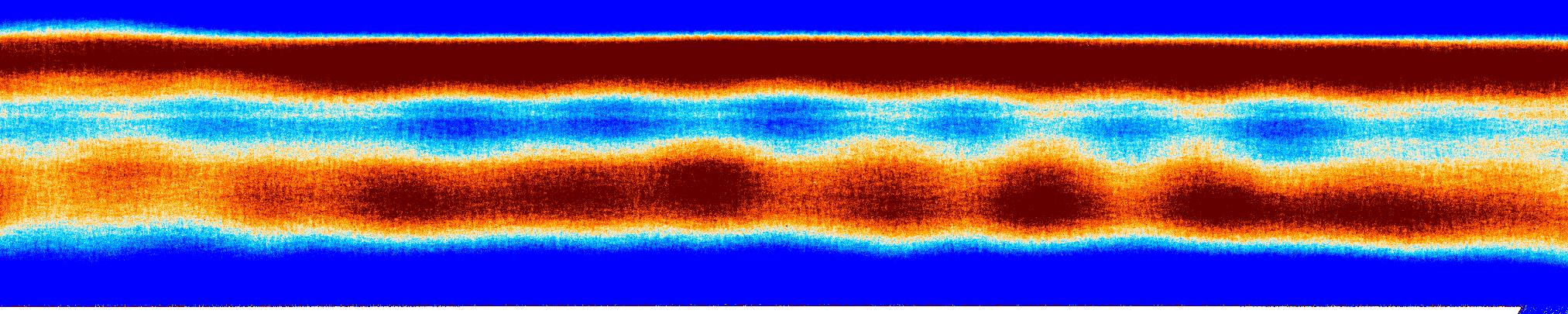}} \\
			\bf U & \raisebox{-0.5\height}{\includegraphics[width=17cm]{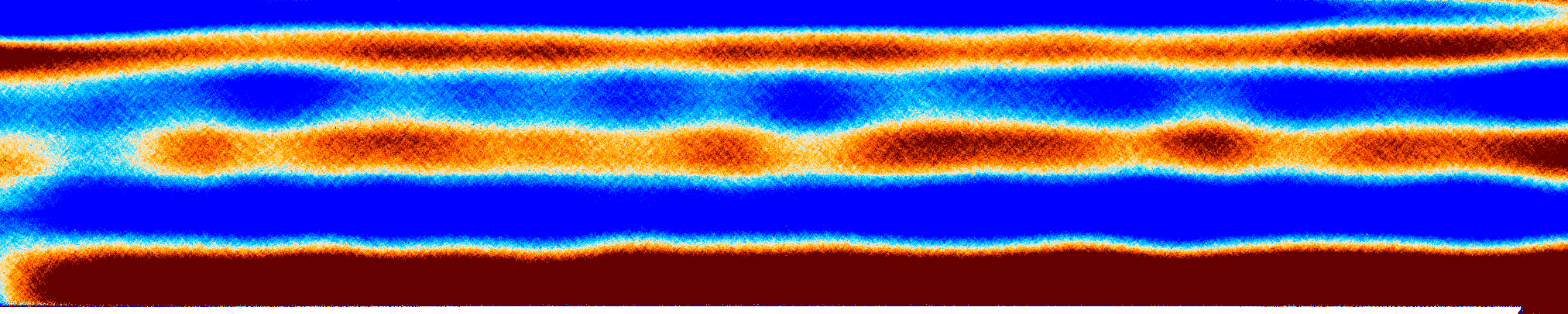}}
	\end{tabular}
	\end{closetabrows} \\
	\medskip
	\dfn{After removing $m<5$}
	\begin{closetabrows}
		\hspace*{-10mm}\begin{tabular}{m{3mm}m{19cm}}
			\bf T & \raisebox{-0.5\height}{\includegraphics[width=17cm]{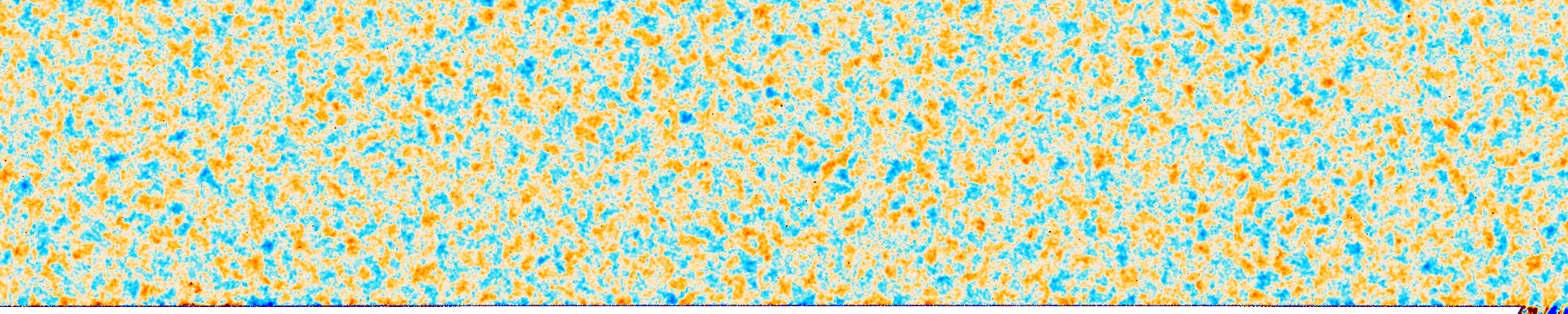}} \\
			\bf Q & \raisebox{-0.5\height}{\includegraphics[width=17cm]{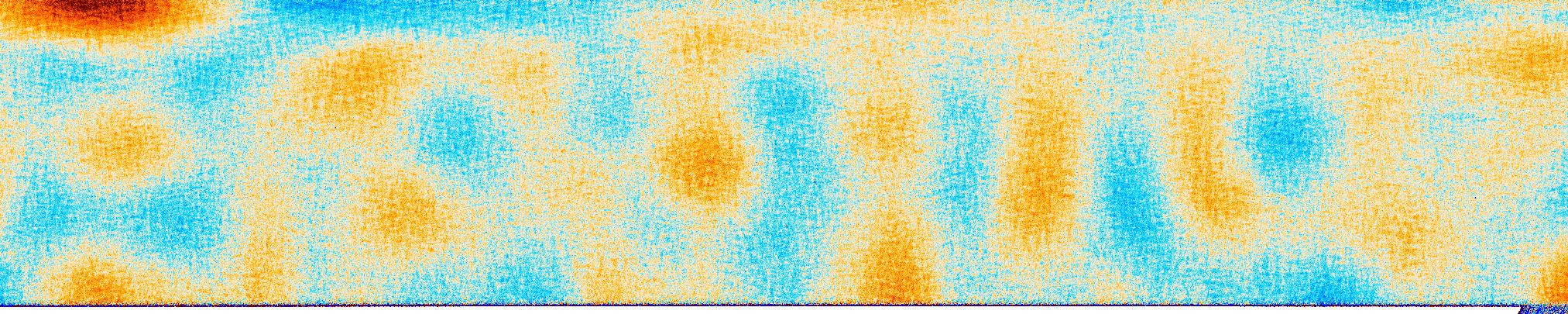}} \\
			\bf U & \raisebox{-0.5\height}{\includegraphics[width=17cm]{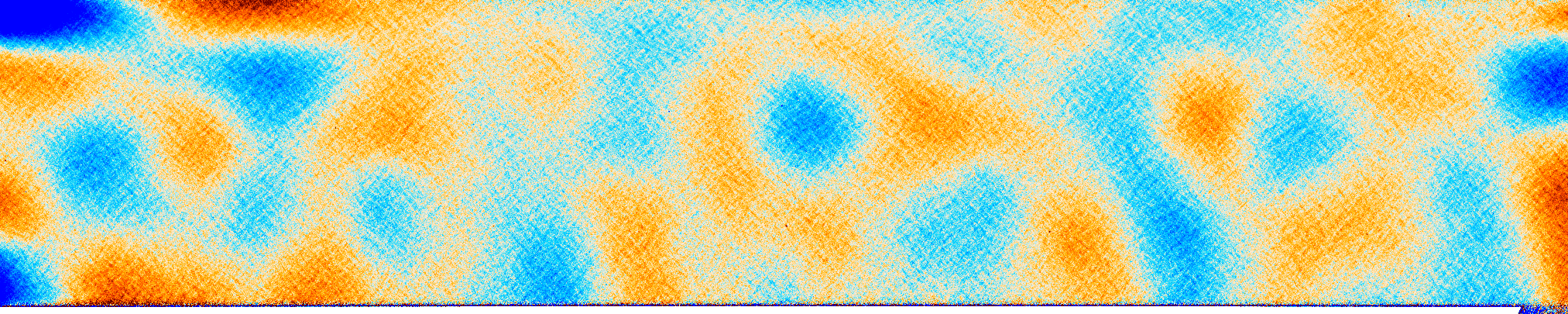}}
	\end{tabular}
	\end{closetabrows}
	\caption{\dfn{Top}: Average of the 4 splits for PA5 f090 night-time data covering the 3600 square degree
	area $255\degree > \text{RA} > 120\degree$, $-5.5\degree < \text{dec} < 21.5\degree$ illustrating the deficit of large-scale power
	in total intensity (\dfn{T}) and the presence of instrumental pickup in all Stokes components. The
	pickup shows up as broad horizontal stripes. The color range is $\pm 500\micro$K in T and $\pm50\micro$K in Q
	and U. \dfn{Bottom}: The same maps after removing spherical harmonics
	coefficients with $m<5$; or equivalently, removing 2D Fourier modes with
	$|\ell_x|<5$. The pickup is no longer visible, revealing
	the CMB T and E-modes, and the correlated noise structure in polarization.}
	\label{fig:map-raw-and-filtered}
\end{figure*}

An example of a raw DR6 map is shown at the top of figure~\ref{fig:map-raw-and-filtered}.
It is visibly dominated by bright horizontal bands of azimuth-synchronous pickup in polarization,
with an amplitude of $\sim 100$ µK. These bands are less visible in total intensity, but have roughly the same amplitude there. Some of this is caused by sidelobes hitting corners of the ground screen, but much of it appears to be pickup of uncertain origin internal to the telescope.

Despite their dire appearance, these stripes do not impact the maps' usefulness much due to occupying a tiny region of harmonic space. We find that a very gentle filter that simply removes spherical harmonics modes with $m<5$, or equivalently $|\ell_x|<5$ in 2D Fourier space, is enough to make the maps visually pickup free (see bottom of figure~\ref{fig:map-raw-and-filtered}). For our power spectrum analysis, however, we found that a stricter cut of $m<77$ ($|\ell_x|<90$) was necessary to pass our null tests.\footnote{The discrepancy beteen $m$ and $\ell_x$ arises because we use the average pixel size in the map to calculate $\ell_x$. 2D Fourier space is based on the flat-sky approximation, so the correspondence between Fourier wavenumbers and $\ell$ is always approximate. For our patch, $\ell_x=1.171 m$.}\fnsep\footnote{The power spectrum analysis also removes $|\ell_y|<50$, but this is motivated by correlated noise, not pickup.} We interpret this as residual pickup too faint to make out by eye. See also section~3.3.1 of \cite{act-dr6-spectra}.

It is because of this relatively simple structure of the pickup in the maps that we chose to forego time-domain pickup subtraction in section~\ref{sec:data-preparation}. The bias introduced by subtraction there would have been much more expensive to characterize, requiring full time-domain simulations, while not necessarily being as effective as the map-level simulation at removing all the pickup. For example, in DR4 we found that TOD-level filtering did not clean the data sufficiently, requiring map-level filtering anyway, in the form of the ``k-space filter''.

\section{Conclusion}
\label{sec:conclude}
We have presented the ACT DR6 maps, based on the 2017--2022 survey with the
AdvancedACT camera. The maps, which cover the three frequency bands
f090, f150 and f220 with an angular resolution of 1.4\arcmin{} at f150,
fall into two categories:
Average sky maps, which cover 45\% of the sky with a median
frequency-combined depth of 10 µK\arcmin{}; and more than ten thousand
single-scan ``depth-1 maps'' with a typical depth of 250 µK ($\sim 25$ mJy)
that are suitable for time domain astronomy.

\begin{figure*}[htp]
	\centering
	\includegraphics[height=0.9\textheight]{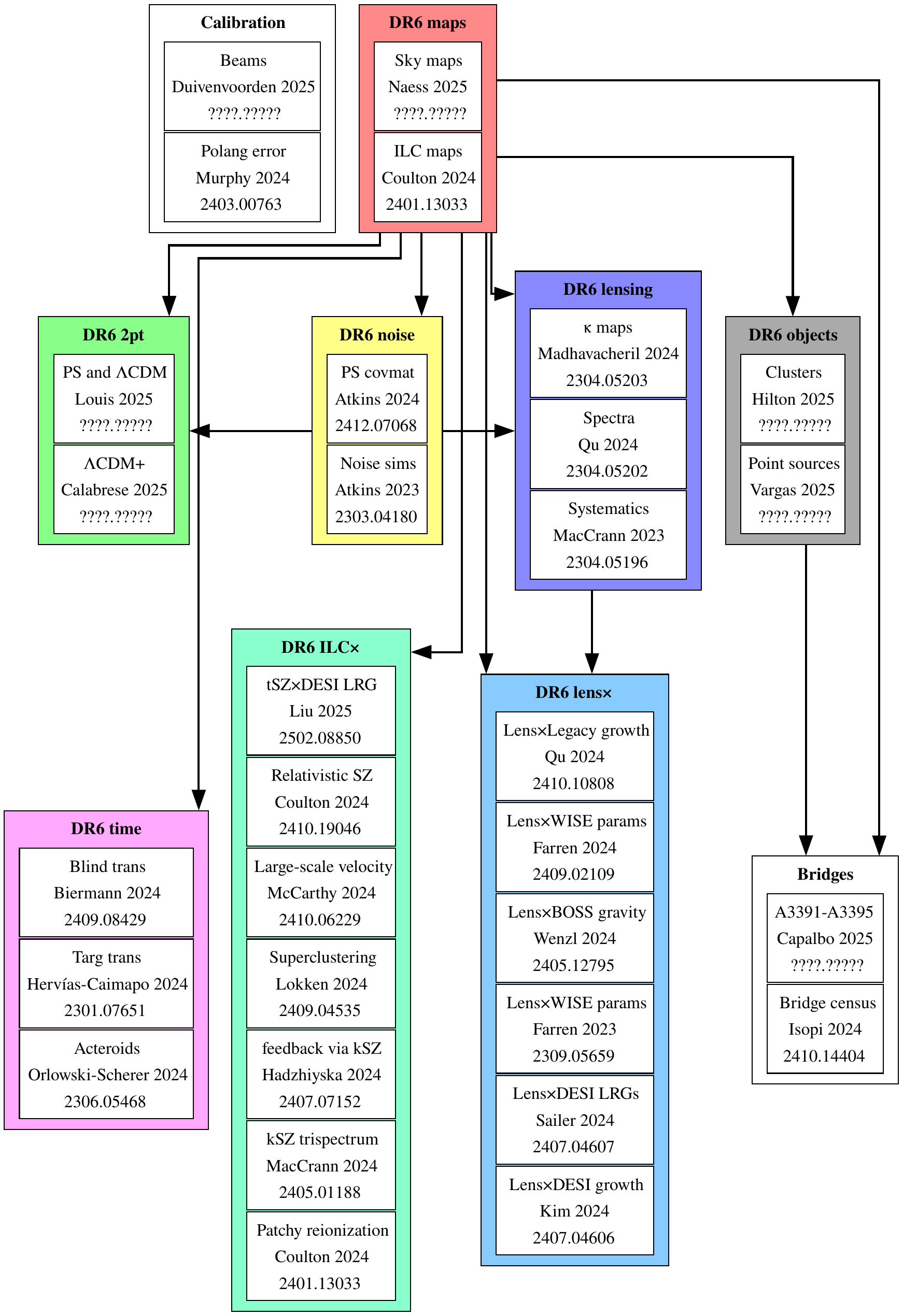}
	\caption{This paper (top of red box) in the context of other ACT DR6 papers.
	Arrows show data dependencies. The maps presented in this paper form the basis of
	the rest of the DR6 analysis. Papers with an arXiv ID of ``????.?????'' have not
	yet been assigned an ID. All papers potentially depend on the ones in the
	Calibration group.}
	\label{fig:dr6-papers}
\end{figure*}

These maps form the basis of the rest of the DR6 analysis effort,
as laid out in figure~\ref{fig:dr6-papers}.
The DR6.02 maps' angular power spectra and fit to \LCDM{} are described in
\citet{act-dr6-spectra}, and extensions to \LCDM{} are evaluated in
\citet{act-dr6-extensions}. \citet{act-dr6-sources} and \citet{act-dr6-clusters}
use the maps to detect around 30,000 point sources at $5\sigma$ and find
$\sim 10\,000$ optically confirmed galaxy clusters ($\sim 6000$ at $5\sigma$). DR6.01 maps were used to produce
weak lensing maps \citep{act-dr6-lens-map} and power spectra \citep{act-dr6-lens-ps},
as well as component-separated maps \citep{act-dr6-ilc}. These in turn
were used in a large number of cross-correlation studies. The depth-1 maps
are used to in blind \citep{depth1-transients} and targeted \citep{dr6-targ-trans}
transient searches, and \citet{acteroids} uses them to study solar system asteroids.

We make the maps available on the Legacy Archive for Microwave Background Data
Analysis (LAMBDA) at \url{https://lambda.gsfc.nasa.gov/product/act/actadv_prod_table.html}
and at NERSC at \texttt{/global/cfs/cdirs/cmb/data/act\_dr6/dr6.02}.
With the exception of the depth-1 maps and the null
test maps, the maps are also made available reprojected to HEALPix. An interactive
web atlas is available at \url{https://phy-act1.princeton.edu/public/snaess/actpol/dr6/atlas/}.
Some of the coadd maps are also published in the
Aladin interactive sky atlas \citep{aladin-sky-atlas}, e.g. \url{https://alasky.cds.unistra.fr/ACT/DR4DR6/color_CMB}.

While the DR6 maps cover most of the AdvancedACT data, there is still
substantial unexploited potential. The PA7 low-frequency array was installed
in 2020 and covered two new frequency bands, f030 and f040. We hope to
publish these maps in a future DR7. Also, as discussed in sections~\ref{sec:corrnoise}
and \ref{sec:tf}, gain calibration issues are implicated in the
low-$\ell$ power loss in total intensity and correlated noise excess in
low-$\ell$ polarization. A future analysis may be able to substantially
improve on this. Beyond ACT, the Simons Observatory Large Aperture Telescope
is expected to begin observations soon. It will cover a superset of the
DR6 sky area in six frequency bands at a combined depth of $\sim 3$ µK\arcmin{}.

\section*{Acknowledgements}
\paragraph{Resources}
Support for ACT was through the U.S.~National Science Foundation through awards AST-0408698, AST-0965625, and AST-1440226 for the ACT project, as well as awards PHY-0355328, PHY-0855887 and PHY-1214379. Funding was also provided by Princeton University, the University of Pennsylvania, and a Canada Foundation for Innovation (CFI) award to UBC. ACT operated in the Parque Astron\'omico Atacama in northern Chile under the auspices of the Agencia Nacional de Investigaci\'on y Desarrollo (ANID). The development of multichroic detectors and lenses was supported by NASA grants NNX13AE56G and NNX14AB58G. Detector research at NIST was supported by the NIST Innovations in Measurement Science program. Computing for ACT was performed using the Princeton Research Computing resources at Princeton University and the Niagara supercomputer at the SciNet HPC Consortium. SciNet is funded by the CFI under the auspices of Compute Canada, the Government of Ontario, the Ontario Research Fund–Research Excellence, and the University of Toronto. This research also used resources of the National Energy Research Scientific Computing Center (NERSC), a U.S. Department of Energy Office of Science User Facility located at Lawrence Berkeley National Laboratory, operated under Contract No. DE-AC02-05CH11231 using NERSC award HEP-ERCAPmp107 from 2021 to 2025. We thank the Republic of Chile for hosting ACT in the northern Atacama, and the local indigenous Licanantay communities whom we follow in observing and learning from the night sky.

We are grateful to the NASA LAMBDA archive for hosting our data. This work uses data from the Planck satellite, based on observations obtained with Planck (http://www.esa.int/Planck), an ESA science mission with instruments and contributions directly funded by ESA Member States, NASA, and Canada. 

SN, MH, SA and DNS acknowledge support from Simons Foundation.
This work was supported by a grant from the Simons Foundation (CCA 918271, PBL).
IA acknowledges support from Fundaci\'on Mauricio y Carlota Botton and the Cambridge International Trust.
ZA and JD acknowledge support from NSF grant AST-2108126.
EC, IH and HTJ acknowledge support from the Horizon 2020 ERC Starting Grant (Grant agreement No 849169).
JD acknowledges support from a Royal Society Wolfson Visiting Fellowship and from the Kavli Institute for Cosmology Cambridge and the Institute of Astronomy, Cambridge.
RD thanks ANID for grants BASAL CATA FB210003, FONDEF ID21I10236 and QUIMAL240004.
SG acknowledges support from STFC and UKRI (grant numbers ST/W002892/1 and ST/X006360/1).
VG acknowledges the support from NASA through the Astrophysics Theory Program, Award Number 21-ATP21-0135, the National Science Foundation (NSF) CAREER Grant No. PHY2239205, and from the Research Corporation for Science Advancement under the Cottrell Scholar Program.
This research was carried out in part at the Jet Propulsion Laboratory, California Institute of Technology, under a contract with the National Aeronautics and Space Administration (80NM0018D0004).
JCH acknowledges support from NSF grant AST-2108536, the Sloan Foundation, and the Simons Foundation, and thanks the Scientific Computing Core staff at the Flatiron Institute for computational support.
MH acknowledges support from the National Research Foundation of South Africa (grant no. 137975).
ADH acknowledges support from the Sutton Family Chair in Science, Christianity and Cultures, from the Faculty of Arts and Science, University of Toronto, and from the Natural Sciences and Engineering Research Council of Canada (NSERC) [RGPIN-2023-05014, DGECR-2023-00180]. JPH (George A. and Margaret M. Downsbrough Professor of Astrophysics) acknowledges the Downsbrough heirs and the estate of George Atha Downsbrough for their support.
JK, MSM and KPS acknowledge support from NSF grants AST-2307727 and  AST-2153201. ALP acknowledges support from a Science and Technology Facilities Council (STFC) Consolidated Grant (ST/W000903/1). ML acknowledges that IFAE is partially funded by the CERCA program of the Generalitat de Catalunya.
TN thanks support from JSPS KAKENHI Grant No. JP20H05859, No. JP22K03682, and World Premier International Research Center Initiative (WPI Initiative), MEXT, Japan.
FN acknowledges funding from the European Union (ERC, POLOCALC, 101096035).
LAP acknowledges support from the Wilkinson and Misrahi Funds.
KKR is supported by an Ernest Rutherford Fellowship from the UKRI Science and Technology Facilities Council (grant number ST/Z510191/1).
NS acknowledges support from DOE award number DE-SC0025309.
CS acknowledges support from the Agencia Nacional de Investigaci\'on y Desarrollo (ANID) through Basal project FB210003.
KS acknowledges support from the NSF Graduate Research Fellowship Program under Grant No.~DGE 2036197, and acknowledges the use of computing resources from Columbia University's Shared Research Computing Facility project, which is supported by NIH Research Facility Improvement Grant 1G20RR030893-01, and associated funds from the New York State Empire State Development, Division of Science Technology and Innovation (NYSTAR) Contract C090171, both awarded April 15, 2010.

\paragraph{Software} We gratefully acknowledge the many publicly available software packages that
were essential for parts of this analysis. They include
\texttt{healpy}~\citep{Healpix1}, \texttt{HEALPix}~\citep{Healpix2},
\texttt{DUCC}~\citep{ducc} and
\texttt{pixell}\footnote{https://github.com/simonsobs/pixell}.
This research made use of \texttt{Astropy}\footnote{http://www.astropy.org},
a community-developed core Python package for Astronomy \citep{astropy:2013,
astropy:2018}. We also acknowledge use of the
\texttt{matplotlib}~\citep{Hunter:2007} package and the Python Image Library
for producing plots in this paper.

\bibliographystyle{act_titles}
\bibliography{refs}

\collaboration{1000}{}
\allauthors

\appendix

\section{General data selection}
\label{app:cuts}
The data selection pipeline generates three types of cuts: per-sample cuts, per-detector cuts, and per-TOD cuts. Per-sample cuts flag a range of samples within each detector's timestream, per-detector cuts flag a subset of detectors within a TOD and cut the entire detector timestream, and per-TOD cuts reject entire TODs.
\subsection{Per-sample cuts}
\label{sec:per-sample-cuts}
To generate per-sample cuts, we begin by identifying glitches in the detector timestreams. As glitches typically occur on short time scales ($\gtrsim 1$~Hz), we apply a high-pass filter with a cutoff frequency of 5~Hz and a Gaussian filter with a full-width-half-maximum (FWHM) of $\sim 0.015$\,s which is the typical time scale of a glitch seen in ACT. We estimate the noise level in each detector timestream using the inter-quartile range and flag samples with signal-to-noise ratio above 10 as glitches. Each identified interval is padded with a buffer of 200 samples ($\sim$ 0.5\,s) on each side. Adjacent flagged intervals with a gap less than 30 samples are automatically merged.

One limitation in the above steps is that some point sources might be sufficiently bright to be flagged as glitches. To avoid this issue, we create a sky mask that excludes pixels within a 3\arcmin{} radius around any bright objects, defined as objects with f090 flux greater than 500 mJy. We project this mask to each TOD based on our pointing model to pinpoint impacted samples and remove them prior to initiating glitch detection.\footnote{These samples are only ignored for the purpose for glitch detection -- they are still present for the actual mapmaking.}

In addition to glitches, we also flag samples based on scan patterns. These include samples during scan turnarounds and samples affected by scan anomalies, such as when the telescope suddenly stops scanning, or scans with a speed that differs from the expected in the middle of an observational run.

\subsection{Per-detector cuts}
\label{sec:per-detector-cuts}
After generating the per-sample cuts, we assess the statistical properties of each detector and flag any that show statistical anomalies. We first calibrate a TOD from the data acquisition unit to a physical unit (pW) using bias step responsivity. Detectors without valid bias step calibration are automatically rejected. Subsequently we examine the performance of each detector through a multi-frequency analysis. Specifically, we divide the power spectrum of each TOD into two frequency windows: a low-frequency window ($\sim$\,0.01\,Hz--0.1\,Hz), which is dominated by atmospheric noise, and a high-frequency window ($\sim$\,10\,Hz--20\,Hz), which is dominated by detector noise.

\textbf{Low-frequency analysis:}
The low frequency band ($\sim$\,0.01\,Hz--0.1\,Hz) of a TOD is dominated by $1/f$ atmospheric noise which acts as a common mode across the detector array. We identify detectors poorly correlated with the atmospheric signal by performing a common-mode analysis with singular value decomposition (SVD). We can decompose each TOD, $d$, with
\begin{equation}
\label{eq:svd}
d = USV^T,
\end{equation}
where both $U$ and $V$ are column-wise orthonormal matrices, and $S$ is a diagonal matrix with non-negative elements. Each element in $S$ corresponds to a common mode, and $U_{ij}$ represents the response of detector $i$ to the common mode $j$. Assuming the strongest common mode is dominated by atmosphere signal in the low-frequency window, $U_{i0}$ therefore provides an assessment of the optical responsiveness of each detector. We refer to it as {\tt gain}, and it is one of the eight statistical parameters that we use to characterize each detector.

Another two parameters that we evaluate for each detector are {\tt norm} and {\tt corr}, where {\tt norm} is the norm of the TOD signal within the low-frequency window, $\|d_i \|$ for the $i$th detector, and {\tt corr} is the correlation between a detector TOD and the strongest common mode, which provides a useful measure of the amount of contaminating modes in the detector timestream.

To reduce the impact from outliers during common mode analysis, we perform the common mode analysis using a pre-selected group of ``well-behaving'' detectors based on the detector-detector correlation matrix. When pre-selection is not possible due to poor data quality, we reject the entire TOD. Note that in ACT DR4 \citep{act-dr4-maps} only the pre-selected detectors are accepted for mapmaking; now we use the pre-selected detectors only for common mode extraction, which improves the detector yield by up to $10$\%.

To ensure our data selection is robust to frequency window choices, the common mode analysis is repeated at ten slightly different frequency bins at steps of 0.05\,Hz; we obtain {\tt gain}, {\tt corr}, and {\tt norm} as averages across these ten bins.

\textbf{High-frequency analysis:}
The high-frequency (10\,Hz--20\,Hz) part of a TOD is dominated by detector noise, which allows us to assess the noise property of detectors. To do that we first use SVD to identify 10 strongest correlated noise modes in this frequency window and deproject them from the detector timestreams. The resulting TOD is now dominated by uncorrelated noise, from which we can evaluate the noise properties of individual detectors. Specifically we measure three statistical parameters, termed {\tt rms}, {\tt skew}, and {\tt kurt}, corresponding to the root-mean-square noise level (standard deviation), skewness, and kurtosis for each detector.

\textbf{Additional statistics:}
In addition to the six pathology parameters derived from low- and high-frequency analysis, we derive two more heuristic parameters: TOD signals may drift slowly on the minute timescale. Excessive slow drift can be a sign of systematic contamination. We estimate the amount of slow drift present in each detector by taking the root-mean-square noise level of the lowpass-filtered TODs below $0.03$~Hz, after deprojecting the three strongest correlated modes. We refer to this parameter as {\tt drift error}. We also calculate the {\tt mid-frequency error} as a similar measure to {\tt drift error} but in a different frequency range (0.3\,Hz--1\,Hz) after deprojecting the eight strongest correlated modes; excessive power at this frequency range may be a sign of thermal contamination.

As a result of the multi-frequency we obtain a total of eight statistical parameters to assess the health of each detector, as summarized in table~\ref{tab:patho}.
\begin{table}[h]
    \centering
    \begin{tabular}{c|c}
       name        &  freq \\\hline
       {\tt gain}  &  $\sim$0.01\,Hz--0.1\,Hz \\
       {\tt corr}  &  $\sim$0.01\,Hz--0.1\,Hz \\
       {\tt norm}  &  $\sim$0.01\,Hz--0.1\,Hz \\\hline
       {\tt rms}   &  10\,Hz--20\,Hz \\
       {\tt skew}  &  10\,Hz--20\,Hz \\
       {\tt kurt}  &  10\,Hz--20\,Hz \\\hline
       {\tt drift error}  & $\leq 0.03$\,Hz \\
       {\tt mid-freq error}  & 0.3\,Hz--1\,Hz \\
    \end{tabular}
    \caption{Eight pathology parameters used to characterize the performance of each detector in each TOD. The right column indicates the frequency range that the corresponding parameter is extracted from.}
    \label{tab:patho}
\end{table}

We compute these eight parameters for all TODs in DR6 and group them based on detector array (PA4, PA5, PA6), frequency band (f090, f150, f220), and observational season (2017, 2018, 2019, etc.). Detectors with outlying statistics are identified by comparing against the seasonal statistical distribution for the given array and frequency, based on a pre-defined threshold determined by manual inspection of the distribution. These detectors are rejected from mapmaking. In addition, we also cut a whole detector when it sees an excessive number of glitches ($> 20000$), or when $>40\%$ of the samples in a detector TOD are cut by per-sample cuts.

\subsection{Per-TOD cuts}
\label{sec:per-tod-cuts}
Finally, we cut an entire TOD when the TOD has fewer than 100 usable detectors after all detectors cuts have been applied, or when the TOD is collected during high optical loading condition with PWV$/\sin(el)>4$\,mm, with $el$ being the elevation angle.

\clearpage

\section{Near sidelobes}
\label{sec:buddies}
\begin{figure}
	\centering
	\begin{tabular}{ccc}
		T & $Q_r$ & $U_r$ \\
		\includegraphics[width=55mm]{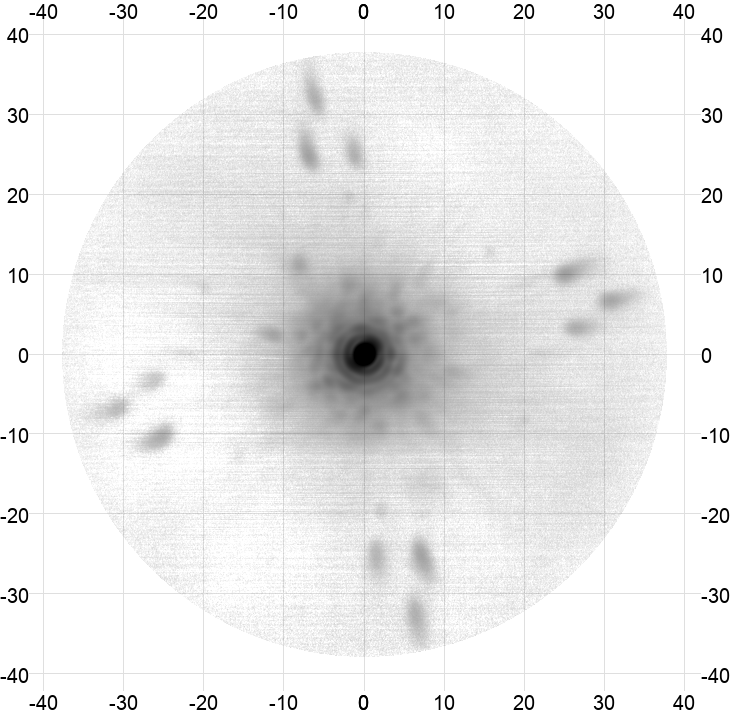} &
		\includegraphics[width=55mm]{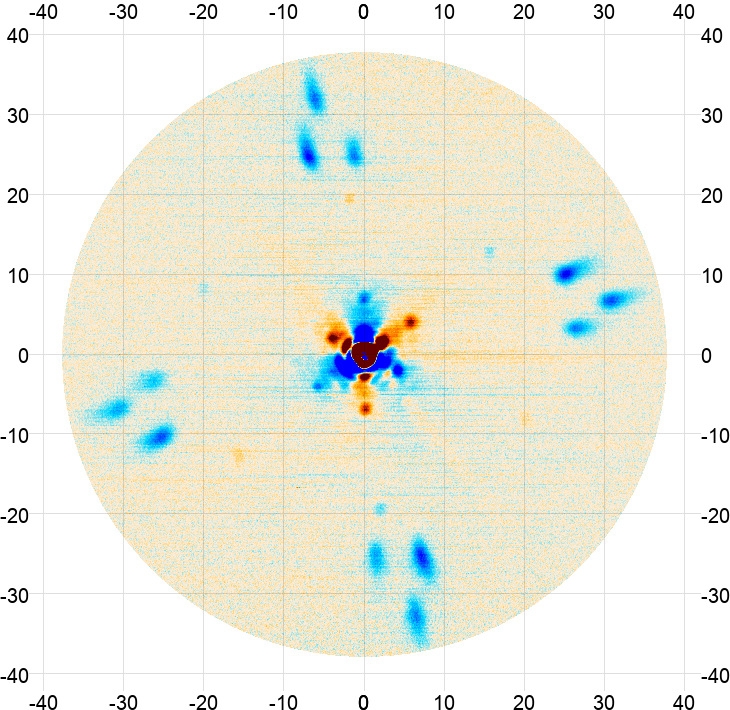} &
		\includegraphics[width=55mm]{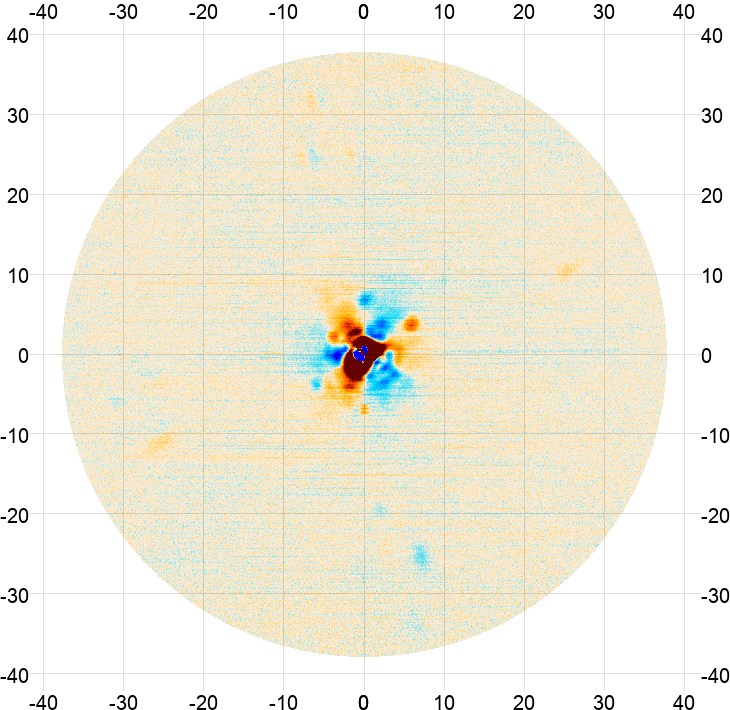}
	\end{tabular}
	\caption{PA5 f150 observations of Saturn, mapping out the central beam and near sidelobes.
	The left panel shows total intensity on an arcsinh color scale to make both the central beam behavior
	and sidelobes visible. The middle and right panels show $Q_r$ and $U_r$ respectively,
	with the colors going from -0.3\% (blue) to +0.3\% (red) of the main beam peak.
	The coordinate grid is in focalplane coordinates in arcminutes. If the telescope were pointed
	at zero elevation, then the $x$ and $y$ axes would correspond to azimuth and elevation respectively.}
	\label{fig:buddies}
\end{figure}

We map out the near sidelobes by projecting targeted observations of Saturn into
Saturn-centered maps. Since Saturn is approximately a point source, this results in
an image of the point-spread function. The high brightness of Saturn means that
small artifacts from model errors near the peak can overwhelm fainter structures
further away when using a normal mapmaker \citep{model-error}. We therefore use
a special planet mapmaker which exploits the compact nature of the planet signal
to filter away much of the atmosphere with almost no effect on the signal itself.
It works as follows:
\begin{enumerate}
	\item Measure the detector-detector noise covariance using data far away from
		the planet.
	\item Define an area with radius $R$ centered on the planet as the target
		region.
	\item For each sample that hits the target region, use concurrent data from
		all detectors that don't currently hit it together with the detector-detector
		covariance to predict what noise it should be observing.
	\item Subtract this prediction from the TOD.
	\item Project the cleaned TOD onto the sky using simple white noise inverse
		variance weighted binning.
\end{enumerate}
To the extent that the signal is entirely contained in the target region,
this procedure will not introduce any bias. But the noise reduction will
only be effective to the extent that the noise is correlated between the target
region and the outside. The method therefore rapidly loses effectiveness as
$R$ grows, but for a bright object like Saturn, the residual correlated noise
is still acceptably low with the $R=36\arcmin$ we need to capture the near sidelobes.
The result is shown for PA5 f150 in figure~\ref{fig:buddies}.

In total intensity the main beam dominates the center, first exhibiting Airy rings and then
fragmenting into blobs as it fades away. Then, at a distance of 22--36\arcmin, a 4-way symmetric
pattern of 3 near sidelobes appears, each somewhat wider than the main beam, elongated by a factor of 2--3
and with a peak amplitude of around 0.3\%. These ``little buddies'' are probably a result of diffractive effects in the free space filters located near the Lyot Stop. These filters are constructed from multiple layers of metal mesh grids with different periodicities oriented along axes that are rotated between layers. This construction leads to a Moiré pattern with centimeter scale periodicities consistent with the observed pattern.

In $Q_r$ and $U_r$ the central region shows a
3-fold T→P leakage pattern, but in terms of the regular $Q$ and $U$ this is simply a dipole
and will therefore mostly cancel when observations taken at different telescope orientations
(e.g. rising and setting) are combined. The near sidelobes, on the other hand, are almost
purely $Q_r$ and near 100\% polarized, resulting in a net T→E polarization leakage regardless
of the telescope orientation, predominantly to $\ell < 1500$. Overall the near
sidelobes have a flux of around 0.8\% that of the main beam for PA5 f150 and less for the
other arrays. We do not detect them at f090 or for PA4.

\subsection{Near sidelobe subtraction}
With 0.8\% of the main beam's flux leaking into E-modes via the near sidelobes, they
represent a small but not negligible systematic, which we handle in two stages.
Firstly, we subtract an approximate model of their signal during mapmaking. Secondly,
we measure the residual T→E leakage and include it in our effective beam model.

If computational resources were not a concern, the sidelobes could be deconvolved
by including them in the pointing matrix, $P = P_\text{main} + P_\text{side}$.
Since we wish to solve for the sky as seen by the main beam, $P_\text{side}$
would include the near sidelobe pattern shown in figure~\ref{fig:buddies} with
the main beam deconvolved. The total $P$ would then be used in the mapmaking
equation, resulting in a sidelobe-free map. But in practice this approach
is much too costly, as each sample in the pointing matrix would need to touch
all pixels with a significant sidelobe signal, making it hundreds of times more
expensive. This expense would be incurred every time $P$ is used, which is
twice per CG step, making the whole mapmaking process forbiddingly slow.

\begin{figure}
	\centering
	\begin{tabular}{ccc}
		data & model & residual \\
		\includegraphics[width=55mm]{figs/coadd_pa5_f150_map_rpol_lin_1.png} &
		\includegraphics[width=55mm]{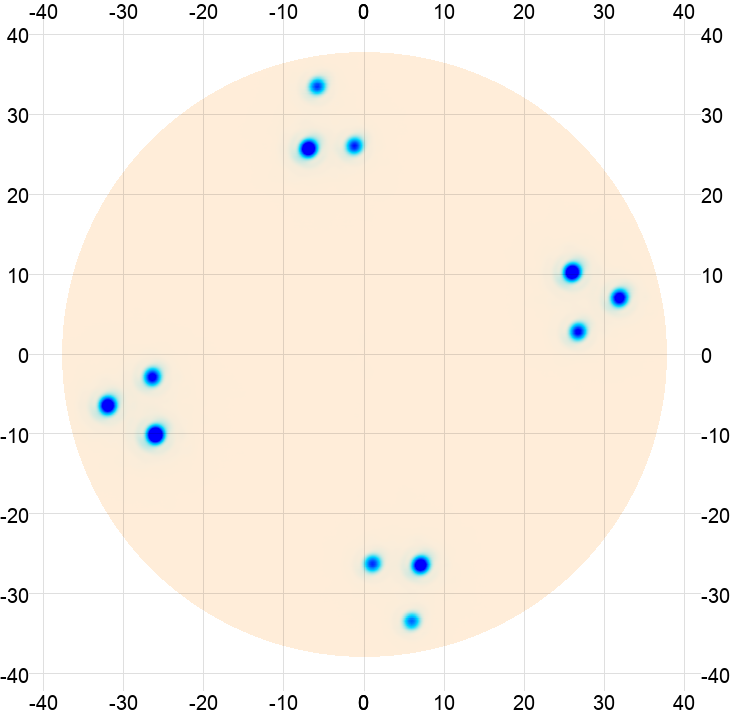} &
		\includegraphics[width=55mm]{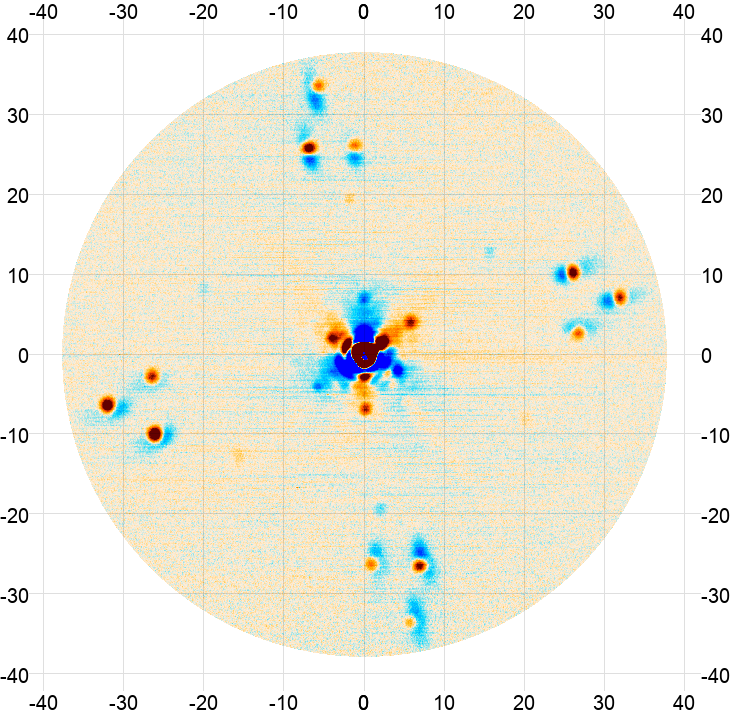}
	\end{tabular}
	\caption{
		\dfn{left}: Near sidelobe map for PA5 f150 $Q_r$.
		\dfn{Middle}: We approximate the near sidelobes as a set of scaled copies of the main
		beam.
		\dfn{Right}: Residual after subtracting the model. The detailed shape of each sidelobe
		is not matched, but its overall flux is subtracted. While residuals are quire large, this approximation is good enough
		to reduce the T→E leakage by a factor of 5 in the $\ell<1500$ range where most of
		the leakage happens.
		All panels have a color range of $\pm 0.3\%$ of the central peak in T.
	}
	\label{fig:buddy-subtraction}
\end{figure}

To keep things practical we make two approximations:
\begin{enumerate}
	\item Instead of including $P_\text{side}$ in the main pointing matrix, we
		use it to clean the TOD: $d = d_\text{raw} - P_\text{side} m_\text{guess}$.
		Here $m_\text{guess}$ is a guess at what the sky map looks like. In our case,
		that's the output map from the previous pass in the multipass mapmaking.
		Since the near sidelobes are an $\mathcal{O}(10^{-3})$ perturbation,
		this approximation is more than good enough.
	\item We approximate $P_\text{side}$ as a sum of 12 scaled and offset versions
		of the main beam, one per near sidelobe. This reduces the cost of $P_\text{side}$
		to be 12 times that of $P_\text{main}$. This is a much rougher approximation
		than the first, but is sufficient for our purposes given the already low level
		of the sidelobes.
\end{enumerate}
Together, these approximations mean that the near sidelobe subtraction can be
done at the cost of 6 CG steps -- less than a percent of our mapmaking expenses.
Figure~\ref{fig:buddy-subtraction} shows how this approximation compares to the
actual sidelobe pattern, and what residual is left. While the residual is visually
quite large, it is mostly limited to high $\ell$. This approximate sidelobe subtraction
succeeds in reducing the near sidelobe T→E leakage by a factor of 5
in the $\ell<1500$ range where most of this leakage happens.

See \citet{act-dr6-beams} for more details on the beams and the
modeling and treatment of T→P leakage.

\section{Far sidelobes}
\label{sec:sidelobe-cuts}

\begin{figure*}[p]
	\centering
	\includegraphics[width=15cm,trim=0mm 120mm 0mm 0mm,clip]{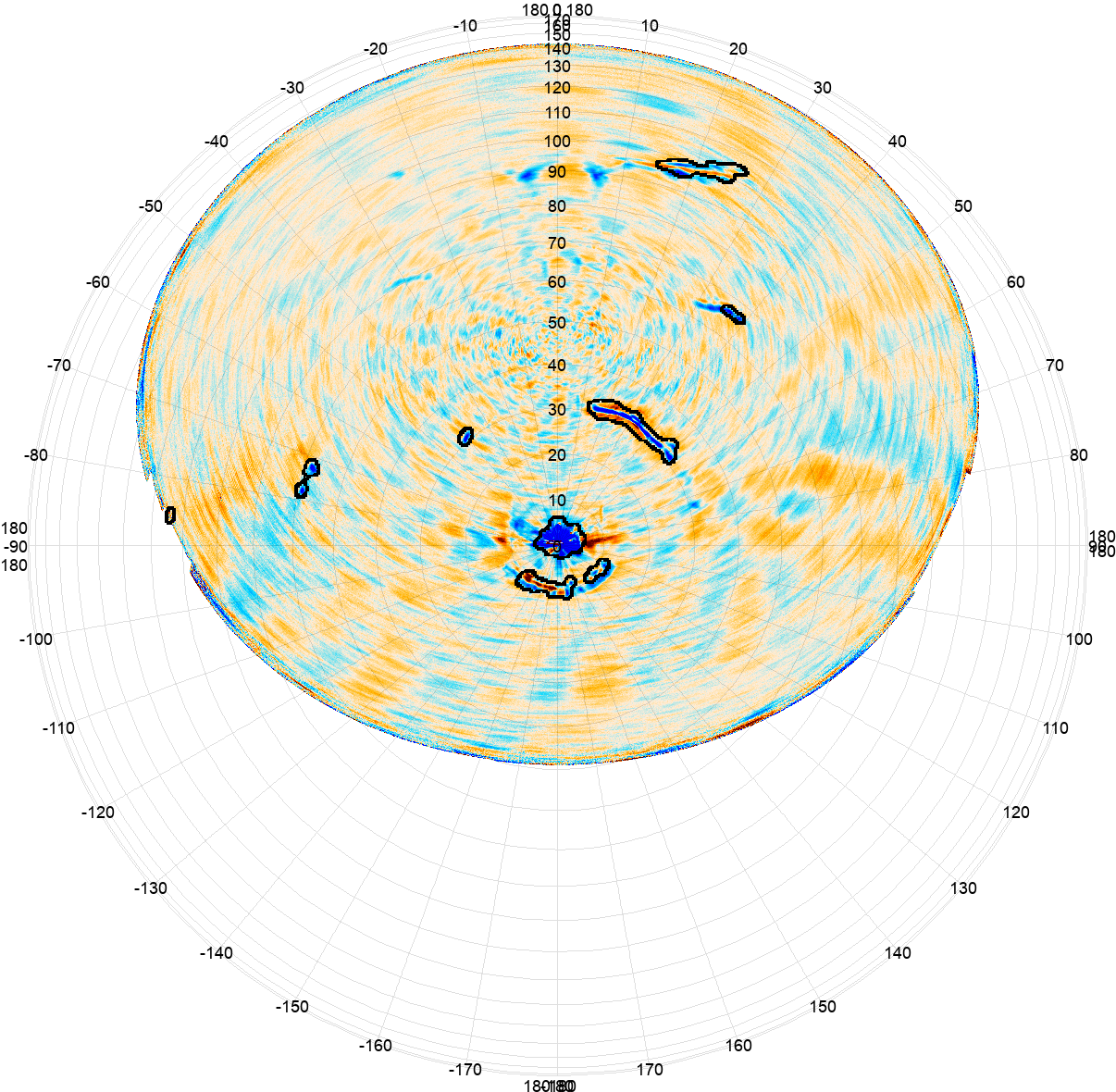} \\
	\includegraphics[width=15cm,trim=0mm 120mm 0mm 0mm,clip]{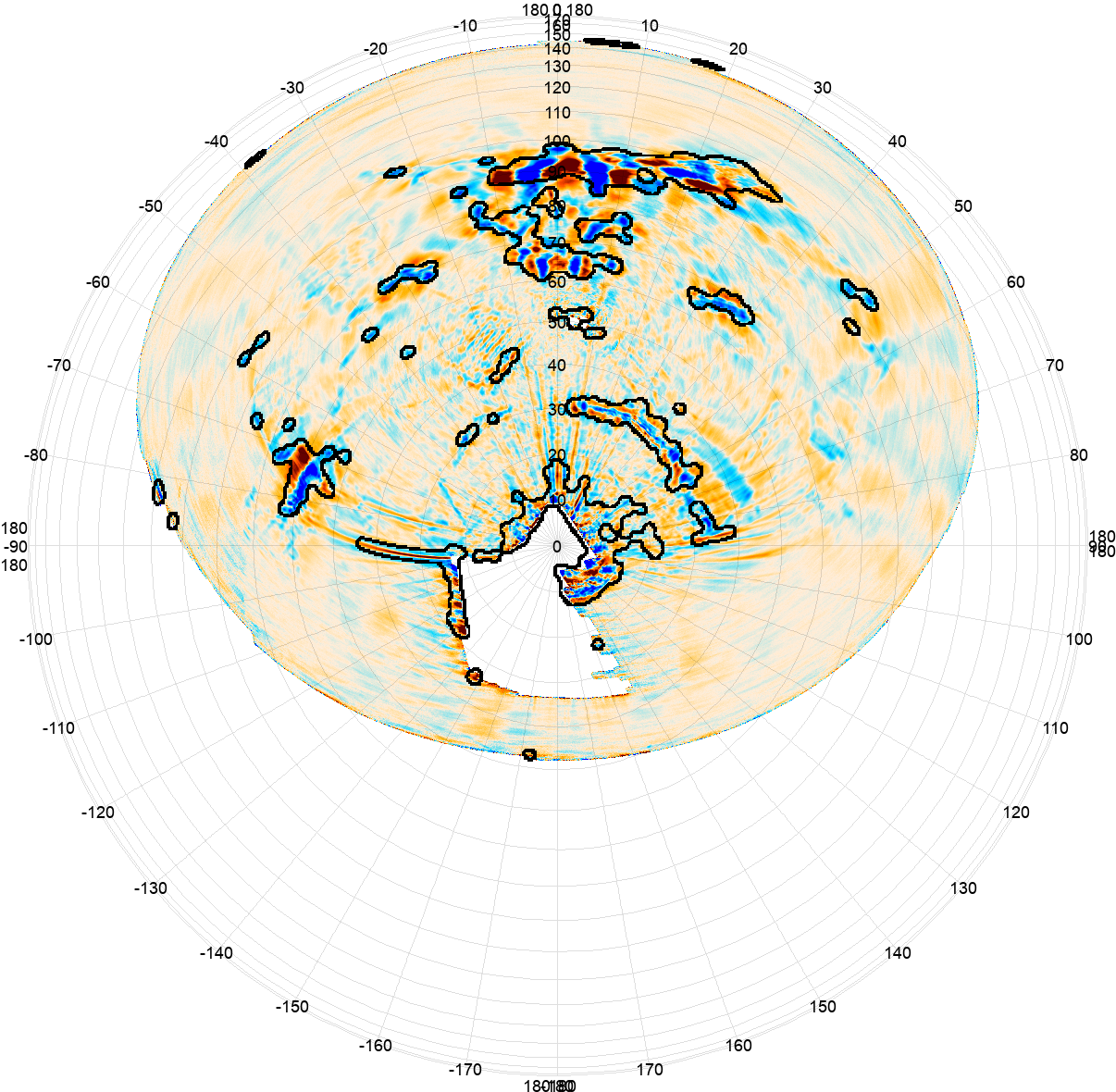}
	\caption{\dfn{Top}: Map of PA5 f150 Moon Q polarization sidelobes, with
	a $\pm 200 \micro$K color scale. Due to their large angular size, the
	sidelobes are easier to map out in polarization than in total intensity,
	which is very noisy on those scales. The radial coordinate is the distance from the Moon.
	The angular coordinate is $180\degree$ for observations below the Moon in the sky, $0\degree$
	for observations above the Moon or on the opposite side of the sky, and $\pm90\degree$ for observations
	left/right of the Moon.
	The black contours show the masked regions.
	Detector samples that fall within these regions while the Moon is above the horizon are cut.
	The pervasive large-scale features are a combination of correlated noise and ground
	pickup.
	\dfn{Bottom}: Same, but for the Sun, and with a $\pm 500 \micro$K color scale.
	Due to the stronger signal the cut is more aggressive, but only applies to
	day-time data.}
	\label{fig:sidelobes}
\end{figure*}

We map out the far sidelobes by projecting our night-time CMB observations into Moon-centered maps and day-time CMB observations into Sun-centered maps. Only observations where the object in question is above the horizon were used. Examples of these maps are shown in figure~\ref{fig:sidelobes}. Overall these sidelobes are quite weak, with the most important ones reaching a few mK when sourced by the 6000 K Sun, corresponding to a $60-70$ dB suppression compared to the main beam. They are relatively extended, often covering tens of degrees in size, but can also be as narrow as the $0.5\degree$ Sun and Moon allow us to resolve.

While the location and overall shape of the sidelobes are robust, the substructure depends strongly on a detector's position in the focalplane. Detectors separated by just a few arcminutes can observe a $180\degree$ phase shift in the substructure. This makes the sidelobes difficult to subtract, and might also be the source of the larger than expected disagreement between the Sun-centered and Moon-centered sidelobe maps in figure~\ref{fig:sidelobes}.

As when mapping the CMB, atmospheric noise makes it difficult to measure total
intensity on scales larger than about a degree, and our sidelobe maps can
therefore only capture the narrowest of features in total intensity. We therefore
base our sidelobe analysis on the polarization maps, which have much higher S/N
on large scales. No features visible in total intensity are missing in the polarized
maps, indicating that all our sidelobes are strongly polarized. This is also expected physically,
as the sidelobes are caused by oblique reflections off flat surfaces surrounding the main mirror
\cite{act-dr6-beams}.

Since they're so weak, we can safely ignore the far sidelobes unless something extremely bright
hits them. We chose to mask samples where the Moon sidelobe signal is brighter
than 200 µK or (for day-time observations) where the Sun sidelobe signal is brighter than 500 µK.
These areas are marked with black contours in figure~\ref{fig:sidelobes}. While a 200/500~µK threshold might seem permissive
compared the CMB's $\sim 100$ µK and $\sim 4$ µK RMS in total intensity and polarization,
this averages down rapidly when observations with different telescope orientations
and Moon/Sun positions are combined. The residual Moon/Sun sidelobe contamination after
this cut therefore ends up being a subdominant noise\footnote{The Moon/Sun sidelobe signal is
a noise source, not a bias source, because the Moon and Sun have time to move significantly
between the observations that go into different parts of our data splits, meaning their signal
does not add up coherently in the cross spectrum.} source compared to the atmosphere.

Overall these cuts remove around 1\% of the samples during the night and 5\% during the day.

\clearpage

\section{Daytime beam}
\label{sec:daytime-beam}
\begin{figure}[ht]
	\centering
	\subfigure[]{
		\hspace{-10mm}\begin{tabular}[b]{c}
			\scalebox{-1}[1]{\includegraphics[width=8cm]{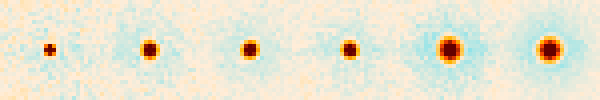}} \\
			\scalebox{-1}[1]{\includegraphics[width=8cm]{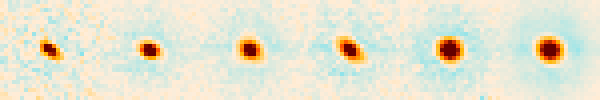}} \\
			\scalebox{-1}[1]{\includegraphics[width=8cm]{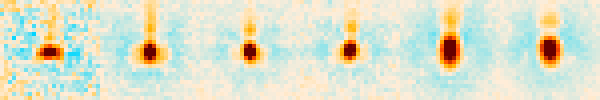}} \\
			\scalebox{-1}[1]{\includegraphics[width=8cm]{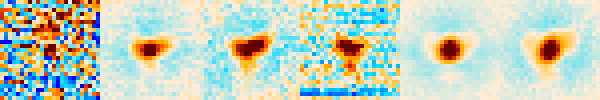}}
		\end{tabular}
	}
	\subfigure[]{\includegraphics[width=8cm]{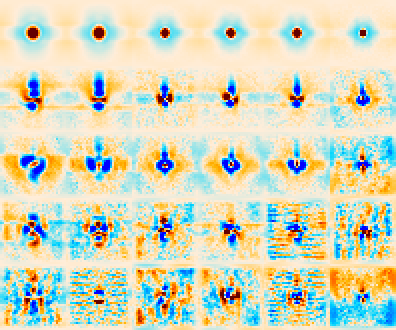}}
	\caption{\dfn{a}: Example of four different beam-shapes seen among
	thousands of day-time observations. The top is the best case, equivalent to
	the night-time beam. The other represent increasingly pathological
	behavior. From left to right we have PA5 f090, PA6 f090,
	PA4 f150, PA5 f150, PA6 f150 and PA4 f220.
	\dfn{b}: While we have thousands of day-time beam examples, there aren't
	thousands of different ways for the beam to vary in practice. Most of
	the beam variability can be explained with just the top five eigenmodes,
	which are shown here for the same arrays as in \dfn{a}.
	The eigenmodes are sorted from the strongest (top) to the weakest
	(bottom).}
	\label{fig:daybeam-eigmaps}
\end{figure}

During the day, the Sun heats ACT's primary mirror, causing it to expand and the
beam to deform. This deformation depends on how much each part of the mirror
or supporting structure expands, and can be quite complicated. Figure~\ref{fig:daybeam-eigmaps}a
shows four examples of ACT day-time beam shapes for each of the detector arrays considered in
DR6. The behavior varies from no worse than the night to strongly deformed and multi-lobed.

\begin{figure}[htpb]
	\centering
	\includegraphics[width=\columnwidth]{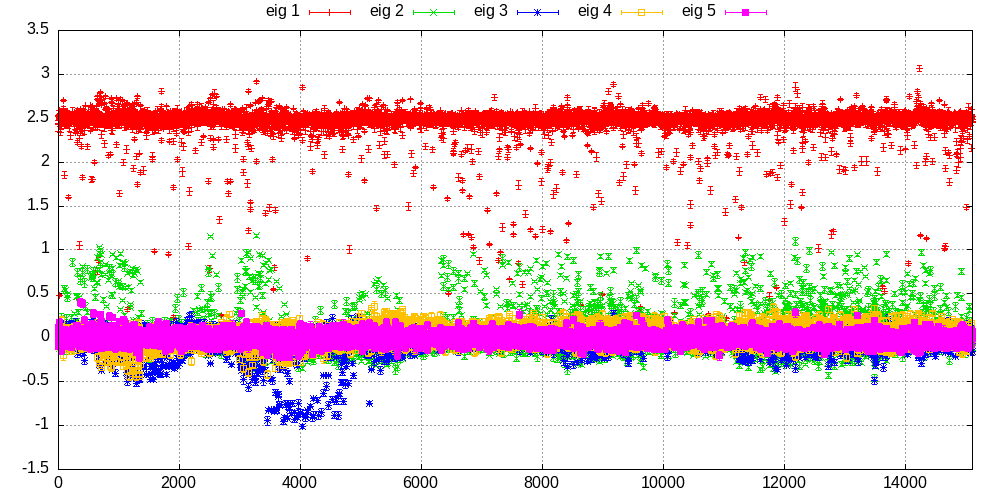}
	\caption{Time-dependence of the amplitude of the top five beam eigenmodes
	for PA5 f090. The horizontal axis is the index of the (roughly hour-long)
	stack of point sources that was used in the measurement. The vertical
	amplitude is the corresponding eigenvalue amplitude for each of the top
	five eigenmodes shown in figure~\ref{fig:daybeam-eigmaps}.
	While some patterns are visible, these become much clearer in figure~\ref{fig:daybeam-time-hour}.}
	\label{fig:daybeam-amps}
\end{figure}

\begin{figure}[htpb]
	\centering
	\includegraphics[width=\columnwidth]{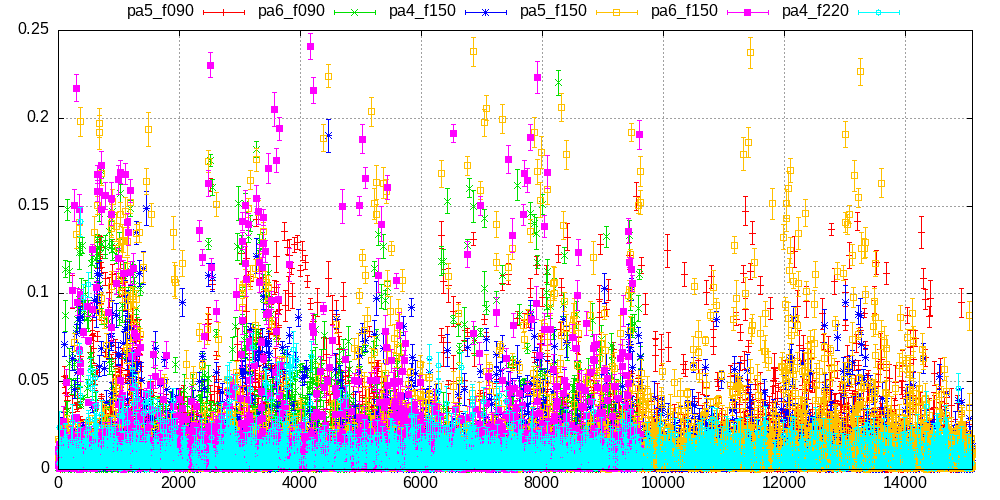}
	\caption{Like figure~\ref{fig:daybeam-amps}, but shows what fraction
	of the eigenmode power is in modes other than the primary one. We take
	this as a measure of the beam badness. See figure~\ref{fig:daybeam-time-hour} for
	a much more readable view of this.}
	\label{fig:daybeam-badness}
\end{figure}

\begin{figure}[htpb]
	\centering
	\includegraphics[width=\columnwidth]{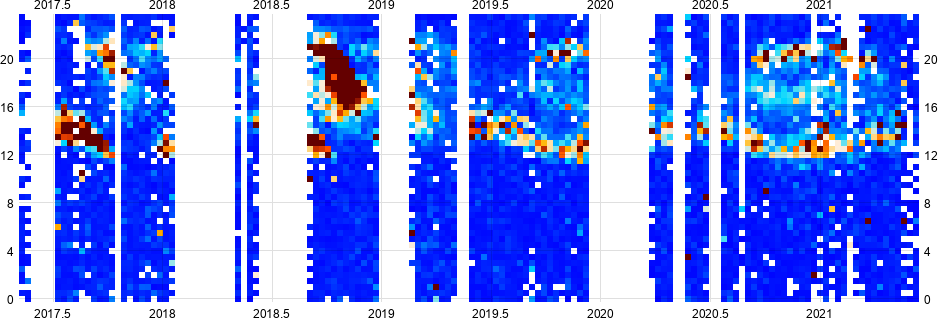}
	\caption{The average beam badness (fraction of eigenmode power outside
	primary mode) for PA5 f090 as a function of date (horizontal axis)
	and UTC hour (vertical axis). Day-time is roughly 11 to 23 UTC.
	The colors go from 0\% badness (dark blue)
	to 7\% badness (dark red). Clear regions are visible, making it
	possible to define a data selection cut in this space. However,
	as we found this pattern to be somewhat azimuth-dependent, we define
	the actual cut separately in four bins of azimuth. We impose
	a 4\% badness cut, which results in a loss of 30\%/34\%/33\% of
	the PA4/PA5/PA6 day-time data. Due to a lack of high S/N point sources
	for PA7, we use the same cut here as for PA5.
	}
	\label{fig:daybeam-time-hour}
\end{figure}

The process of constructing the per-TOD pointing model also produces a large number of
point source images, which we used to find the eigenmodes for the beam variability.
Most of the beam variance can be explained by the top five eigenmodes, which are shown in
figure~\ref{fig:daybeam-eigmaps}b.
Figure~\ref{fig:daybeam-amps} shows the amplitude $\alpha_i$ of the eigenmodes for each such point
source image, sorted by time. The first eigenmode corresponds to the ACT night-time
beam, while the others represent deviations from this ideal. We can construct a badness
statistic $\beta = 1-(\sum_{i=2}^5 \alpha_i^2)/(\sum_{i=1}^5 \alpha_i^2)$ which is plotted in
figure~\ref{fig:daybeam-badness}. This looks quite chaotic, but patterns emerge when plotted
vs. both day-of-year and hour-of-day, as shown in figure~\ref{fig:daybeam-time-hour}. Most
of the day-time region (11 < UTC < 23) has a relatively well-behaved beam, with relatively
compact areas of badness. This becomes even more apparent when split into four groups by
the central azimuth of the scan. We use this to define the following beam badness cut:
For each TOD, use its date and UTC hour to look up its predicted beam badness $\beta$ in
the azimuth-split version of figure~\ref{fig:daybeam-time-hour}, and exclude it from
mapmaking if $\beta>4\%$. This cut excludes 30\%/34\%/33\% of the PA4/PA5/PA6 day-time data.
The night-time data are not impacted.

\clearpage

\section{Per-TOD pointing correction}
\label{sec:pointing-details}
The DR4 pointing fit was performed using a TOD-level likelihood search
which was significantly more expensive than the CMB mapmaking, itself
one of the heaviest steps in the whole analysis. To avoid this,
DR6 uses a new pointing model that aims to be faster, higher time resolution
and easier to debug by doing the pointing fit in pixel space instead of time
domain. It proceeds as follows:

\begin{enumerate}
	\item Build a catalog of bright point sources with known locations.
		We crossmatched a preliminary catalog of 5345 point sources detected at
		$>5\sigma$ and $>20$ mJy in the first two seasons of ACT DR6 f150 data
		vs. bright subsets\footnote{
			Many of these catalogs are very large (e.g. hundreds of millions of
			objects for WISE), which would lead to a high chance of false association
			in the crossmatch. We therefore restrict to brighter subsets, with
			size given by the second number in the parentheses. This was done
			somewhat arbitrarily, but the details have little impact for the
			$>20$ mJy subset of ACT sources considered here.
		} of the following external catalogs:
		ALMA (1503 matches out of 3354 objects considered = 1503/3354) \citep{alma_calib},
		AT20G (2103/5890) \citep{at20g-catalog},
		CLASS (1401/14353) \citep{class-radio-sources},
		2MASS (12/34322) \citep{2mass-catalog},
		WISE (38/6823) \citep{allwise-data-release},
		HyperLEDA (117/21591) \citep{hyperleda-catalog},
		PKS (984/8265) \citep{pks-catalog},
		PMN (1329/50818) \citep{pmn-catalog},
		LQAC 6cm (1183/5340) \citep{lqac-catalog},
		BSC5 (8/9110) \citep{bsc5-catalog},
		NVSS (4164/581097 \citep{nvss-catalog})
		and a small list of manually entered objects (7/7). After rejecting 149 non-pointlike
		objects and false positives we were left with 4958 objects matching at least one
		external catalog. The reason for only including $>20$ mJy objects was that
		they need to be bright enough to be detectable or close to detectable in
		a depth-1 map, which is much shallower than our full sky maps.
	\item Make depth-1 maps for all the CMB observations. Depth-1 maps are
		more generally useful for time domain science, and are described in
		section~\ref{sec:depth1}. The depth-1 maps used in the pointing correction
		differ from those by using a rough pointing model based on planet observations
		from the beginning of AdvancedACT (since the final pointing model wasn't ready
		yet), and by using a rotated coordinate system to save memory by minimizing
		the size of the bounding box. The result is a set of maps across
		our bands and arrays. These are generally not written to disk, but intead
		are passed directly to the next step.
		Figure~\ref{fig:pointing-map} shows an example of such map.
		\begin{figure}[h]
			\centering
			\includegraphics[width=\columnwidth]{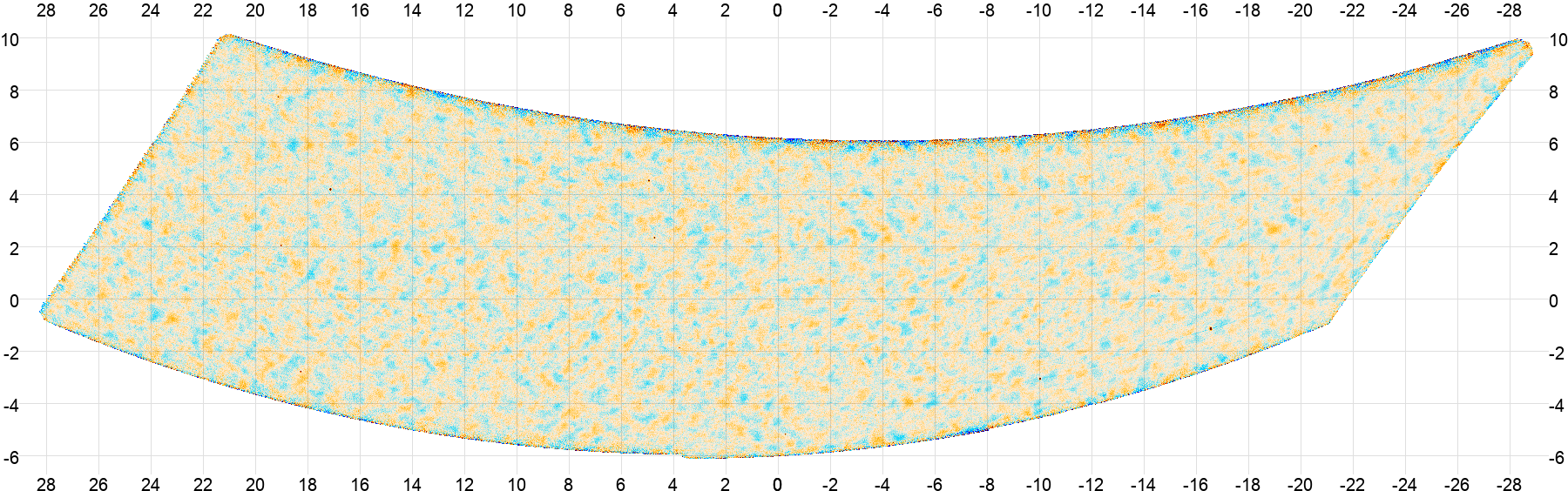}
			\caption{Example depth-1 for pointing measurement, representing
			around one hour of data for a single detector array. Point sources
			are visible as tiny red dots. Pointing errors
			can change significantly in one hour, but it only takes the array
			around 4 minutes to drift across each point in this map, so there
			is no appreciable pointing smearing. This plot is downsampled by a
			factor of 3 from its full 0.5\arcmin{} resolution. Note that this map is
			in a different coordinate system than the standard depth-1 maps. The
			coordinate axes are in degrees, but do not correspond to any standard
			coordinates.}
			\label{fig:pointing-map}
		\end{figure}
	\item Extract $30\arcmin\times 30\arcmin$ thumbnails around each such source in each depth-1 map.
		See figure~\ref{fig:pointing-thumbs} (top) for an example of what these look like.
		\begin{figure}[h]
			\centering
			\includegraphics[width=\columnwidth,trim=0 128mm 0 0,clip]{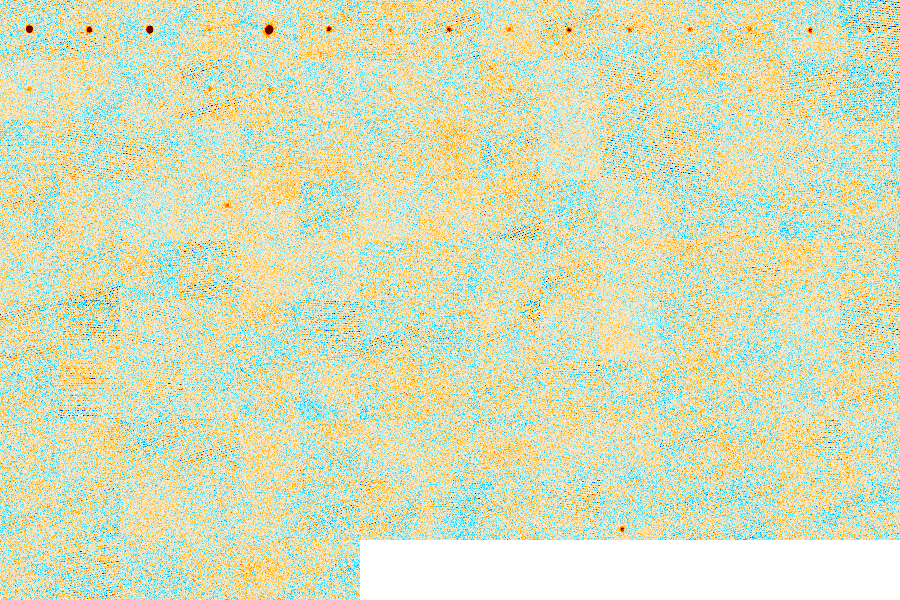} \\
			\includegraphics[width=\columnwidth,trim=0 106mm 0 0,clip]{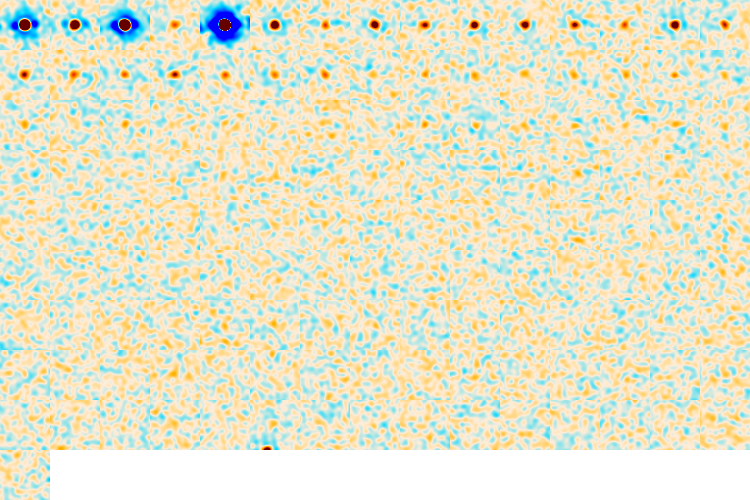} \\
			\caption{\dfn{Top}: $30\arcmin\times 30\arcmin$ thumbnails around the first 60 of the 141 point
			sources considered for the map in figure~\ref{fig:pointing-map}.
			Many of them are too faint to be seen by eye, but can still contribute
			to the pointing fit when modeled together. The few objects that
			appear to be off-center are simply bright neighbors. \dfn{Bottom}: Thumbnails
			for the same map after rotating to boresight coordinates, applying a matched filter,
			and cutting objects that are not at least 5 times as bright
			as any others within a radius of 10\arcmin{}.}
			\label{fig:pointing-thumbs}
		\end{figure}
	\item Reject objects that are not at least 5 times as bright as the others within
		a 10 arcminute radius since the pointing fit assumes isolated sources.
	\item Rotate to the boresight coordinate system where we define
		our pointing offsets and apply a matched filter to maximize S/N.
		Coadd the matched filter maps for the two bands of each detector array,
		since an array's pointing offset should be frequency-independent. This is
		illustrated in figure~\ref{fig:pointing-thumbs} (bottom).
	\item Do a preliminary fit of the position and flux of each object. For
		objects detected at $<6\sigma$ use the ACT catalog flux instead of the one
		measured here. Using this flux, measure the objects' S/N. Objects with
		$S/N > 8$ are fit individually. Other objects are grouped with others
		observed close in time until their combined S/N exceeds 8.
	\item Fit the position of each group, resulting in several measurements of the
		horizontal and vertical pointing offsets per array per TOD. An example of
		this is shown in figure~\ref{fig:pointing-data}. We see a strong time-of-day
		dependence for the pointing offset: During the night (roughly 23-11 UTC)
		the pointing is relatively stable, but during the day (roughly 11-23 UTC)
		it fluctuates with an amplitude of around 3 arcminutes.
		\begin{figure}[h]
			\centering
			\hspace*{-5mm}\includegraphics[width=1.1\columnwidth]{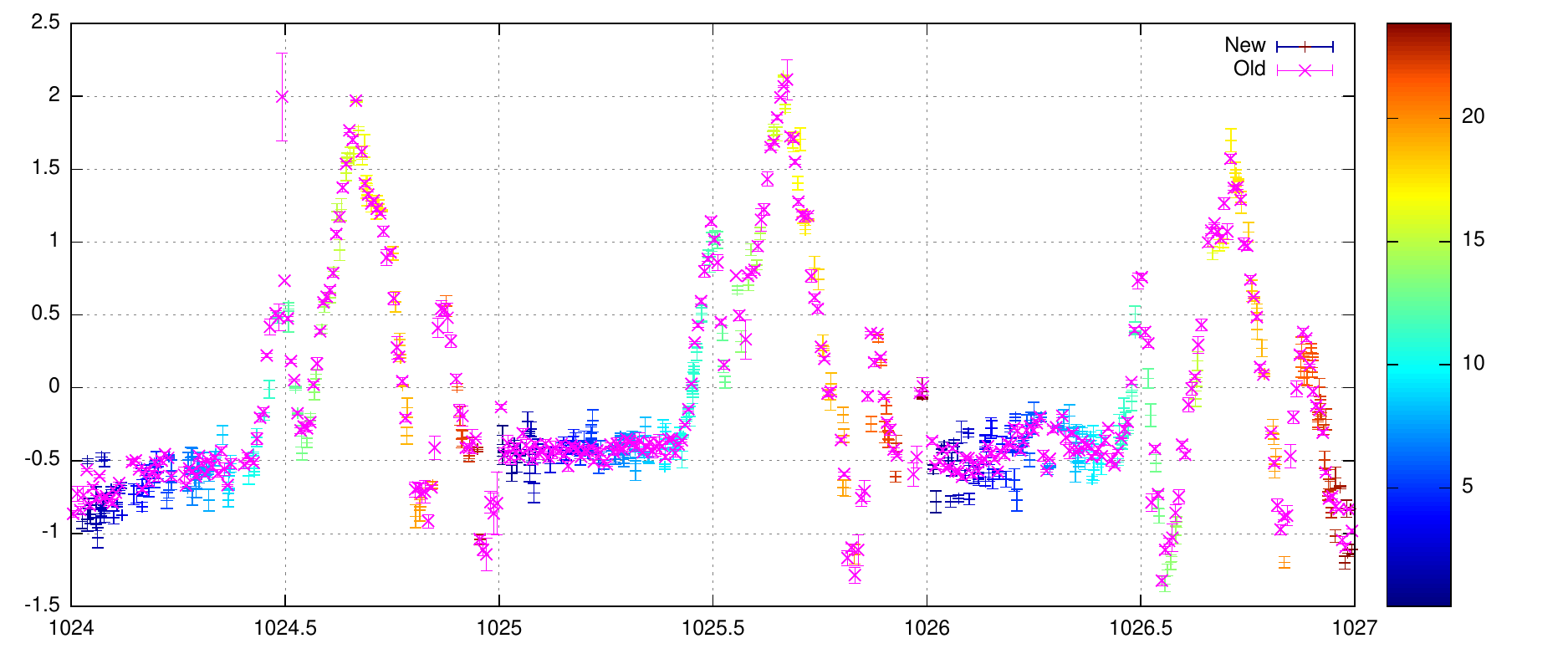} \\
			\hspace*{-5mm}\includegraphics[width=1.1\columnwidth]{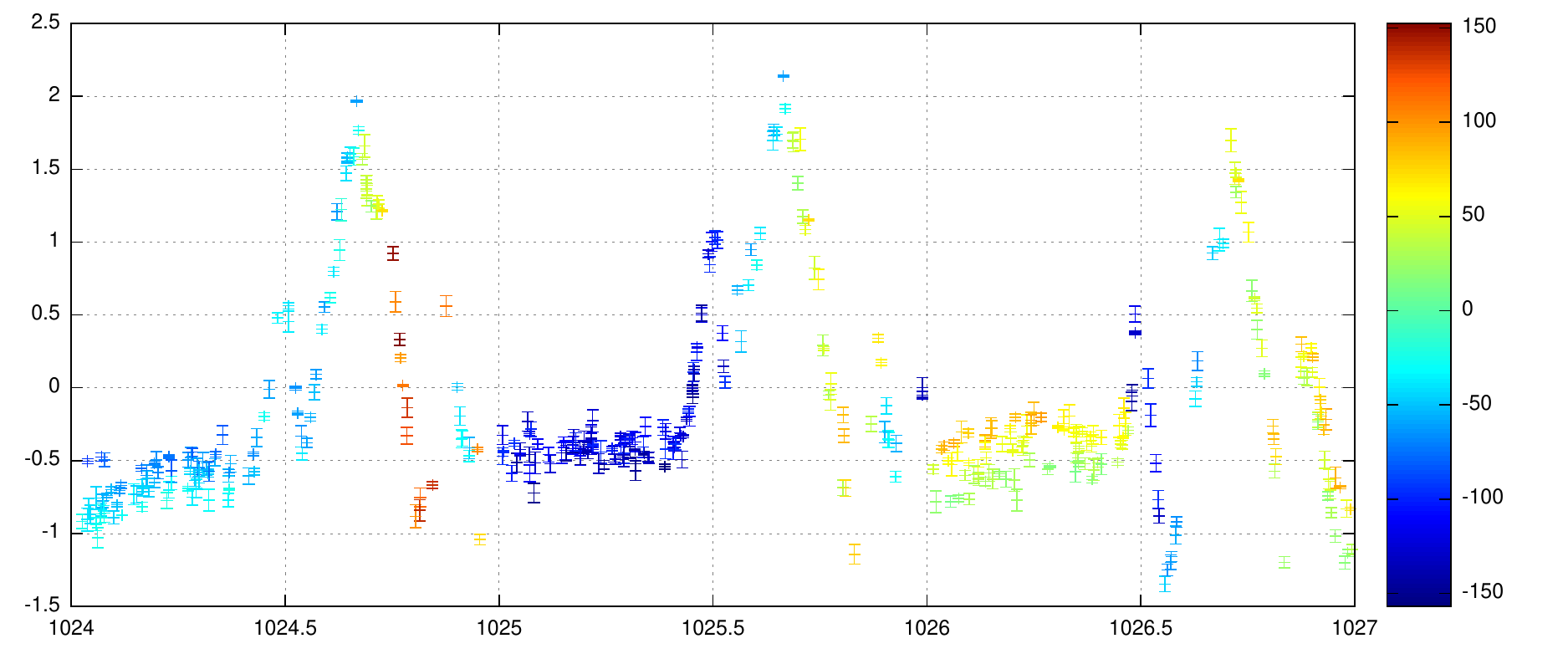} \\
			\caption{\dfn{Top}: Measured vertical pointing displacement, in arcminutes, as a
			function of time, in days since 2017-01-01. The points are colored by the UTC hour,
			with 11-23 being the rough day-time period. The magenta points show the result of the
			DR4 method for comparison. These have smaller error bars but lower time resolution
			due to averaging over all sources in a TOD. This averaging also masks the azimuth-dependent
			pointing offset, which we were not aware of in DR4.
			\dfn{Bottom}: The same data, this time colored by azimuth. The high scatter for
			$1026<t<1026.5$ is strongly correlated with azimuth.}
			\label{fig:pointing-data}
		\end{figure}

		New compared to DR4 is a surprising amount of short-term scatter even during the night,
		with a peak-to-peak amplitude of up to 0.5\arcmin{}.
		This scatter turns out to be strongly correlated with azimuth, and was
		not discovered in DR4 because we didn't have sub-TOD resolution. The pointing
		offset measured in DR4 was the average over all the sources inside that TOD,
		some of which would have had low and some high azimuth. DR4 therefore had an
		uncorrected pointing offset of up to 0.25\arcmin{} at the ends of its widest scans.
		In practice this was not a big concern there because 1) the offset would usually
		be less than 0.25\arcmin{} at the ends, 2) the average offset inside a scan would be
		about half that at the edges, 3) when averaging across multiple observations
		it adds in quadrature to the effective beam, 4) only a small subset of
		the DR4 data (2016 only) used as wide scans as these, and 5) we measure
		an empirical beam correction factor from the final maps that would absorb this.
		Even if all the TODs has been maximally affected by this, the expected increase
		in beam size at f150 would be just 0.4\%.
	\item Both because the pointing offset varies within a TOD (due to the
		azimuth effect) and because a TOD in some cases has no usable measurements
		due to a lack of bright point sources, we can't use these measurements
		directly to correct the pointing. We instead fit a smooth model that can
		be interpolated for each az and TOD. We build a separate model for the
		data points from each depth-1 map, with each model consisting of a piecewise
		linear function in time plus a constant slope in azimuth. This is done
		both for the vertical and horizontal pointing offset, for a total of
		$2 n_p = 2 n_t + 2$ free parameters per depth-1 map. We adjust $n_p$
		to minimize $\chi^2/(n_d-n_p) + n_p T/(500\text{s})$, where $n_d$
		is the number of data points from the depth-1 map and $T$ is the
		duration they span. Typically we end up with $\sim50$ data points per
		degree of freedom in the fit.
		\begin{figure}[h]
			\centering
			\hspace*{-5mm}\includegraphics[width=1.1\columnwidth]{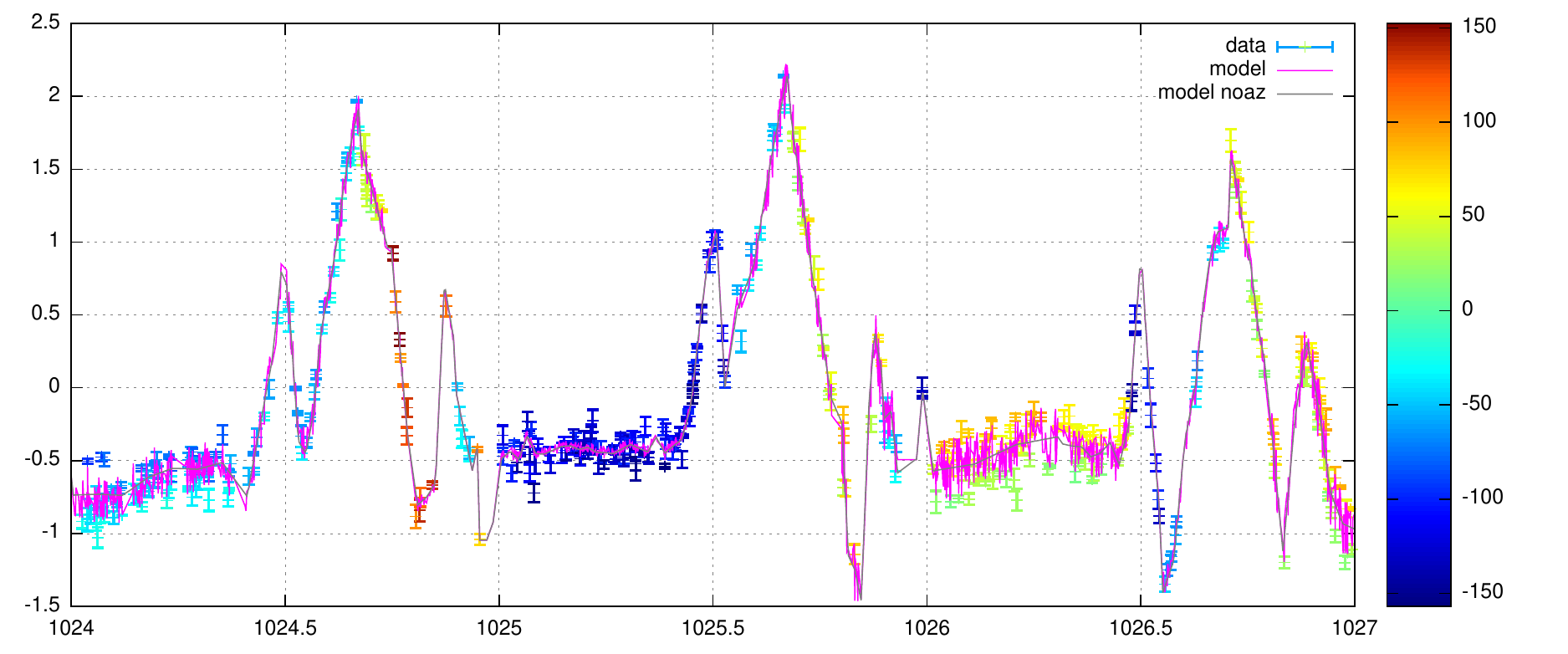} \\
			\hspace*{-5mm}\includegraphics[width=1.1\columnwidth]{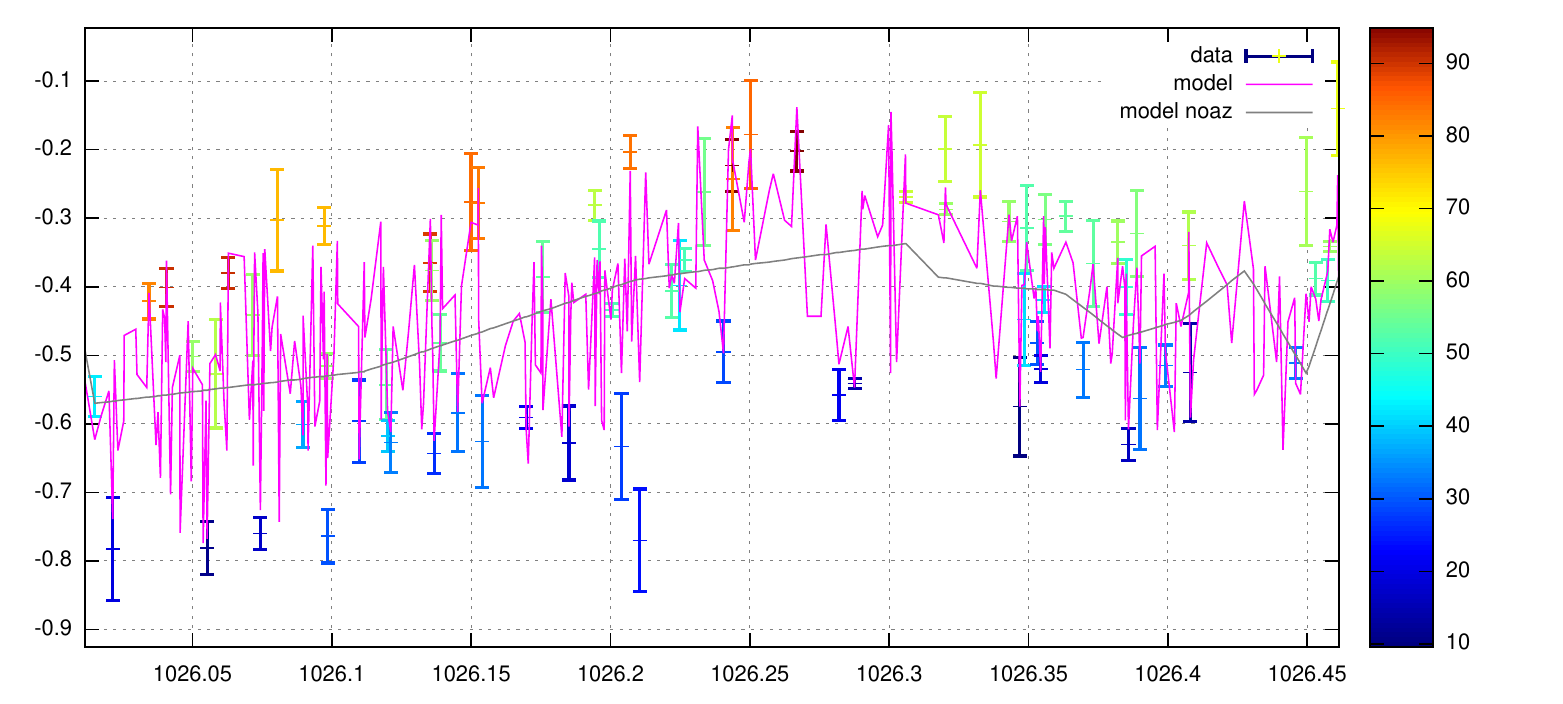} \\
			\caption{\dfn{Top}: The best-fit pointing model (magenta) compared to the same measurements
			as in figure~\ref{fig:pointing-data}, again with the vertical pointing offset in
			arcminutes on the y axis, days since 2017-01-01 on the x axis and azimuth in degrees
			in the colorbar. The model successfully captures both the time- and azimuth-dependence
			of the pointing. It is easiest to see around day 1026, where the azimuth-dependence is particularly strong. It can be hard distinguish in regions with low azimuth-dependence. The gray curve shows the model with the azimuth-dependence disabled. \dfn{Bottom}: Zoom in on the messy region $1026<t<1026.5$, illustrating
			how the model captures the apparent scatter. The model appears to jump erratically
			up and down here, but this is simply a consequence of each point having its own azimuth
			-- the model would look smooth if plotted in a 3D time-az-displacement graph.}
			\label{fig:pointing-model}
		\end{figure}

		The resulting model is plotted in figure~\ref{fig:pointing-model}.
	\item For each TOD in DR6, evaluate the model at the TOD's halfway time,
		resulting in a vertical and horizontal pointing offset along with a
		vertical and horizontal azimuth slope. The latter enter into a correction
		of the horizontal coordinate system as
		\begin{align}
			\text{az}'(t) &= \text{az}(t) + \frac{d\text{az}'}{d\text{az}}(\text{az}-\text{az}_0) &
			\text{el}'(t) &= \text{el}(t) + \frac{d\text{el}'}{d\text{az}}(\text{az}-\text{az}_0)
		\end{align}
		where $\text{az}',\text{el}'$ are the corrected versions of $\text{az},\text{el}$
		respectively, and $\text{az}_0$ is the azimuth of the mid-point of the scan for
		this TOD. Figure~\ref{fig:pointing-test} tests the round-trip performance of
		the pointing correction by measuring the pointing offsets from new depth-1 maps
		built after applying the pointing correction. The new pointing model has similar
		scatter as the old DR4, but fewer outliers due to capturing the azimuth slope.

		The pointing slope is similar to the effect of a tilt of the telescope's azimuth
		axis from the vertical, but cannot be explained that way because it is not
		constant in time. We have so far not gotten to the bottom of what causes it.
\end{enumerate}

\begin{figure}
	\centering
	\includegraphics[width=0.4\columnwidth]{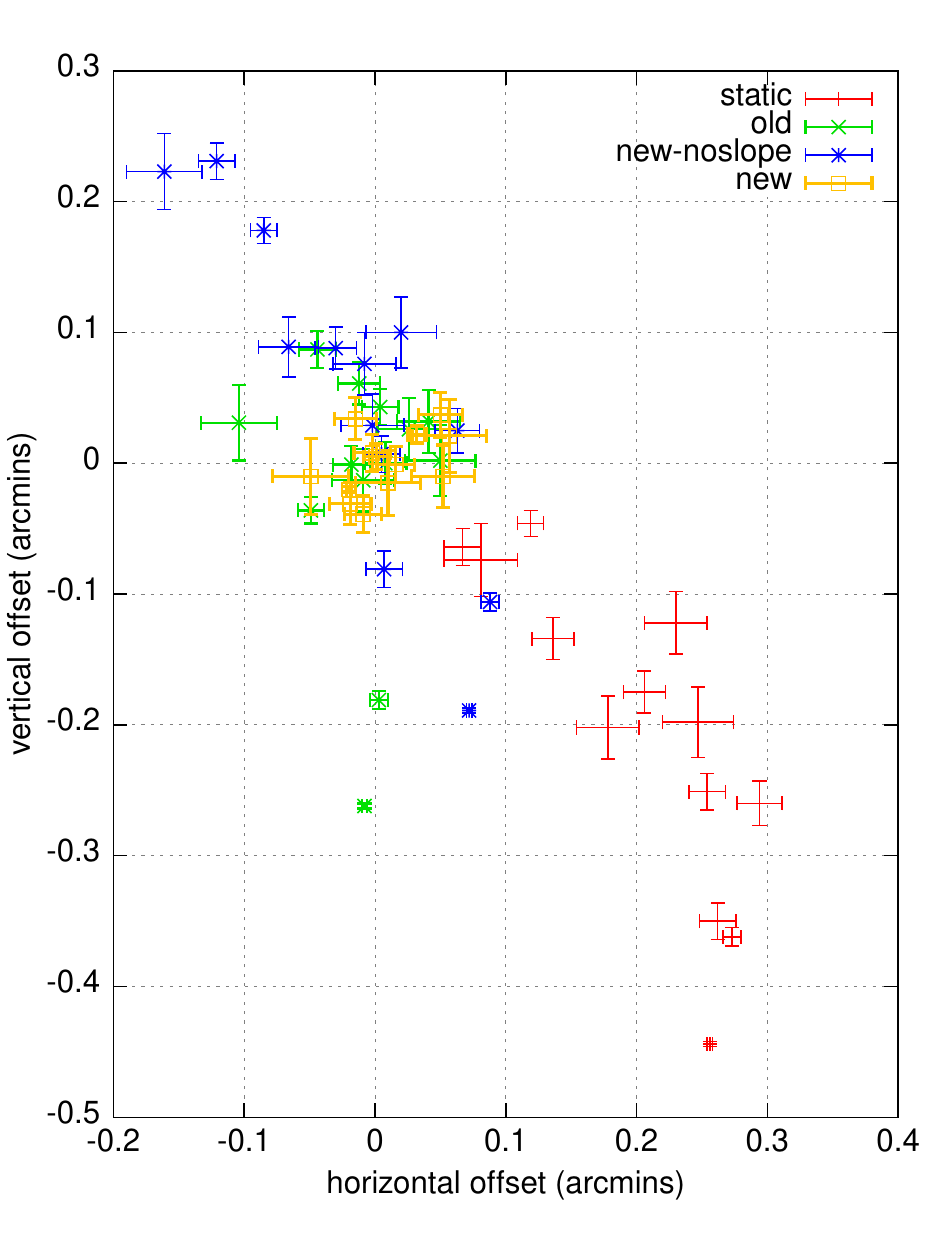}
	\caption{Comparison of the performance of the initial (static)
	pointing model (red), the DR4 pointing model (green) and
	the new pointing model without (blue) and with (yellow) the
	azimuth-dependent pointing offset, tested on a small but representative
	set of TODs. For most of these the new model is only a modest improvement
	on the old one, but the azimuth slope eliminates a few outliers.}
	\label{fig:pointing-test}
\end{figure}

Overall the DR6 pointing model is a moderate improvement over the DR4 pointing model,
but this is countered by DR6 having a much higher fraction of high-amplitude azimuth
scans that are more difficult to model. As described in section~\ref{sec:beam},
the beam size inferred from point sources in our CMB maps is slightly larger than
that directly measured from planet observations. If interpreted as being entirely
due to uncorrected pointing jitter, then DR6 has an average jitter of $0.15\pm0.04$
arcmin, compared to $0.11\pm0.08$ for DR4, so it's around 30\% worse.

\clearpage

\section{Spatially dependent noise}
\label{sec:spatdep}
The atmosphere acts as a strongly spatially and temporally correlated noise
component for ground-based CMB telescopes like ACT. When projected onto the
sky via the telescope scanning pattern, this results in a map where not
just the overall noise level is spatially dependent (as shown in figure~\ref{fig:depthmap}),
but also the correlation length, directions of stripiness, and so on.

\begin{figure}[htp]
	\centering
	{\Large TT} \\
	\includegraphics[width=\textwidth]{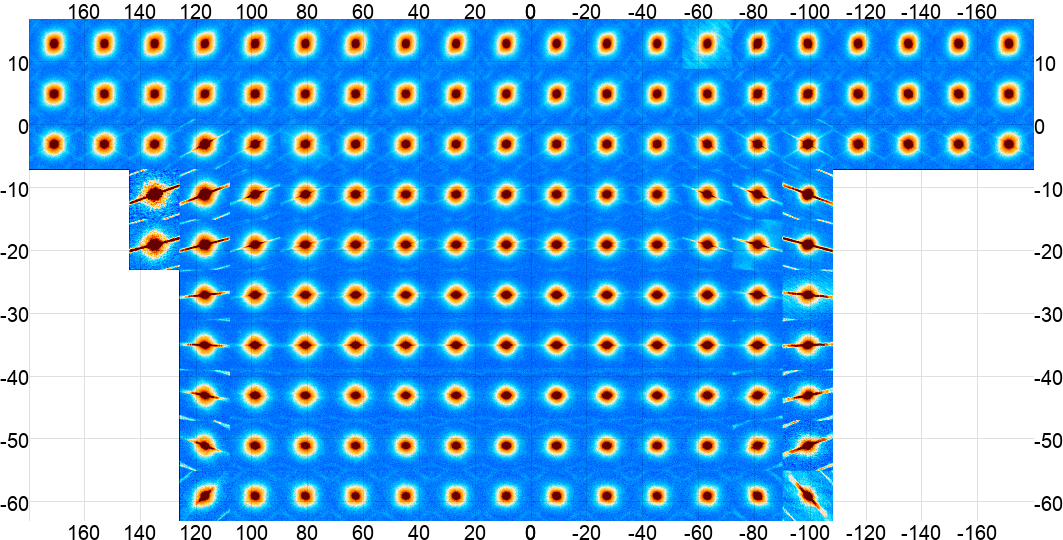} \\
	\vspace{5mm}
	{\Large EE} \\
	\includegraphics[width=\textwidth]{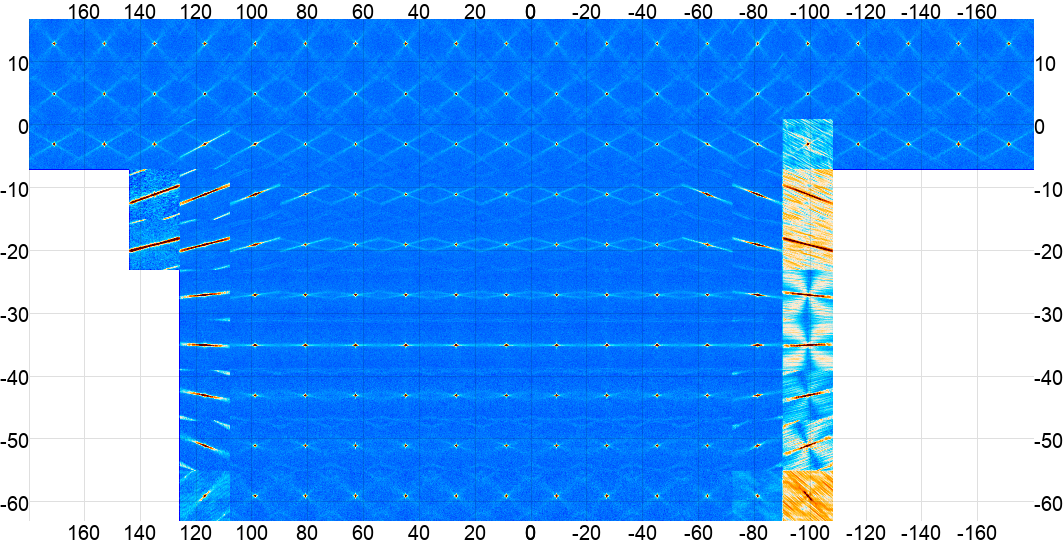}
	\caption{Spatial dependence of the 2D noise power spectrum for PA5 f150.
	The ACT sky coverage is split into tiles, and for each tile the 2D noise
	spectrum is plotted out to $\pm 5000$ for $\ell_y$ and $\ell_x$,
	with $\ell_y=\ell_x=0$ being where each tile's noise spectrum peaks
	(center of the red blobs). The tiles are shown in standard celestial
	coordinates, so the horizontal and vertical axes are RA and dec.
	\dfn{Top}: Total intensity. \dfn{Bottom}: polarization.
	The color range is logarithmic, going from dark blue at 0.3 times the
	white noise level and dark red at 100 times the white noise level.
	Other arrays and
	bands are qualitatively similar, but with different $\ell_\textrm{knee}$,
	see figure~\ref{fig:lknee-map-TT}, figure~\ref{fig:lknee-map-EE} and figure~\ref{fig:lknee-dec}.}
	\label{fig:nspec-2d-map}
\end{figure}

One way of visualizing this by splitting the sky into tiles, and measuring the
2D noise power spectrum for each. These 2D noise spectra are shown in
figure~\ref{fig:nspec-2d-map}. The dominant
feature here is the low-$\ell$ atmospheric noise, which is much more prominent
in total intensity than in polarization. From this central peak two thin
lines of higher noise power extend outwards in a strongly
declination-dependent X-shape along the two dominant scanning directions.
Unexpectedly, this X-shape has similar width and amplitude for polarization
and total intensity, but that amplitude is low enough, at around two times
the white noise floor, that it can probably be ignored in most contexts.

\begin{figure}[htp]
	\centering
	\begin{closetabcols}
	\begin{tabular}{cc}
		PA5 f090 & PA6 f090 \\
		\includegraphics[width=0.5\textwidth,trim=0 0 12mm 0]{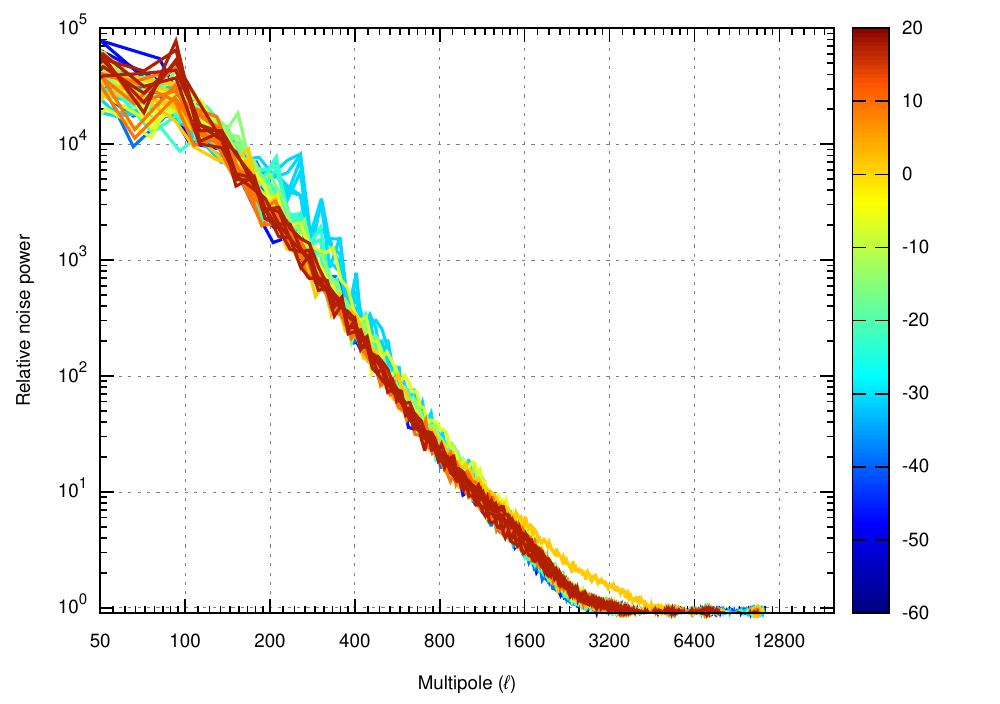} &
		\includegraphics[width=0.5\textwidth,trim=0 0 12mm 0]{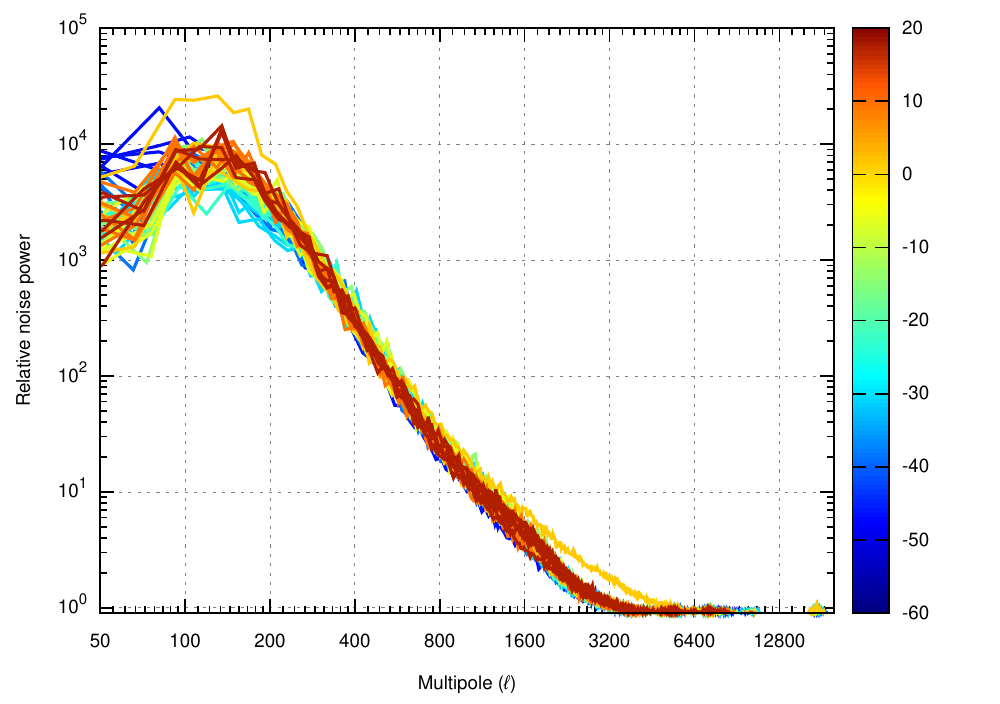} \\
		PA5 f150 & PA6 f150 \\
		\includegraphics[width=0.5\textwidth,trim=0 0 12mm 0]{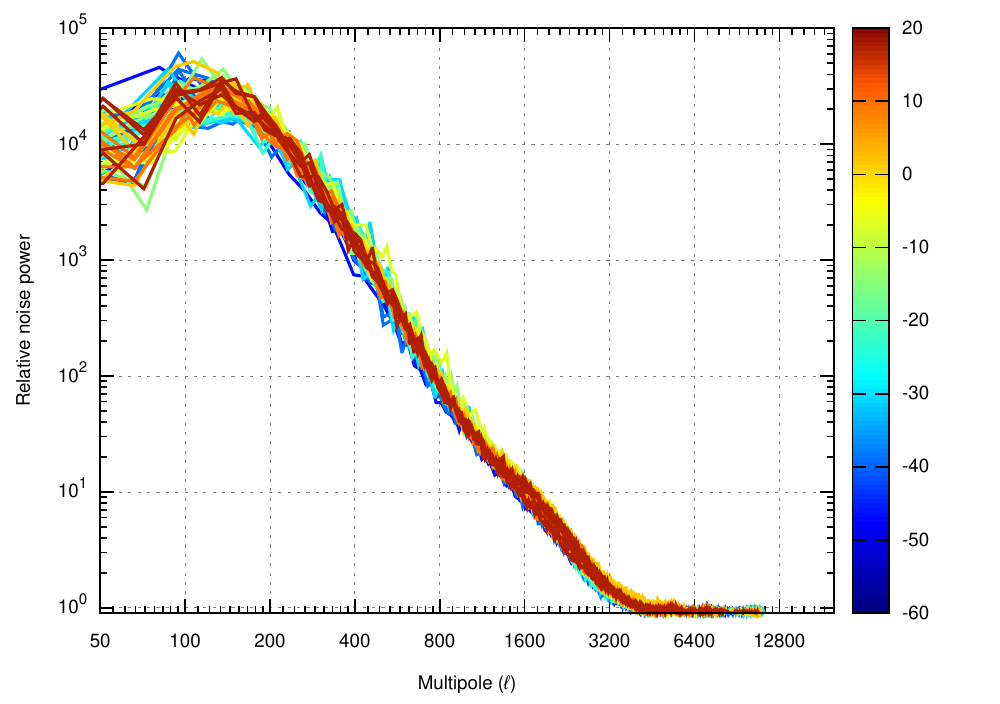} &
		\includegraphics[width=0.5\textwidth,trim=0 0 12mm 0]{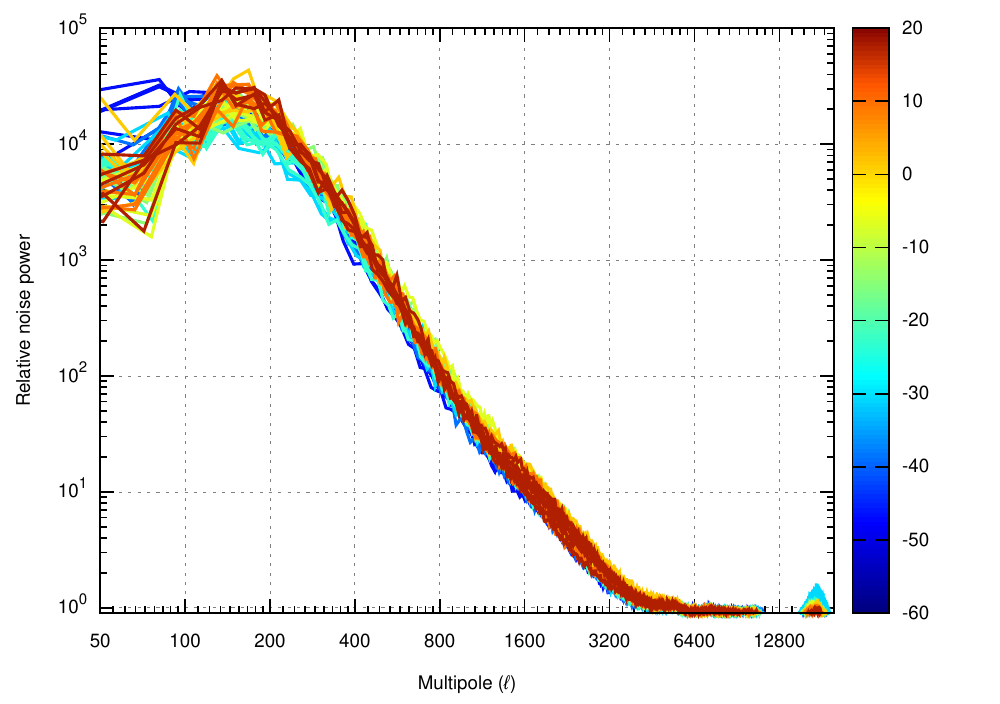} \\
		PA4 f150 & PA4 f220 \\
		\includegraphics[width=0.5\textwidth,trim=0 0 12mm 0]{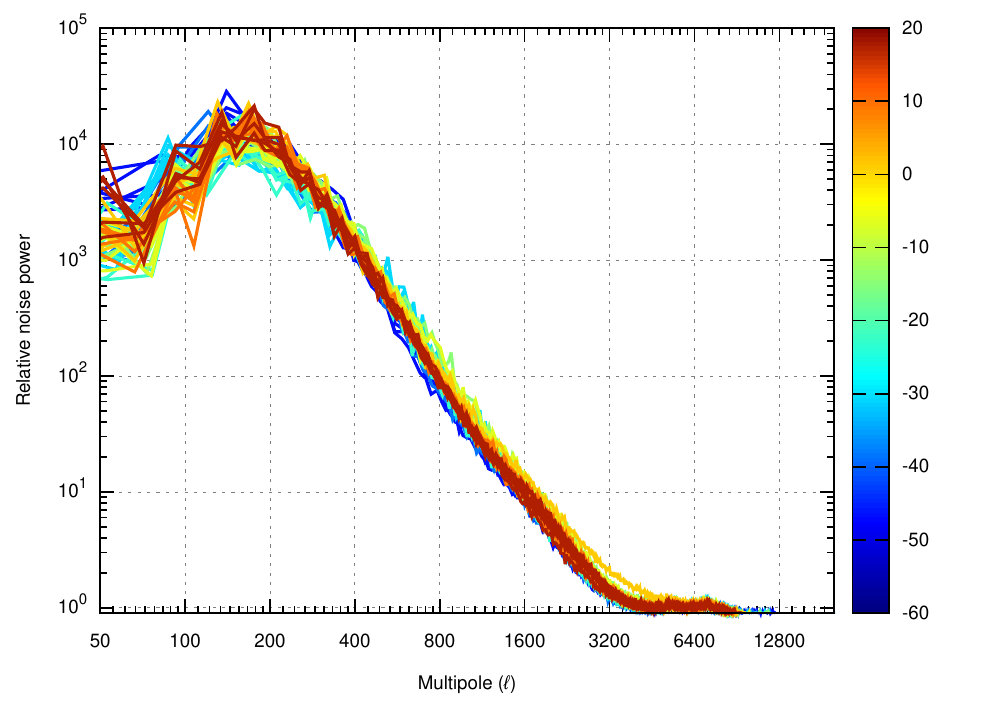} &
		\includegraphics[width=0.5\textwidth,trim=0 0 12mm 0]{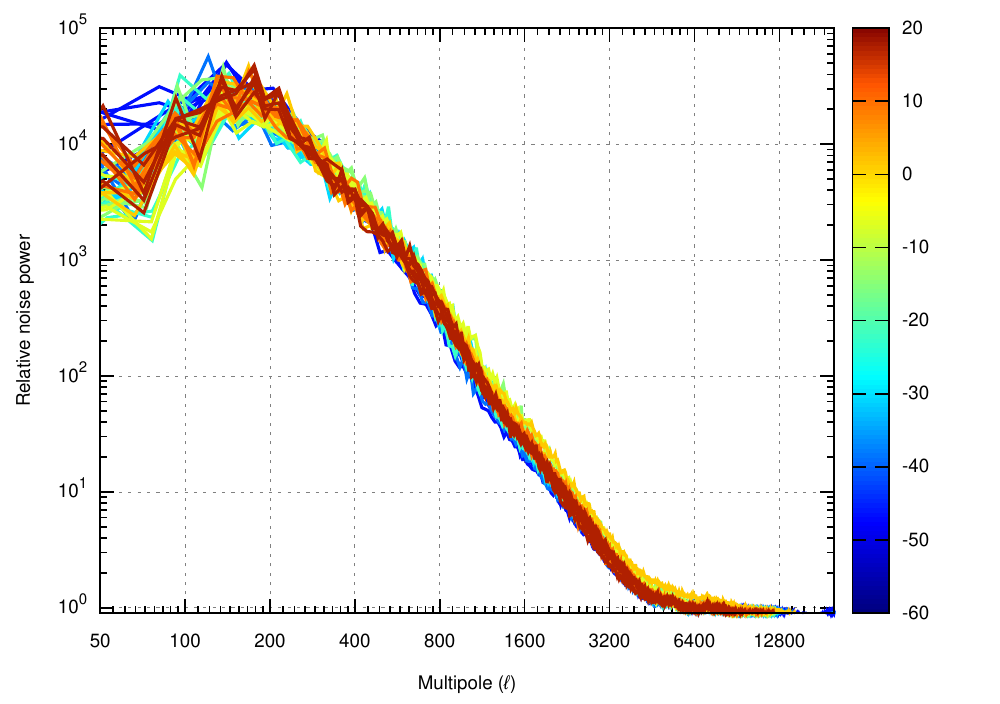}
	\end{tabular}
	\end{closetabcols}
	\caption{TT noise spectra for each detector array for the tiles in
	figure~\ref{fig:nspec-2d-map} for $-40\degree<RA<60\degree$, colored by declination.
	The turnaround for $\ell \lesssim 200$ is due to the mapmaker transfer
	function. Overall there is little spatial dependence for the TT spectrum.}
	\label{fig:nspec-1d-var-TT}
\end{figure}

\begin{figure}[htp]
	\centering
	\begin{closetabcols}
	\begin{tabular}{cc}
		PA5 f090 & PA6 f090 \\
		\includegraphics[width=0.5\textwidth,trim=0 0 12mm 0]{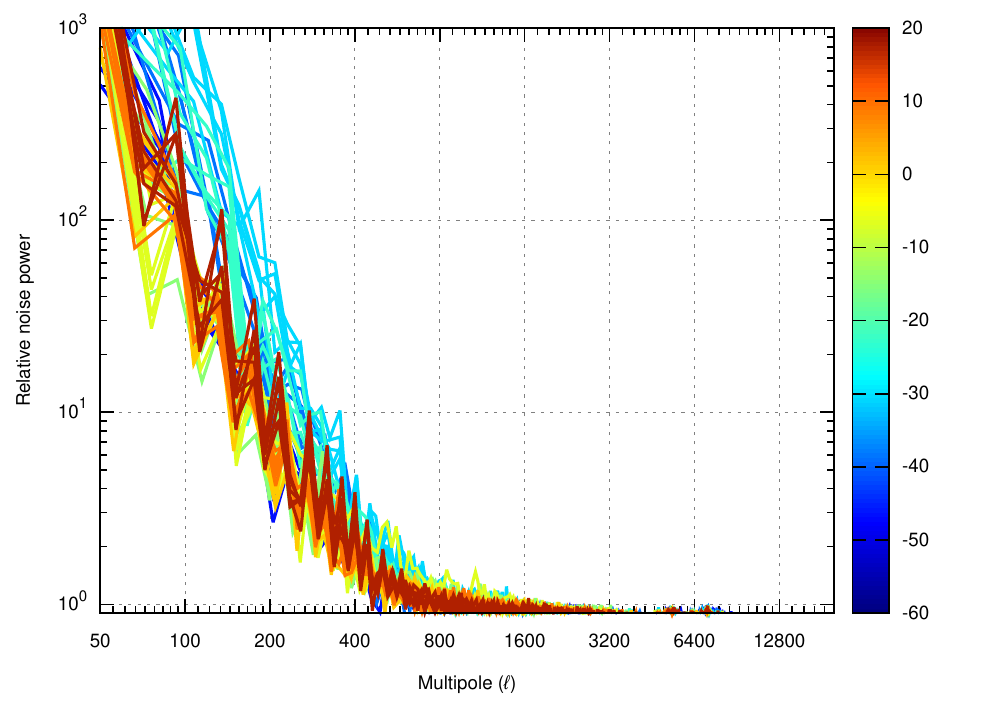} &
		\includegraphics[width=0.5\textwidth,trim=0 0 12mm 0]{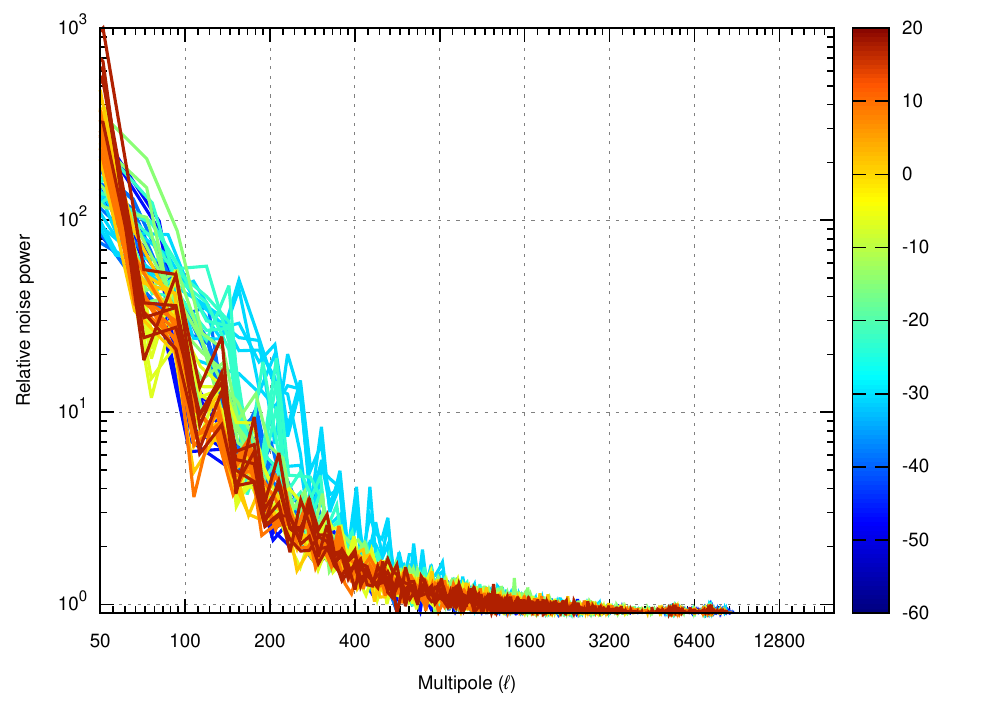} \\
		PA5 f150 & PA6 f150 \\
		\includegraphics[width=0.5\textwidth,trim=0 0 12mm 0]{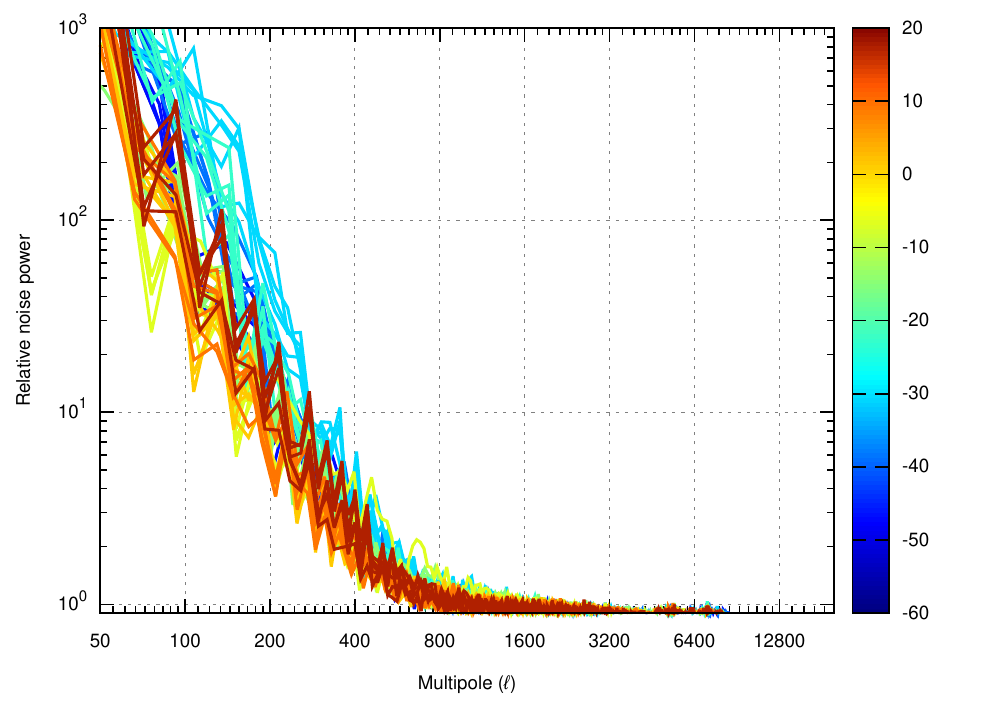} &
		\includegraphics[width=0.5\textwidth,trim=0 0 12mm 0]{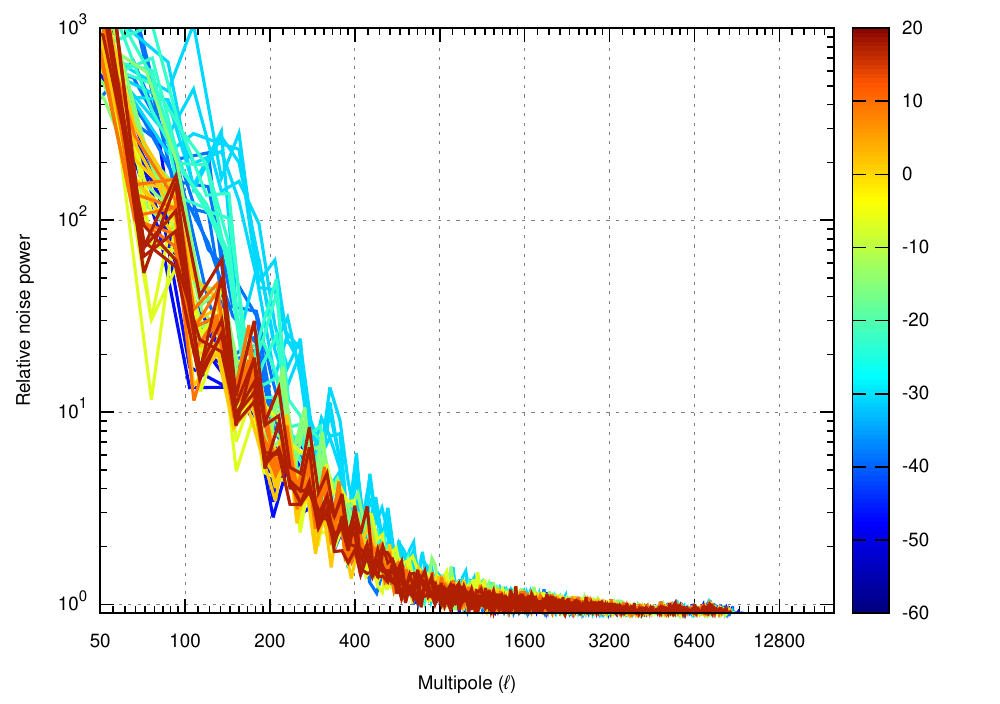} \\
		PA4 f150 & PA4 f220 \\
		\includegraphics[width=0.5\textwidth,trim=0 0 12mm 0]{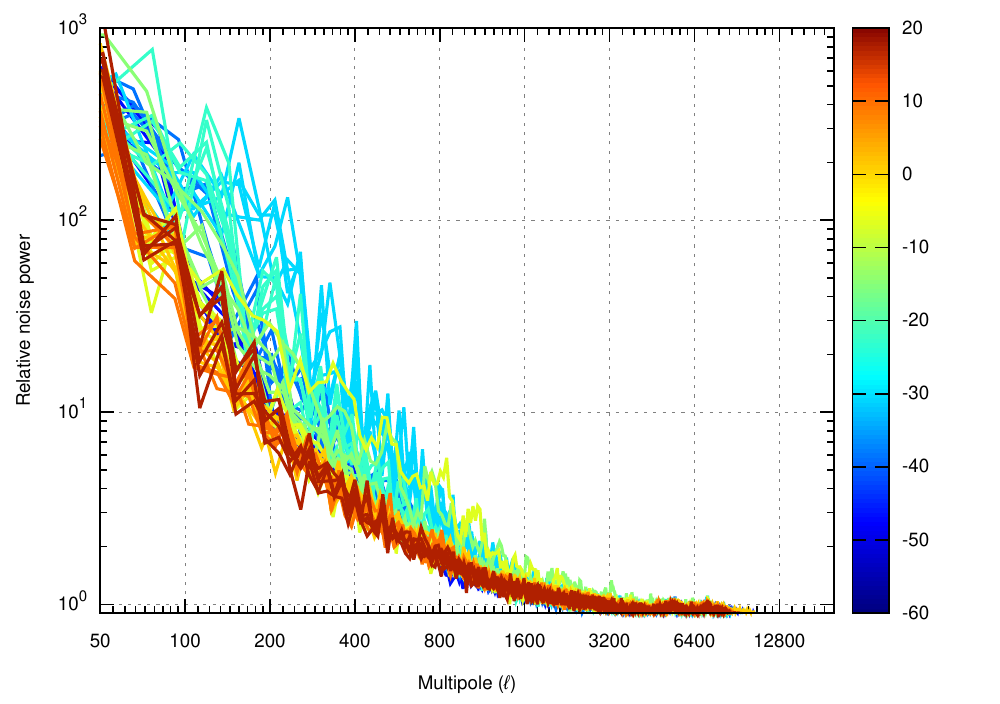} &
		\includegraphics[width=0.5\textwidth,trim=0 0 12mm 0]{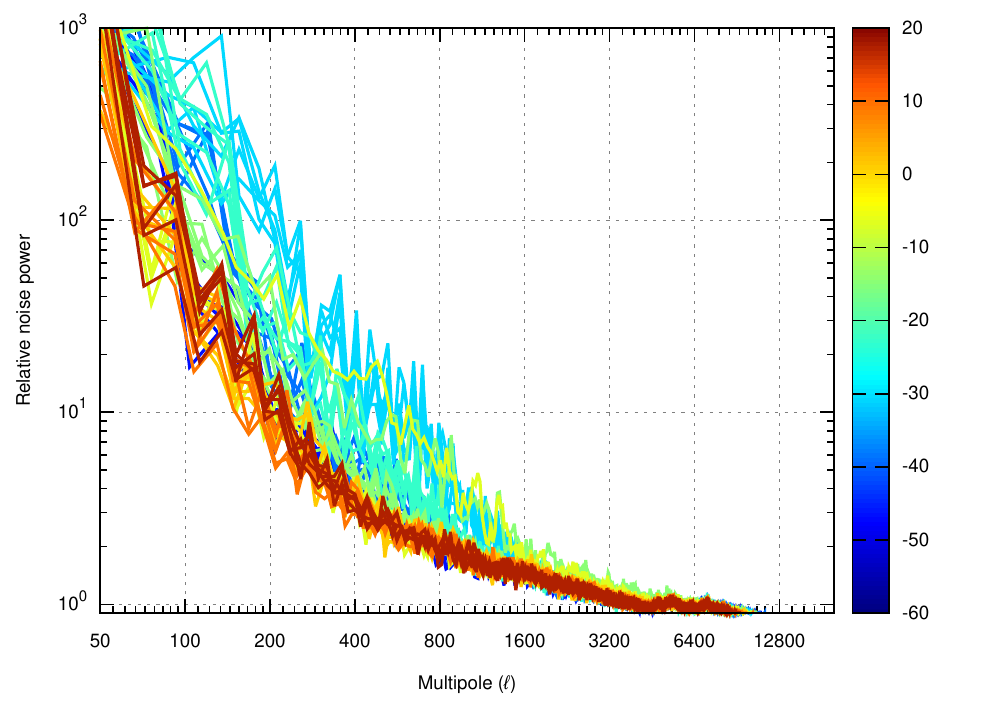}
	\end{tabular}
	\end{closetabcols}
	\caption{EE noise spectra for each detector array for the tiles in
	figure~\ref{fig:nspec-2d-map} for $-40\degree<\textrm{RA}<60\degree$, colored by declination.
	The jaggedness of the curves at low $\ell$ is a moireé effect from the
	angular averaging of a pixelated anisotropic 2D noise spectrum. The EE
	spectrum depends mainly on declination, with the non-crosslinked region
	$\textrm{dec}\sim -35\degree$ having higher noise, especially for PA4.}
	\label{fig:nspec-1d-var-EE}
\end{figure}

The symmetric part of the total intensity and polarization noise spectra
are shown in figures~\ref{fig:nspec-1d-var-TT} and \ref{fig:nspec-1d-var-EE}
respectively. The spectra roughly follow the 1/f profile
$\sigma^2(1+(\ell/\ell_\textrm{knee})^\alpha$, with $\alpha\sim-3$, but
there are some clear deviations. At $\ell \lesssim 200$ the TT noise spectrum
turns over due to the mapmaker transfer function, and in polarization
the slope steepens below $\ell\sim 400$ for some of the arrays. The different
tile spectra are quite consistent in TT. In EE, on the other hand, there is a
strong declination dependence with up to ten times higher noise at low $\ell$
in the non-crosslinked region $\textrm{dec}\sim-35\degree$.

Figure~\ref{fig:nspec-1d-avg} compares the tile-averaged noise spectra
for the different detector arrays. The TT noise spectra behave as expected,
getting progressively more atmospheric noise as we go to shorter wavelengths,
and with the arrays being consistent with each other for each bandpass.
The situation is more complicated in polarization. Here individual array
differences dominate, with PA6 f090 having significantly less correlated
noise than PA5 f090, and both bands of PA4 having much more correlated
noise for $\ell>200$ than the other arrays.

\begin{figure}[htp]
	\centering
	\begin{closetabcols}
	\begin{tabular}{cc}
		PA5 f090 & PA6 f090 \\
		\includegraphics[width=0.5\textwidth,trim=0 0 0 0]{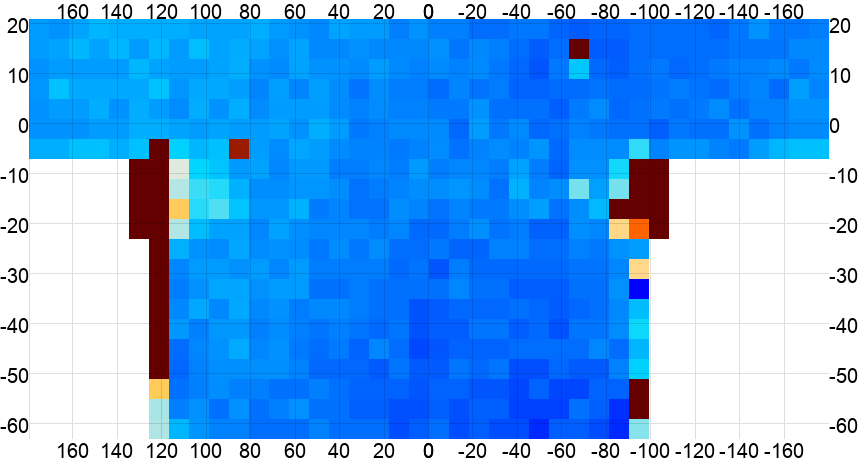} &
		\includegraphics[width=0.5\textwidth,trim=0 0 0 0]{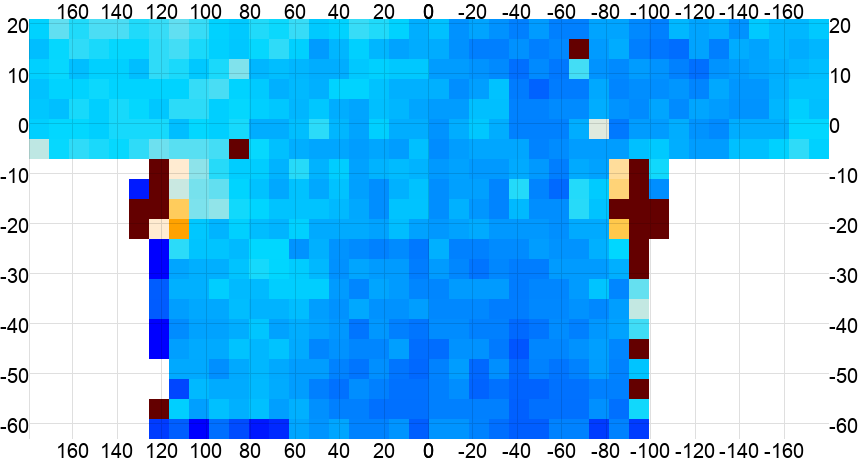} \\
		PA5 f150 & PA6 f150 \\
		\includegraphics[width=0.5\textwidth,trim=0 0 0 0]{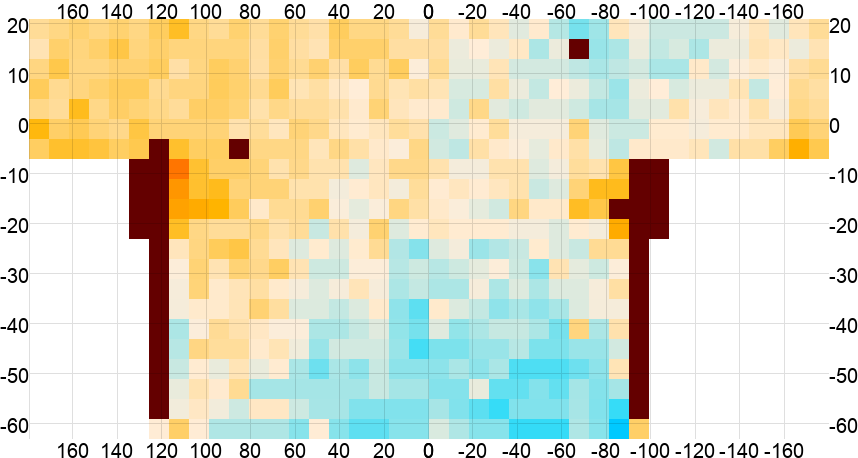} &
		\includegraphics[width=0.5\textwidth,trim=0 0 0 0]{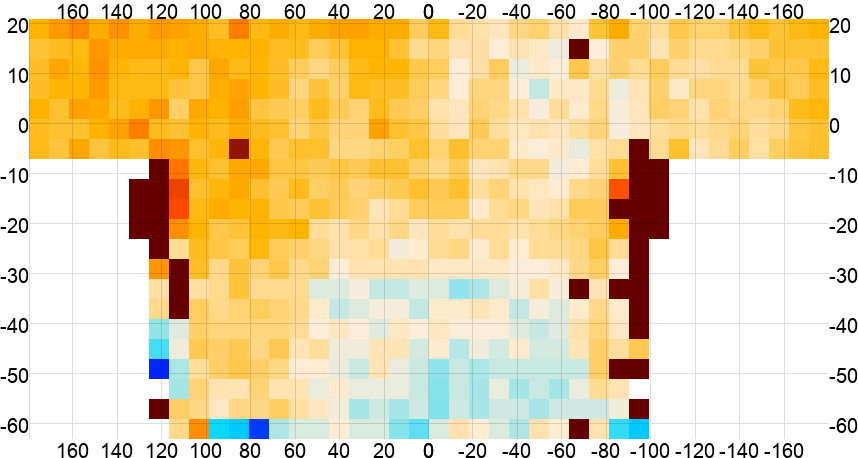} \\
		PA4 f150 & PA4 f220 \\
		\includegraphics[width=0.5\textwidth,trim=0 0 0 0]{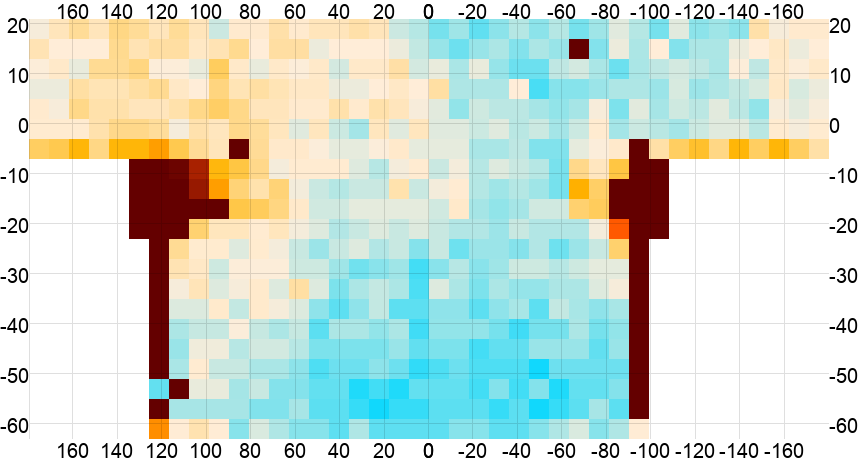} &
		\includegraphics[width=0.5\textwidth,trim=0 0 0 0]{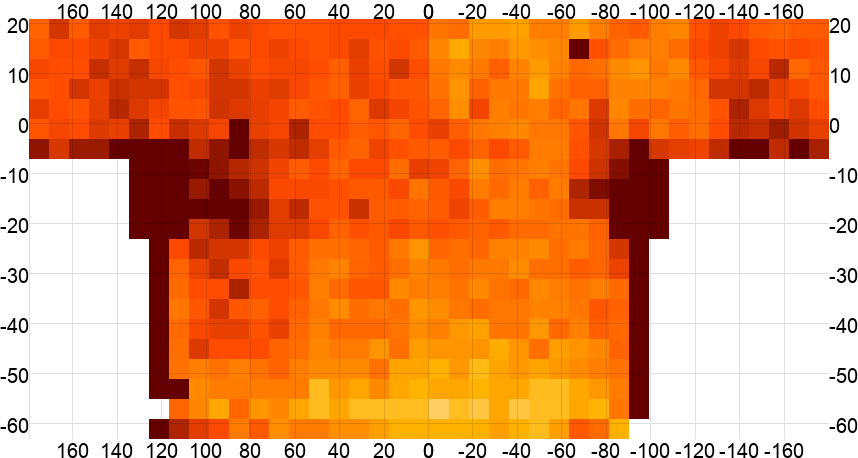}
	\end{tabular}
	\end{closetabcols}
	\caption{Spatial dependence of $\ell_\textrm{knee}$ in total intensity.
	The color range goes from 1500 (dark blue) to 4500 (dark red). The horizontal
	and vertical axes are RA and dec in degrees.}
	\label{fig:lknee-map-TT}
\end{figure}

\begin{figure}[htp]
	\centering
	\begin{closetabcols}
	\begin{tabular}{cc}
		PA5 f090 & PA6 f090 \\
		\includegraphics[width=0.5\textwidth,trim=0 0 0 0]{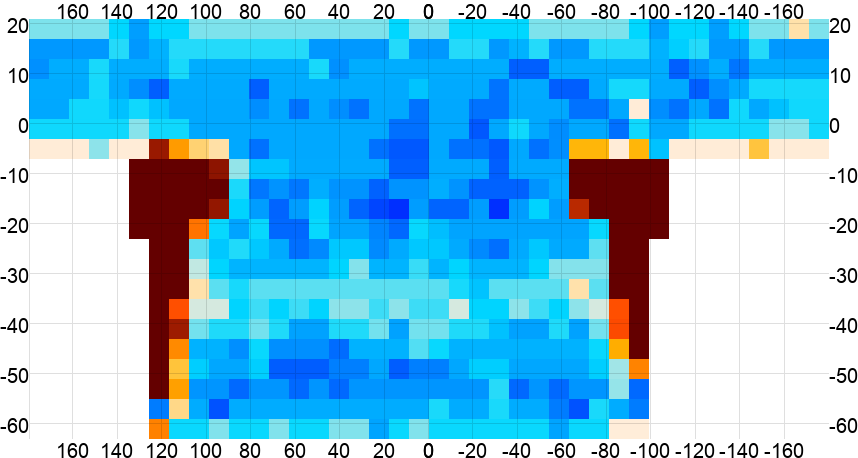} &
		\includegraphics[width=0.5\textwidth,trim=0 0 0 0]{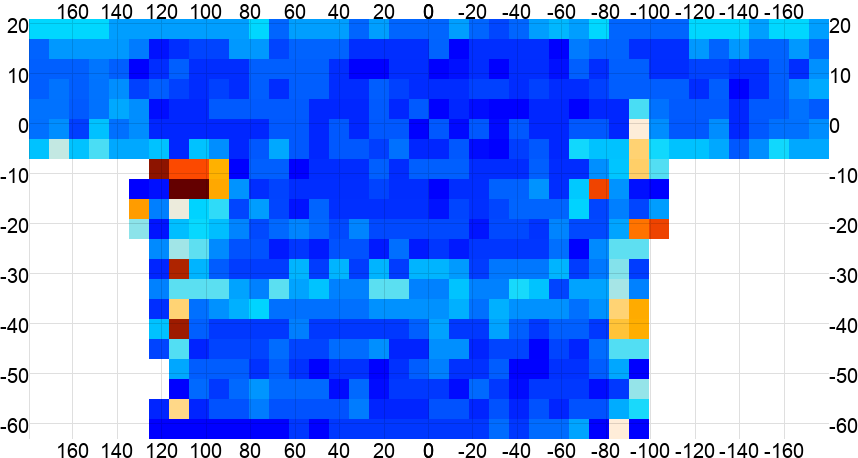} \\
		PA5 f150 & PA6 f150 \\
		\includegraphics[width=0.5\textwidth,trim=0 0 0 0]{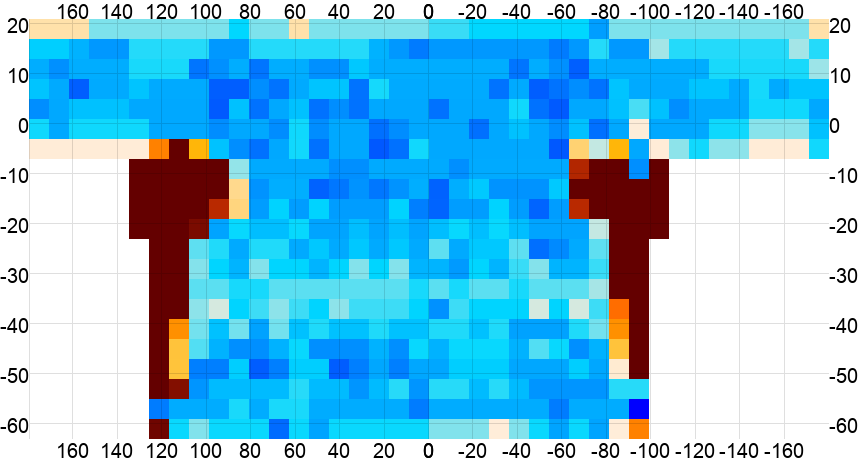} &
		\includegraphics[width=0.5\textwidth,trim=0 0 0 0]{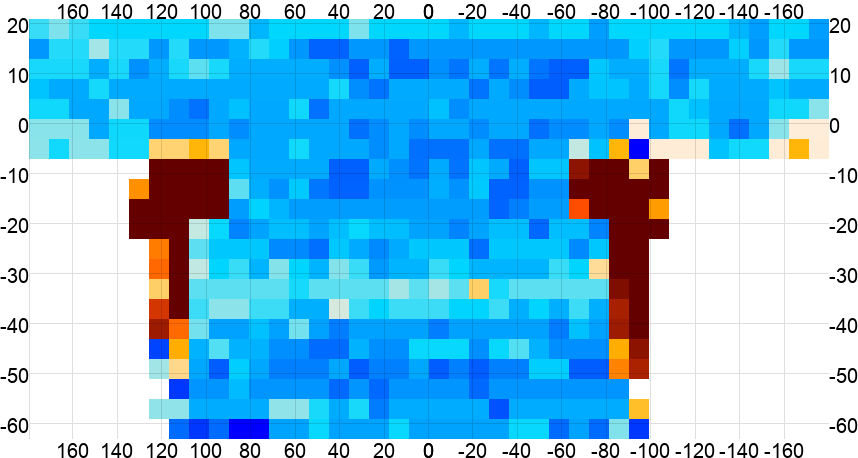} \\
		PA4 f150 & PA4 f220 \\
		\includegraphics[width=0.5\textwidth,trim=0 0 0 0]{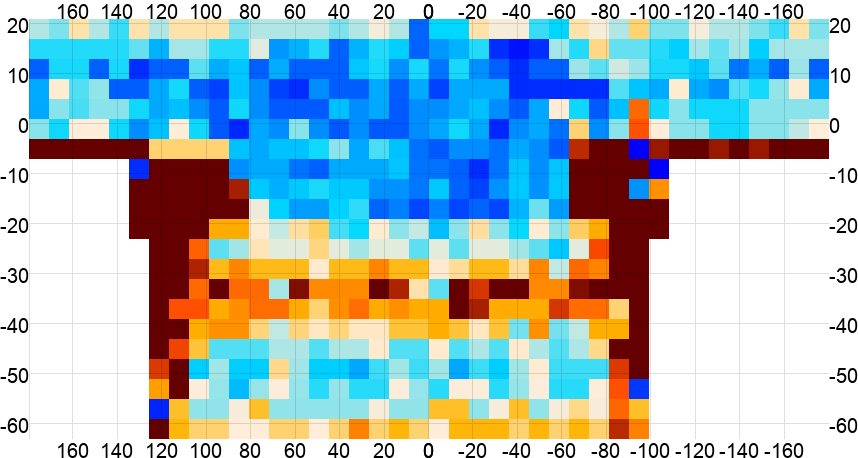} &
		\includegraphics[width=0.5\textwidth,trim=0 0 0 0]{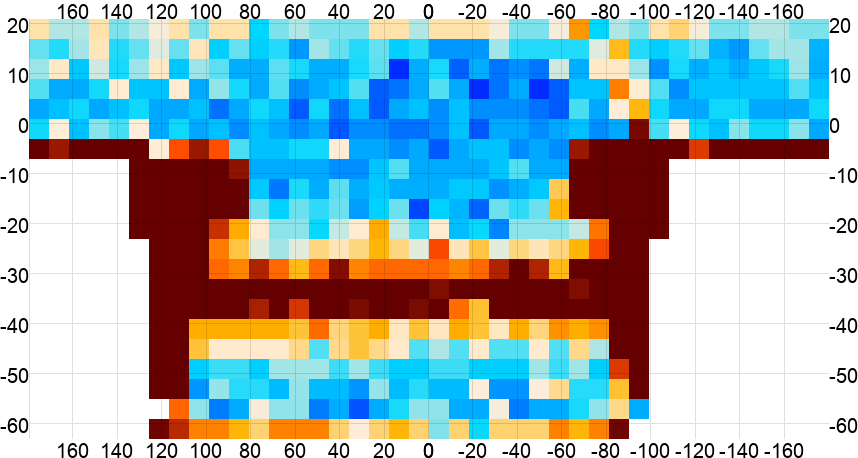}
	\end{tabular}
	\end{closetabcols}
	\caption{Spatial dependence of $\ell_\textrm{knee}$ in polarization.
	The color range goes from 200 (dark blue) to 1200 (dark red). The horizontal
	and vertical axes are RA and dec in degrees.}
	\label{fig:lknee-map-EE}
\end{figure}

\begin{figure}[htp]
	\centering
	\includegraphics[width=0.7\textwidth]{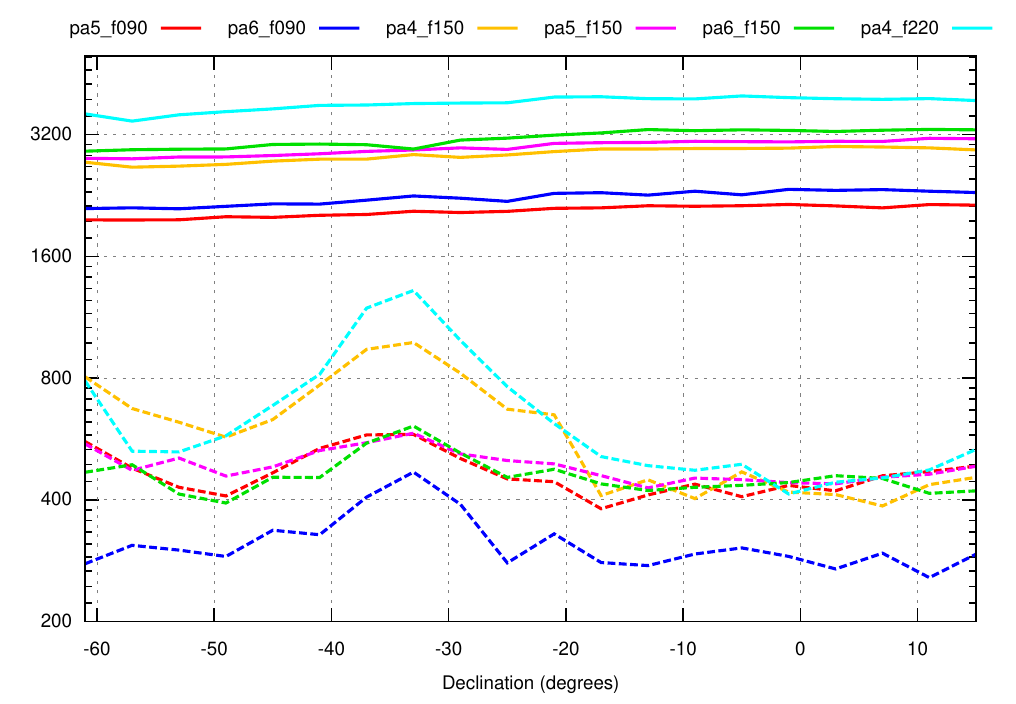}
	\caption{Declination-dependence of $\ell_\textrm{knee}$ measured
	for $-40\degree<\textrm{RA}<60\degree$. Solid lines are total intensity, dashed
	lines polarization. Total intensity shows little declination dependence,
	and $\ell_\textrm{knee}$ is mainly determined by the band. In polarization
	the individual detector array dependence is much greater, as is the declination
	dependence. The peak for $\textrm{dec}\sim-35\degree$ corresponds to an area
	with no crosslinking. The generally higher $\ell_\textrm{knee}$ for PA5 f090
	than PA6 f090 in polarization is probably a sign of stronger atmospheric T to P leakage in
	PA5 f090.}
	\label{fig:lknee-dec}
\end{figure}

Figures~\ref{fig:lknee-map-TT} and \ref{fig:lknee-map-EE} show the
spatial distribution of $\ell_\textrm{knee}$, measured as the point
where the symmetric part of the noise spectrum falls below twice the
noise floor. Aside from some edge areas with very few observations,
the variations are declination-dominated, though there is also some
RA-dependence in total intensity. Figure~\ref{fig:lknee-dec}
shows the average $\ell_\textrm{knee}$ as a function of declination,
and here we again see the qualitatively different behavior in
total intensity and polarization. In total intensity $\ell_\textrm{knee}$
is mainly a function of bandpass, while in polarization is depends
strongly on the detector array and crosslinking angle. In the extreme
case of PA4 f220 $\ell_\textrm{knee}$ is more than 3 times as high
at $\textrm{dec}\approx-33\degree$ as at $\textrm{dec}\approx 0\degree$.

The inconsistent behavior of the correlated noise in polarization
is consistent with this noise being sourced by array-dependent
T to P leakage. We suspect that relative gain miscalibration is
the main culprit, as it has already been implicated in causing
the low-$\ell$ power loss in the mapmaker transfer function. The
difference in $\ell_\textrm{knee}$ between PA5 f090 and PA6 f090
gives a lower bound on the noise improvements that would be possible
with better calibration, but the total room for improvment is probably
much greater.

It is hard to improve on this in practice for the ACT data. We perform
regular bias steps, but these have proven to be even less reliable
for relative gain calibration than the atmospheric common mode
calibration we currently use. In theory it would be possible to
solve for the per-detector gains jointly with the sky map. We have
tested this for small toy examples, but we know of no algorithm that
can solve this in reasonable time for realistic data sets.

ACT's successor, the Simons Observatory Large Aperture Telescope,
should improve on this situation by having a built-in optical
stimulator that will provide frequent calibration, though even
this approach is vulnerable to percent-level bandpass mismatch between
different detectors.

See \cite{act-map-noise-sims} for more in-depth discussion of
the noise properties of the ACT maps.

\clearpage

\section{Transfer function and $\epsilon$}
\label{sec:epsilon}
\begin{figure}
	\centering
	\begin{tabular}{cc}
		Nearest neighbor subpixel errors & Relative gain errors \\
		\includegraphics[width=0.5\textwidth,trim=13mm 0 0 0]{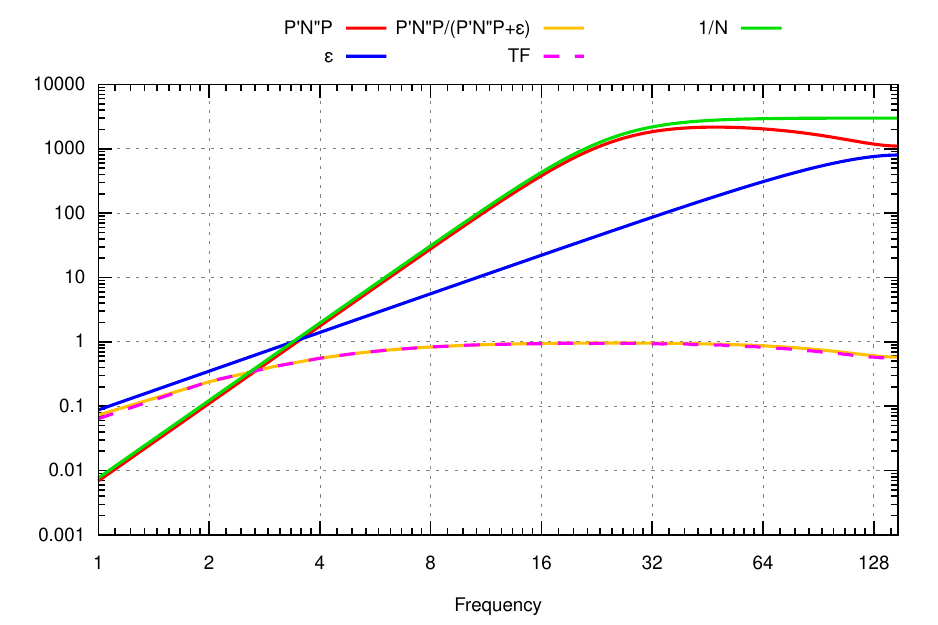} &
		\includegraphics[width=0.5\textwidth,trim=13mm 0 0 0]{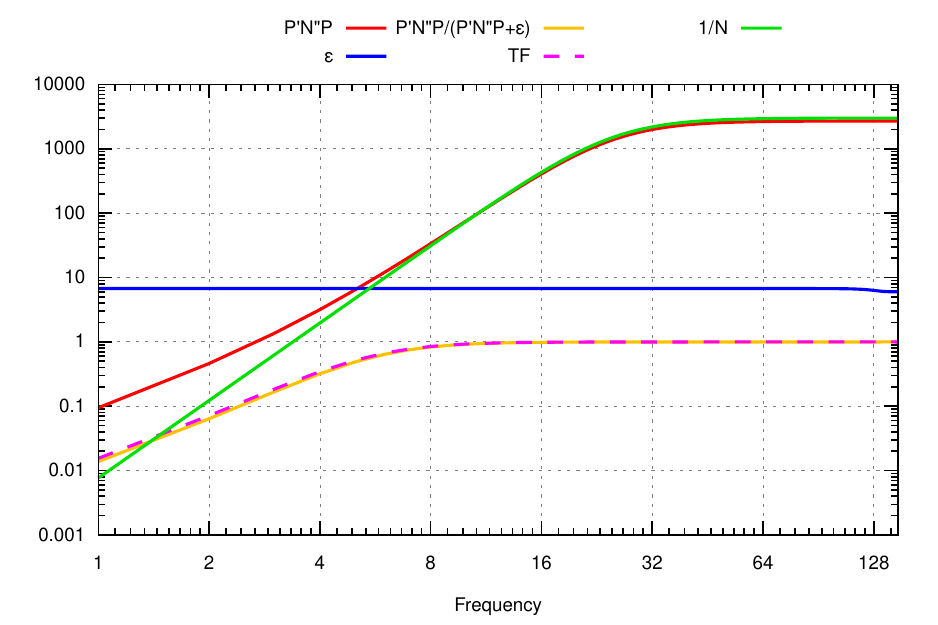}
	\end{tabular}
	\caption{1D toy example showing the relationship between $P^TN^{-1}P$,
	$\epsilon$ and the transfer function. The frequency $f$ is analogous
	to $\ell$. \dfn{Left}: The situation for subpixel errors introduced by
	approximation a smooth signal as a step function with a nearest
	neighbor pointing matrix. $P^TN^{-1}P$ (red) has a
	similar shape as $N^{-1}$ (green), while $\epsilon$ (blue) is shallower,
	causing it to dominate at low $\ell$. The yellow curve implements the
	approximation for the transfer function in equation~\ref{eq:tf}. It
	is an almost perfect match for the exact transfer function (dashed magenta).
	\dfn{Right}: The same but with gain errors instead of subpixel errors.
	Here $\epsilon$ is almost completely flat, making the transfer function
	``turn on'' more suddenly than for subpixel errors. The gain errors simulated
	here varied from sample to sample, simulating the effect of multiple
	co-observing detectors with different calibration.}
	\label{fig:epsilon}
\end{figure}

To illustrate the different scale-dependence of $P^TN^{-1}P$ and $\epsilon$
in equation~\ref{eq:tf}, we implemented a simple 1D toy example consisting of
900 samples scanning at constant speed across 300 pixels. We considered two
cases.
\begin{enumerate}
	\item The true instrument response is given by a linear interpolation
		pointing matrix while the analysis is done assuming a nearest neighbor
		pointing matrix. This mismatch results in sub-pixel model errors.
	\item Both use nearest neighbor (so there is no sub-pixel mismatch
		in this case), but each row of the pointing matrix
		used for the analysis is scaled by 0.9 for even samples and 1.1 for
		odd samples, representing gain miscalibration. These rapid gain
		changes emulate the effect of multiple detectors with relative
		gain errors observing the same spot in the sky in rapid succession.
\end{enumerate}
The results are shown in figure~\ref{fig:epsilon}, and show that
$\epsilon$ has a shallower scale dependence than $P^TN^{-1}P$.
This means that it can be negligible on small scales while still
being important at large scales. The figure also shows that
\begin{align}
	\textrm{TF} \approx \frac{P^TN^{-1}P}{P^TN^{-1}P + \epsilon}
\end{align}
from equation~\ref{eq:tf} is a good approximation for the transfer
function. We recognize the general behavior from figure~\ref{fig:tf-sim}.
The nearest neighbor transfer function turns on more gradually than
the gain error transfer function, consistent with the smaller slope
difference between $P^TN^{-1}P$ and $\epsilon$ for that case.

See \cite{model-error2} for more detailed discussion about model error
induced transfer functions in CMB mapmaking, including how it affects
all mapmaking methods, not just maximum-likelihood.

\section{Peak stacking}
\label{sec:stacking}
We produced the stacks in figures~\ref{fig:stack-T} and \ref{fig:stack-E} by
first finding peaks in a reference map (e.g. the E-mode map
when stacking on E), and then stacking the target map on these locations.
We considered four data sets:

\begin{enumerate}
	\item \dfn{Planck}: Planck 2018 SMICA CMB-only maps
	\item \dfn{ACT}: A naive, per-pixel inverse variance weighted coadd of ACT DR6
		point source subtracted night-only f090 and f150 maps with no beam reconvolution.
		The effective beam is an average of the f090 and f150 beams,
		Beam details on this scale have negligible impact on the
		stack plots.
		\item \dfn{ACT+Planck}: Naive, per-pixel inverse variance weighted coadd of the
		point source subtracted f090 and f150 ACT+Planck coadd maps. Similar beam as ACT-only.
		\item \dfn{Simulation}: A beam-free, noiseless simulation.
\end{enumerate}

The same data set was used both for peak finding and peak stacking,
e.g. when stacking ACT+Planck E on T, peak-finding was done on
ACT+Planck E, and ACT+Planck T was then stacked on these locations.

\subsection{Peak finding}
We find peaks by masking high-foreground and high-noise regions,
apodizing the edge by $0.25\deg$, and low-pass filtering to
remove noise-dominated modes. We then found local maxima and
minima as using \verb|scipy.ndimage.maximum_filter| and
\verb|scipy.ndimage.minimum_filter| with a 2-pixel radius
circular footprint. We label these as ``positive'' and ``negative''
peaks respectively.\footnote{This refers to the sign of a peak
relative to its local surroundings, not the absolute value of the
map at this spot.} The band-pass filter was a Butterworth
profile
\begin{align}
	F(\ell) &= (1+(\ell/\ell_\textrm{knee})^{5})^{-1} .
\end{align}
For ACT and the simulation, we used an $\ell_\textrm{knee}$ of
3000/2000 for T/E, while for Planck we used 2000/800 due to
lower S/N. No beam deconvolution was performed. The result
was a list of around 300\,000 peaks in T and 450\,000 peaks
in E.

\subsection{Peak stacking}
We apply the same mask to the map to stack. For each peak,
we reproject a $4\deg\times4\deg$ thumbnail around it
to RA=$0\deg$, dec=$0\deg$ using non-uniform FFT interpolation\footnote{
	When interpolating it's important that there's at least one output
	pixel for every input pixel to avoid losing information. The maximal
	resolution in the input map is $0.5\arcmin/cos(-60\deg) = 0.25\arcmin$,
	so we reproject to $0.25\arcmin$ resolution.}
while taking into account the per-pixel polarization angle rotation
caused by the parallel transport. We define the positive/negative
stack map as the plain, unweighted average of the positive/negative
reprojected thumbnails. The full stack map is then
\begin{align}
	\textrm{stack}_\textrm{full} &= \frac12(\textrm{stack}_\textrm{pos}-\textrm{stack}_\textrm{neg})
\end{align}
This stack map suffers from large-scale horizontal striping with two causes:
\begin{enumerate}
	\item The ACT-only map contains horizontal stripes from azimuth-synchronous
		pickup (see figure~\ref{fig:map-raw-and-filtered}).
	\item The ACT+Planck map lacks power for $|\ell_x|<5$ due to a
		pickup filter applied to the ACT contribution to the coadd. This
		results in a bias towards detecting positive peaks in areas with
		mostly negative signal on the same declination. When stacking
		even an unfiltered map using these peaks, the result is a negative
		horizontal stripe around a stack on positive peaks.
\end{enumerate}
These horizontal stripes are a distraction from the baryon-acoustic
osciallation physics the stacks illustrate, so we remove them by
subtracting the average value in each row of the map, excluding
the $|x|<1.67\deg$ region to avoid biasing the BAO signal itself.

For plotting purposes, we convert the linear polarization Stokes parameters
Q and U into E and B. We also show the related quantities $Q_r$ and $U_r$,
which are Simply Q and U defined relative to the stacking center instead
of the North pole \citep{cmb-polarization-1997}.

Finally, we normalize each plot T plot by dividing it by the
95\% quantile of the absolute value of the T signal for $0.9\deg<r<1.3\deg$.
Similarly, $Q_r$ and $U_r$ are normalized using $Q_r$ and E and B
using E.

\clearpage

\end{document}